\documentclass[11pt,a4paper]{article}
\usepackage[T1]{fontenc}    % Accents cods dans la fonte.
\usepackage{graphicx}
\usepackage{pstricks,pst-coil,pst-fill,pst-plot}
\usepackage[fleqn]{amsmath}    % Les symboles les plus frquents.
\usepackage{amssymb}    % Des symboles.
\usepackage{amsfonts}   % Des fontes, eg pour \mathbb.
\usepackage{verbatim}   % Pour les codes sources en informatique.
\usepackage{mathrsfs}   % Des lettres majuscules cursives (\mathscr).
\usepackage{dsfont}
\usepackage{euscript}
\usepackage{yfonts}
\usepackage{enumerate}     % met en place l'environement enumerate pour les listes
\usepackage{txfonts}
\usepackage{marvosym}
\usepackage{vmargin}        % Rgler la taille de la feuille.

\usepackage[normalem]{ulem}
\usepackage{microtype}

\setmarginsrb{1.8cm}{2cm}{1.8cm}{2cm}{1cm}{1cm}{1cm}{1.6cm}
 \makeatletter
 \@addtoreset{equation}{section}
 \makeatother

\providecommand{\bysame}{\leavevmode\hbox to3em{\hrulefill}\thinspace}
\providecommand{\MR}{\relax\ifhmode\unskip\space\fi MR }
\providecommand{\href}[2]{#2}

\newcommand{\bra}[1]{\big< \,#1\,\big|}
\newcommand{\ket}[1]{\big|\,#1\, \big>}

\let\ua=\uparrow
\let\da=\downarrow
\let\tend=\rightarrow

\long\def\symbolfootnote[#1]#2{\begingroup%
\def\thefootnote{\fnsymbol{footnote}}\footnote[#1]{#2}\endgroup}

\newtheorem{theorem}{Theorem}[section]
\newtheorem{prop}{Proposition}[section]

\newtheorem{lemme}{Lemma}[section]

\def\Proof{\medskip\noindent {\it Proof --- \ }}

\def\qed{\hfill\rule{2mm}{2mm}}

\newcommand\beq{\begin{equation}}
\newcommand\enq{\end{equation}}
\newcommand\bem{\begin{multline}}
\newcommand\enm{\end{multline}}

\def\beqa{\begin{eqnarray}}
\def\eeqa{\end{eqnarray}}
\def\ba{\begin{array}}
\def\ea{\end{array}}
\def\det{\operatorname{det}}

\newcommand{\f}[2]{{\ensuremath{%
    \mathchoice%
    {\dfrac{#1}{#2}}
    {\dfrac{#1}{#2}}
    {\frac{#1}{#2}}
    {\frac{#1}{#2}}
}}}
\newcommand{\tf}[2]{\ensuremath{#1/#2}}
\def\a{\alpha}

\def\ga{\gamma}
\def\Ga{\Gamma}

\def\de{\delta}

\def\De{\Delta}
\def\eps{\epsilon}
\def\veps{\varepsilon}
\def\la{\lambda}

\def\sg{\sigma}
\def\vsg{\varsigma}

\def\Ups{\Upsilon}
\def\th{\theta}

\def\om{\omega}
\def\vp{\varphi}

\def\vsg{\varsigma}

\def\i{\mathrm{i}}

\newcommand{\mc}[1]{\ensuremath{\mathcal{#1}}}
\newcommand{\mf}[1]{\ensuremath{\mathfrak{#1}}}
\newcommand{\msc}[1]{\ensuremath{\mathscr{#1}}}

\newcommand{\bs}[1]{\ensuremath{\boldsymbol{#1}}}
\newcommand{\ov}[1]{\ensuremath{\overline{#1}}}
\newcommand{\wt}[1]{\ensuremath{\widetilde{#1}}}
\newcommand{\wh}[1]{\ensuremath{\widehat{#1}}}

\newcommand{\Int}[2]{\ensuremath{\int\limits_{#1}^{#2}}}
\newcommand{\Oint}[2]{\ensuremath{\oint\limits_{#1}^{#2}}}
\newcommand{\Fint}[2]{\ensuremath{\fint\limits_{#1}^{#2}}}

\newcommand{\sul}[2]{\ensuremath{\sum\limits_{#1}^{#2}}}
\newcommand{\pl}[2]{\ensuremath{\prod\limits_{#1}^{#2}}}

\newcommand{\R}{\ensuremath{\mathbb{R}}}
\newcommand{\Cx}{\ensuremath{\mathbb{C}}}

\newcommand{\Dp}[1]{\ensuremath{\partial_{#1}}}
\newcommand{\ex}[1]{\ensuremath{\e{e}^{#1}}}

\newcommand{\norm}[1]{\ensuremath{|| #1 ||}}

\newcommand{\moy}[1]{\ensuremath{\langle #1 \rangle}}

\newcommand{\dd}{\mathrm{d}}
\newcommand{\e}[1]{\ensuremath{\mathrm{#1}}}

\newcommand{\intff}[2]{\ensuremath{ [  #1 \,; #2 ] }}
\newcommand{\intfo}[2]{\ensuremath{ [  #1 \,; #2 [ }}
\newcommand{\intof}[2]{\ensuremath{ ]  #1 \,; #2 ] }}
\newcommand{\intoo}[2]{\ensuremath{ ]  #1 \,; #2 [ }}

\newcommand{\intn}[2]{\ensuremath{[\![ \, #1 \,;\, #2 \,]\!]}}

\newcommand{\qprod}[1]{(#1;q^4)_\infty}
\newcommand{\qtriprod}[1]{(#1;q^4,q^4)_\infty}

\begin{document}

%\begin{flushright}

%\end{flushright}
%\par \vskip .1in \noindent

%\vspace{14pt}

\begin{center}
\begin{LARGE}
{\bf On form-factor expansions for the XXZ chain
in the massive regime}
\end{LARGE}

\vspace{30pt}

\begin{large}

{\large
Maxime Dugave\footnote{e-mail: dugave@uni-wuppertal.de},
Frank G\"{o}hmann\footnote{e-mail: goehmann@uni-wuppertal.de}}%
\\[1ex]
Fachbereich C -- Physik, Bergische Universit\"at Wuppertal,\\
42097 Wuppertal, Germany.\\[2.5ex]

{\large Karol K. Kozlowski\footnote{e-mail: karol.kozlowski@u-bourgogne.fr}}%
\\[1ex]
IMB, UMR 5584 du CNRS,
Universit\'e de Bourgogne, France. \\[2.5ex]

{\large Junji Suzuki\footnote{e-mail: sjsuzuk@ipc.shizuoka.ac.jp}}%
\\[1ex]
Department of Physics, Shizuoka University,\\
Ohya 836, Shizuoka,  Japan.

\par

\end{large}

\vspace{40pt}

\centerline{\bf Abstract} \vspace{1cm}
\parbox{12cm}{\small
We study the large-volume-$L$ limit of form factors of the longitudinal
spin operators for the XXZ spin-$\tf{1}{2}$ chain in the massive regime.
We find that the individual form factors decay as $L^{-n}$, $n$ being
an even integer counting the number of physical
excitations -- the holes -- that constitute the excited state. 
Our expression  allows us to derive  
the form-factor expansion of two-point spin-spin correlation functions
in the thermodynamic limit $L\tend +\infty$. The staggered magnetisation
appears naturally as the first term in this expansion. We show that all other
contributions to the two-point correlation function are exponentially
small in the large-distance regime.}

\end{center}

\vspace{40pt}

\section*{Introduction}

The study of form factors in integrable massive quantum field theories  goes back to the late `70s when the
bootstrap program was proposed by Karowski and Weiss
\cite{KarowskiWeiszFormFactorsFromSymetryAndSMatrices}.
It was then supplemented  with an additional axiom  by Kirillov and Smirnov 
in \cite{KirillovSmirnovFirstCompleteSetBootstrapAxiomsForQIFT}.
The resolution of this bootstrap program within the off-shell Bethe Ansatz resulted in multiple-integral based representations for the densities of form
factors in numerous models (see \cite{SmirnovFormFactors} and references therein).
These led to  series of multiple integrals for the two- and multi-point correlation
functions which appeared to be efficient tools for extracting the exponentially
decaying long-distance asymptotic behaviour of multi-point correlators.

Later, in the mid `90s, Jimbo and Miwa  proposed a representation-theoretic approach to the
calculation of form factors of the spin-$\frac{1}{2}$ XXZ chain in its massive regime \cite{JimboMiwaFormFactorsInMassiveXXZ}.
They employed $q$-vertex operators in the diagonalisation of the infinite XXZ chain and applied them to the evaluation of the form factors of the operators
$\sg^z$ and $\sg^{\pm}$. The $q$-vertex operator approach has been used to
calculate form factors and correlation functions of higher-rank and
higher-spin lattice models and was also successfully applied to the
calculation of form factors of the XYZ chain \cite{LashkevichElemBlocksEightVertex,LashkevichPugaiFormFactorsEightVertex}. The latter result enabled the use of form factors in the
calculation of correlation functions of critical lattice models
\cite{CauxKonnoSorrelWestonFFofMasslessXXZfromXYZResults,LukyanovTerrasSpinSpinAmplitudesBetterTreatementXXZ}. So far the
 $q$-vertex operator approach is limited in that it does not
allow one to calculate form factors and correlation functions at finite
magnetic fields or finite lengths or temperatures.

The latter is possible within the algebraic Bethe
Ansatz approach. A key element of this approach is the Slavnov formula \cite{SlavnovScalarProductsXXZ}
for the scalar product between an on-shell and an off-shell Bethe vector. It paved the way  for the computation of
form factors in  quantum integrable lattice models. 
%
%Also, the so-called Slavnov formula \cite{SlavnovScalarProductsXXZ} for the scalar product between an on-shell and an off-shell Bethe vector
%paved the way to the computation of form factors in lattice quantum integrable models starting from the algebraic Bethe Ansatz. 
%Indeed, this result allowed
It was  this formula that allowed Slavnov to derive a representation for the form  factors of the current operator
in the non-linear Schr\"{o}dinger model \cite{SlavnovFormFactorsNLSE}.
These form factors were represented as ratios of determinants
whose size grows with the volume $L$. Slavnov performed
the large-$L$ analysis of his expression and extracted the inverse power-law  in $L$  
behaviour  of the form factors in the $L\tend + \infty$ limit. 
 He also proved the cancellation of divergent
terms in $L$ in the form-factor series up to the first correction
in the interaction strength. 

A first calculation of form factors of a quantum integrable
lattice model by means of the algebraic Bethe Ansatz approach became possible
after the resolution of the so-called quantum-inverse scattering problem by Kitanine,
Maillet and Terras \cite{KMTFormfactorsperiodicXXZ} in 1999. These authors derived
finite-size determinant representations
for the form factors of local operators in the finite-length spin-$\frac{1}{2}$ XXZ chain. 
In the same year Izergin, Kitanine, Maillet and Terras \cite{IzerginKitMailTerSpontaneousMagnetizationMassiveXXZ}
were able to analyse the large-$L$ limit of a special form
factor representing the spontaneous staggered magnetisation
\cite{BaxterStaggeredPolarisationInFModel,JimboMiwaFormFactorsInMassiveXXZ}
of the XXZ spin-$\frac{1}{2}$ chain in the massive regime. More recently
form factors for the SU(3)-invariant spin chain \cite{BelliardPakuliakRagoucySlavnovNestedBAScalarProductsSU(3)} and for the cyclic solid-on-solid
model \cite{Levy-BenchetonTerrasMultiPtLocHeightProbForCSOSinABA} were calculated within
the algebraic Bethe Ansatz approach.

The interest in form factors of quantum integrable lattice models was recently renewed after the authors of
\cite{KozKitMailSlaTerEffectiveFormFactorsForXXZ,KozKitMailSlaTerThermoLimPartHoleFormFactorsForXXZ} managed to
extract the large-volume behaviour of the particle-hole form factors of the XXZ spin-1/2 chain in the massless regime.
These formulae, along with so-called restricted sum summation, led to a novel approach to the
large-distance and long-time asymptotic behaviour of correlation
functions in massless quantum integrable models
\cite{KozKitMailSlaTerRestrictedSums,KozKitMailSlaTerRestrictedSumsEdgeAndLongTime,KozKitMailTerMultiRestrictedSums}. 
The same method works in infinite volume but at finite temperatures
\cite{KozDugaveGohmannThermaxFormFactorsXXZ,KozDugaveGohmannThermaxFormFactorsXXZOffTransverseFunctions,KozMailletSlaLowTLimitNLSE}. 
%
%---------------------------------------------------------------------

In the present work we address the problem of the calculation of the large-$L$ behaviour of the form factors of local operators in the XXZ spin-$\frac{1}{2}$ chain
in its massive regime by the algebraic Bethe Ansatz approach.
We utilize a finite-determinant representation of the form factors of the finite chain which was derived in
\cite{KozKitMailSlaTerXXZsgZsgZAsymptotics} for the purpose of the large-$L$ analysis of form factors in the massless regime.
This representation follows from \cite{KMTFormfactorsperiodicXXZ} and turns out to be useful in the massive regime as well.

Recall that the ground state $|G.S.\rangle$ of the XXZ spin-$\frac{1}{2}$
chain in the massive regime belongs to the zero magnetisation sector
\cite{Yang-YangXXZproofofBetheHypothesis}. In the large $L$ limit, the
excited states $\ket{ \{\chi_a\}_1^{n_{\chi}} ; \{\nu_{h_a}\}_1^{n_h}  }$
having a \textit{finite} excitation energy above the ground state 
are parametrised by `hole'-parameters $\{\nu_{h_a}\}_1^{n_h}$ and
complex roots $\{\chi_a\}_1^{n_{\chi}}$ which solve a set of higher-level
Bethe Ansatz equations \cite{BabelondeVegaVialletStringHypothesisWrongXXZ,%
VirosztekWoynarovichStudyofExcitedStatesinXXZHigherLevelBAECalculations} of the form
\beq
\mc{Y}_{0}\big(\, \chi_a\mid  \{\chi_a\}_1^{n_{\chi}} ;  \{\nu_{h_a}\}_1^{n_h} \big) \, = \, 0 \; , \;  a=1,\dots, n_{\chi} \; . 
\enq
Note that the complex roots $\{\chi_a\}_1^{n_{\chi}}$ arise as a reduced
number of variables parametrising the string-like and wide-pair solutions
of the original Bethe equations \cite{BetheSolutionToXXX}. 
 As a result of our asymptotic analysis of the determinant expressions for the form factors we obtain
the leading large-$L$ behaviour of form factors of spin operators. 
The resulting expression admits an interpretation in terms of a form-factor density
$\mc{F}^{(z)}\big( \{\nu_{h_a}\}_1^{n_h} ; \{\chi_a\}_1^{n_{\chi}} \big)$:
\beq
\Big|\big< G.S.\big| \sg^z \big| \{\chi_a\}_1^{n_{\chi}} ;  \{\nu_{h_a}\}_1^{n_h}  \big> \Big|^2 \, = \, \pl{a=1}{n_h}\bigg\{ \f{1}{L p^{\prime}(\nu_{h_a} ) } \bigg\}
\cdot \f{ \Big( \mc{F}^{(z)}\big( \{\nu_{h_a}\}_1^{n_h} ; \{\chi_a\}_1^{n_{\chi}} \big)\, \Big)^2  }
{ \det_{n_{\chi}} \Big[  \f{ \Dp{} }{ \Dp{} u_b } \mc{Y}_{0}\big( u_a\mid  \{u_c\}_1^{n_{\chi}} ;  \{\nu_{h_a}\}_1^{n_h} \big)  \Big]_{\mid u_a=\chi_a}  } \cdot \Big(1\, + \, \e{O}\big( L^{-1} \big)  \Big)\;. 
\enq
Here the function $p^{\prime}$ represents the density of real roots in the interval $\intff{-\tf{\pi}{2}}{\tf{\pi}{2}}$,
while $p$ itself is the dressed momentum of the hole excitations. 
The Jacobian of the function $\mc{Y}_{0}$ generating the higher-level
Bethe Ansatz equations can be thought of as a higher-level expression
for the norm of an excited state. 

Our approach provides a form-factor expansion of two-point time- and space-dependent correlation functions in
the infinite-volume limit. The first term arising in the series corresponds to the staggered magnetisation
\cite{BaxterStaggeredPolarisationInFModel,JimboMiwaFormFactorsInMassiveXXZ}.
Interestingly,  the form-factor series we obtain takes a form that is structurally
different from the one obtained within the vertex-operator approach. 
The equivalence of these expressions is yet to be proven.
 A numerical comparison, however, indicates that the form-factor density we
obtain for $n_h=2$ and $n_{\chi}=1$ does match with that obtained within  the vertex-operator approach. 
We believe that the equality holds, in fact, for arbitrary numbers of hole excitations $n_h$.

The paper is organised as follows.  In Section \ref{Section analysis of BAE} we revisit the analysis
\cite{BabelondeVegaVialletStringHypothesisWrongXXZ,VirosztekWoynarovichStudyofExcitedStatesinXXZHigherLevelBAECalculations}
of the large-$L$ behaviour of the Bethe roots parametrising the
low-lying excitations above the ground state.  We use the non-linear integral equation that drives the  counting function associated with
 a given excited state to carry out a careful analysis of the $\tf{1}{L}$ corrections and to characterise the extra
Bethe roots describing the excited states in terms of a system of higher-level Bethe Ansatz equations. This allows
us to clarify some controversial fine point in the original literature \cite{BabelondeVegaVialletStringHypothesisWrongXXZ,VirosztekWoynarovichStudyofExcitedStatesinXXZHigherLevelBAECalculations}.
In Section \ref{Section Large L behaviour of Form Factors} we compute the large-volume behaviour of form factors. Finally, in Section \ref{Section FF series expansion}, we apply our results to write down
the form-factor expansion of two-point functions at distance $m$. We identify the first term of
the series with the staggered magnetisation and argue that, in the large-$m$ limit, all other terms 
in the form-factor series are exponentially small. We also provide some details related to the comparison of our formulae with those
obtained by using the vertex-operator formalism. In Appendix \ref{Appendix Explicit resolution LIE}, we gather technical results related to the special functions that arise in the
description of the thermodynamic limit of the XXZ spin-$\frac{1}{2}$ chain in its
massive regime.

%%%%%%%%%%%%%%%%%%%%%%%%%%%%%%%%%%%%%%%%%%%%%%%%%%%%%%%%%%%%%%%%%%%%%%%%%%%%%%%%%%%%%%%%%%%%%%%%%%%%%%%%%%%%%%%%%%%%%%%%%%%%%%%%%%%%%%%%%%%%%%%%%%%%%%%%%%%%%%%%%%%%%%%%%
%%%%%%%%%%%%%%%%%%%%%%%%%%%%%%%%%%%%%%%%%%%%%%%%%%%%%%%%%%%%%%%%%%%%%%%%%%%%%%%%%%%%%%%%%%%%%%%%%%%%%%%%%%%%%%%%%%%%%%%%%%%%%%%%%%%%%%%%%%%%%%%%%%%%%%%%%%%%%%%%%%%%%%%%%

%%%%%%%%%%%%%%%%%%%%%%%%%%%%%%%%%%%%%%%%%%%%%%%%%%%%%%%%%%%%%%%%%%%%%%%%%%%%%%%%%%%%%%%%%%%%%%%%%%%%%%%%%%%%%%%%%%%%%%%%%%%%%%%%%%%%%%%%%%%%%%%%%%%%%%%%%%%%%%%%%%%%%%%%%
%%%%%%%%%%%%%%%%%%%%%%%%%%%%%%%%%%%%%%%%%%%%%%%%%%%%%%%%%%%%%%%%%%%%%%%%%%%%%%%%%%%%%%%%%%%%%%%%%%%%%%%%%%%%%%%%%%%%%%%%%%%%%%%%%%%%%%%%%%%%%%%%%%%%%%%%%%%%%%%%%%%%%%%%%

%%%%%%%%%%%%%%%%%%%%%%%%%%%%%%%%%%%%%%%%%%%%%%%%%%%%%%%%%%%%%%%%%%%%%%%%%%%%%%%%%%%%%%%%%%%%%%%%%%%%%%%%%%%%%%%%%%%%%%%%%%%%%%%%%%%%%%%%%%%%%%%%%%%%%%%%%%%%%%%%%%%%%%%%%
%%%%%%%%%%%%%%%%%%%%%%%%%%%%%%%%%%%%%%%%%%%%%%%%%%%%%%%%%%%%%%%%%%%%%%%%%%%%%%%%%%%%%%%%%%%%%%%%%%%%%%%%%%%%%%%%%%%%%%%%%%%%%%%%%%%%%%%%%%%%%%%%%%%%%%%%%%%%%%%%%%%%%%%%%

%%%%%%%%%%%%%%%%%%%%%%%%%%%%%%%%%%%%%%%%%%%%%%%%%%%%%%%%%%%%%%%%%%%%%%%%%%%%%%%%%%%%%%%%%%%%%%%%%%%%%%%%%%%%%%%%%%%%%%%%%%%%%%%%%%%%%%%%%%%%%%%%%%%%%%%%%%%%%%%%%%%%%%%%%
%%%%%%%%%%%%%%%%%%%%%%%%%%%%%%%%%%%%%%%%%%%%%%%%%%%%%%%%%%%%%%%%%%%%%%%%%%%%%%%%%%%%%%%%%%%%%%%%%%%%%%%%%%%%%%%%%%%%%%%%%%%%%%%%%%%%%%%%%%%%%%%%%%%%%%%%%%%%%%%%%%%%%%%%%

\section{The structure of excited states revisited}
\label{Section analysis of BAE}

The XXZ spin-$\frac{1}{2}$ model is described by the Hamiltonian

\beq
H \; = \; J \sul{n=1}{L} \Big\{  \sg_n^x \sg_{n+1}^{x} \, + \, \sg_n^y \sg_{n+1}^{y} \, + \,  \De \sg_n^z \sg_{n+1}^{z} \Big\}
\; - \; \f{ h }{ 2 } \sul{n=1}{L} \sg_n^z \;. 
\enq
Here $J>0$ is a coupling constant measuring the strength of the exchange interaction, $\De$ is the longitudinal anisotropy 
in the couplings, $h$ is an external magnetic field and the $\sg_n^{a}$ correspond to Pauli matrices  understood as 
acting non-trivially on the $n^{\e{th}}$-quantum space $V_n \simeq \Cx^2$ in the tensor product decomposition 
$\otimes_{n=1}^{L} V_n$ of the Hilbert space on which $H$ acts. We will focus on the massive regime of the chain $\De>1$ and $0<h<h_c$
with $h_c$ given by \eqref{definition chp mah critique} and corresponding to the critical value of the magnetic field above which the model becomes massless. 
We shall assume that the length $L$ of the chain is even. In
this way we avoid having to distinguish ground states of total spin projection 1/2 and -1/2.

The eigenvectors of this Hamiltonian were first constructed within the coordinate
Bethe Ansatz \cite{BetheSolutionToXXX,OrbachXXZCBASolution} and  later also by means of
the algebraic Bethe Ansatz \cite{BabelondeVegaVialletStringHypothesisWrongXXZ,FaddeevTakhtadzhanXXXgeneralOverwiev}.

\subsection{The Bethe equations of the massive XXZ chain and the counting function}

Within the Bethe Ansatz approach, the eigenvectors $\ket{ \psi\big( \{\mu_a\}_1^N\big) }$ of the XXZ Hamiltonian are 
parametrised by a set of $N$ roots $\{ \mu_a \}_1^N$. The integer $N$ is related to the spin sector to which the 
eigenvector belongs. 
It is well known that these roots satisfy a set of algebraic equations, that are now referred to as the Bethe Ansatz equations \cite{BetheSolutionToXXX,OrbachXXZCBASolution}. 
In the massive regime of the XXZ spin-$\tf{1}{2}$ chain, $\De=\cosh(\eta) > 1$ with $\eta>0$,
these equations take the form 
\beq
-\ex{2 \i \pi \a} \; = \; \bigg( \f{ \sin\big(\mu_a - \i \tf{\eta}{2} \big)  }{ \sin\big(\mu_a + \i \tf{\eta}{2} \big) } \bigg)^L
\cdot \pl{b=1}{N}  \f{ \sin\big(\mu_a -\mu_b + \i \eta \big)  }{ \sin\big(\mu_a  -\mu_b - \i \eta \big) } \; .
\label{ecriture BAE}
\enq
%
%
%
%where one should set $\a=0$. 
%
The equation \eqref{ecriture BAE}, \textit{per se}, contains the so-called twist parameter  $\ex{2\i\pi \a}$.
This system of equations is often referred to as the system of $\a$-twisted Bethe Ansatz equations.  
The additional parameter $\alpha$ is zero for the original problem. However, it will turn
out to be useful for the calculation of form factors. It has been established in \cite{VladimirovProofConjInvSolBetheEqns} that, for $\a \in \R $, any solution $\{ \mu_a \}_1^N$ to the Bethe Ansatz
equations \eqref{ecriture BAE} is invariant under complex conjugation, \textit{viz}. $\{ \ov{\mu}_a \}_1^N \, = \, \{ \mu_a \}_1^N$. 
Also, in the following, we shall assume that the roots $\mu_a$ are always pairwise distinct. This property is known to hold in integrable models
with repulsive interactions (see, \textit{e}.\textit{g}. \cite{BogoliubiovIzerginKorepinBookCorrFctAndABA})
such as the $\de$-function Bose gas \cite{LiebLinigerCBAForDeltaBoseGas} in the repulsive regime. Still, to the best of our knowledge, it remains an
open question whether such property holds in the case of the XXZ chain.

Given a set of Bethe roots $\{ \mu_a \}_1^N$ satisfying \eqref{ecriture BAE},
it is convenient to introduce its associated counting function
\beq
\wh{\xi}_{\mu}(\om) \; = \; \f{ p_0(\om) }{ 2\pi }  \; - \; \f{1}{2 \i \pi L } \sul{k=1}{N} \th(\om-\mu_k)
\; + \; \f{ 1-2\a  }{ 2L } \;. 
\label{ecriture representation originale pour xi kappa}
\enq
%
%For later use, we also introduce another  counting function $\wh{\xi}_{\la}(\om)$ associated to the ground state
%Bethe roots $\{ \l_a \}_1^N$ which solve (\ref{ecriture BAE}) with $\alpha=0$.
%
%\beq
%\wh{\xi}_{\la}(\om) \; = \; \f{ p_0(\om) }{ 2\pi }  \; - \; \f{1}{2 \i \pi L } \sul{k=1}{N} \th(\om-\la_k)
%
%\; + \; \f{ 1  }{ 2L } \;. 
%\enq
%
%
The expression for $\wh{\xi}_{\mu}$ involves two auxiliary functions: the bare phase $\th$ and the  bare momentum $p_0$ whose definitions read
\beq
\th(\la) \; = \; 2\i \pi \Int{ - \tf{\pi}{2} }{ \la } K(\mu) \cdot \dd \mu \qquad \e{and} \qquad
p_0(\la) \; = \;  \Int{ - \tf{\pi}{2} }{ \la } p_0^{\prime}(\mu) \cdot \dd \mu \;. 
\label{definition phase et moment nu}
\enq
The integration in the definition of $\th$ and $p_0$ runs along the oriented segments
\beq
\intff{-\tf{\pi}{2} }{ \i \Im(\la) - \tf{\pi}{2}  } \cup \intff{ \i \Im(\la) - \tf{\pi}{2} }{ \la }  
\enq
and the integrands are given by 
\beqa
K(\mu) & = & \f{ \sinh(2\eta) }{ 2\pi \sin(\mu + \i\eta) \sin(\mu - \i \eta)  } \; = \; 
\f{\cot\big(\mu - \i \eta)  \, - \, \cot\big(\mu + \i \eta) }{2 \i \pi} \; ,  \\
p_0^{\prime}(\mu) & = & \f{ \sinh(\eta) }{ \sin(\mu + \i \tf{\eta}{2}) \sin( \mu - \i \tf{\eta}{2})  } \; = \; 
\f{\cot\big(\mu - \i \tf{\eta}{2})  \, - \, \cot\big(\mu + \i \tf{\eta}{2}) }{ \i } \;. 
\eeqa
%
%
%
%Note that a 
A different choice of the integration contour may change the values of $\th$ and $p_0$ by multiples of $2\i\pi$. 
Still, the above definitions of $\th$ and $p_0$ do imply that 
\begin{itemize}
\item for $-\eta < \Im(\la) < \eta $ (resp. $-\tf{\eta}{2} < \Im(\la) < \tf{\eta}{2} $) 
the function $\th$ (resp. $p_0$) is quasi-periodic $\th(\la +  \pi) = \th(\la) + 2 \i \pi $ and quasi-odd $\th(-\la)=2\i \pi - \th(\la)$
(resp. $p_0(\la +  \pi) = p_0(\la) + 2\pi $ and $p_0(-\la ) =  2\pi - p_0(\la)  $);
\item  for $|\Im(\la)|> \eta $ (resp. $|\Im(\la)|> \tf{\eta}{2} $), 
the function $\th$ (resp. $p_0$) becomes periodic $\th(\la +  \pi) = \th(\la)$ and odd $\th(-\la) = - \th(\la)$ 
(resp. $p_0(\la +  \pi) = p_0(\la)$ and $p_0(-\la ) =  - p_0(\la)  $).  
\end{itemize}

\noindent The quasi-periodicity properties of the bare phase and bare momentum ensure that, for any $x \in \R$, 
\beq
\wh{\xi}_{\mu}\big(x+\tf{\pi}{2} \big) \, - \, \wh{\xi}_{\mu}\big( x -\tf{\pi}{2} \big) \; = \; 
\f{ N+n_w}{ L} 
\label{ecriture quasiper prop counting fct}
\enq
where we restricted ourselves to the zero total longitudinal spin sector corresponding to $N=\tf{L}{2}$.
The number of Bethe roots $\mu_a$ such that $|\Im(\mu_a)|>\eta$ is denoted by $n_w$.

The main advantage of the counting function is that it enables us to recast the Bethe Ansatz equations \eqref{ecriture BAE} 
in a very simple form 
\beq
\ex{2 \i \pi L \wh{\xi}_{\mu}(\mu_a) } \; = \; 1 \;. 
\label{ecriture eqn Bethe ac fct cptge}
\enq
The important feature is that, on the level of \eqref{ecriture eqn Bethe ac fct cptge}, one deals with an equation in one variable. 
Hence,  once a characterization of $\wh{\xi}_{\mu}$ alternative to \eqref{ecriture representation originale pour xi kappa} is available, equation \eqref{ecriture eqn Bethe ac fct cptge} provides 
an effective means for computing the roots $\mu_a$. By following the techniques pioneered in 
\cite{BatchelorKlumperFirstIntoNLIEForFiniteSizeCorrectionSpin1XXZAlternativeToRootDensityMethod,DeVegaWoynarowichFiniteSizeCorrections6VertexNLIEmethod}
and further developed in  \cite{DestriDeVegaAsymptoticAnalysisCountingFunctionAndFiniteSizeCorrectionsinTBAFirstpaper,KlumperBatchelorPearceCentralChargesfor6And19VertexModelsNLIE}
we shall set forth a non-linear integral equation satisfied by $\wh{\xi}_{\mu}$. This equation, with some additional input on $\wh{\xi}_{\mu}$ can be
easily solved in the $L\tend +\infty$ limit. In particular, its structure determines the large-$L$ form of the distribution of the roots $\{\mu_a\}_1^N$. 

\subsection{\boldmath Large-$L$ behaviour of the counting function}
\label{Subsection Asympt behaviour counting function}

We shall carry out the large-$L$ asymptotic analysis of $\wh{\xi}_{\mu}$
under the following hypotheses:
\begin{itemize}
\item
the set $\{\mu_a\}_1^N$ contains precisely $n$, $n$ being fixed and independent
of $L$, complex roots $\{z_a\}_1^n$ with non-zero imaginary part; 
%
%
%\item \textcolor{blue}{\sout{these complex roots are such that $\min_{a}
%|\Im(z_a)|>\tau >0$ for some $L$-independent constant $\tau$;}
%[fg: I suggest to delete this, since it is a
%consequence of the other two points (see (1.24))]}
%
\item the counting function $\wh{\xi}_{\mu}$ given in \eqref{ecriture representation originale pour xi kappa}
is strictly increasing on $\intff{-\tf{\pi}{2} }{ \tf{\pi}{2} }$
and its derivative $\wh{\xi}_{\mu}^{\prime}$
is bounded from below by an $L$-independent constant $\varkappa$, %\textit{viz}. 
$\wh{\xi}_{\mu}^{\prime}(\la) > \varkappa >0$  for any $\la \in \intff{-\tf{\pi}{2}}{\tf{\pi}{2}}$. 
\end{itemize}

The first hypothesis simply expresses
the fact that we restrict the analysis to some subset of solutions to the
Bethe Ansatz equations. The second hypothesis ensures that the real roots
can be unambiguously parametrised in terms of integers. It also guarantees
that there exists a small but L-independent neighbourhood $U$ of $\R$ in
$\Cx$  such that $ \wh{\xi}_{\mu}$ maps $U\cap \mathbb{H}^{\pm}$ into
$\mathbb{H}^{\pm}$, hence guaranteeing that there is no complex roots
$z_a$ in U. In other words, there exists an $L$-independent constant $\tau$
such that $\min_{a} |\Im(z_a)|>\tau >0$. We conjecture that, in fact, 
the second hypothesis is a consequence of the first one. We have not been
able to prove such a statement. We have checked, however, that the above
hypotheses hold \textit{a posteriori}, on the level of the answer we obtain. 

The quasi-periodicity of $\wh{\xi}_{\mu}$ adjoined to its strict
increase on $\intff{ -\tf{\pi}{2} }{ \tf{\pi}{2} }$ implies that the counting
function is strictly increasing on $\R$. As a consequence, for every
$\iota \in {\mathbb Z}$ there exists a unique $x_{\mu}$ such that 
\beq
\wh{\xi}_{\mu}\big( x_{\mu}-\tf{\pi}{2} \big)=\frac{1 - 2 \iota}{2L}.
\label{firstroot} 
\enq
Likewise, it follows from the above hypotheses and from the
quasi-periodicity \eqref{ecriture quasiper prop counting fct} 
that the equation $\ex{  2 \i \pi L \wh{\xi}_{\mu}(\om) } = 1$
admits $N+n_{w}$ real roots $\{\nu_a\}_1^{N+n_w}$ belonging to
the interval $\intff{ x_{\mu}-\tf{\pi}{2} }{ x_{\mu}+\tf{\pi}{2} }$ of length $\pi$, 
where $\nu_a$ is the unique real solution to 
\beq
\label{nextroots} 
\wh{\xi}_{\mu}(\nu_a) \, = \, \f{a - \iota}{L} \;. 
\enq

We fix $\iota$ uniquely by demanding that $\{\nu_a\}_1^{N+n_w} \subset
\intff{ x_{\mu}-\tf{\pi}{2} }{ x_{\mu}+\tf{\pi}{2}} \cap
\intfo{-\tf{\pi}{2}}{\tf{\pi}{2}}$. The Bethe Ansatz equations 
(\ref{ecriture BAE}) determine the Bethe roots $\mu_a$ only modulo $\pi$.
For this reason we may assume that $\Re (\mu_a) \in \intfo{-\tf{\pi}{2}}{\tf{\pi}{2}}$,
$a = 1, \dots, N$.
Since the roots $\{ \mu_a \}_1^{N}$ are pairwise distinct,
it follows that $N-n$ among the roots $\{\nu_a\}_1^{N+n_w}$ coincide with the real
roots which arise in the solution $\{ \mu_a \}_1^{N}$. Namely, there exists
integers $h_1 < \dots < h_{n_h}$, $h_a \in \intn{1}{N+n_w}$ and $n_h=n+n_w$ such that
\beq
\{ \mu_a \}_1^{N} \; = \; \{ z_a \}_1^n \cup \Big\{ \{\nu_a\}_1^{N+n_w} \setminus  \{\nu_{h_a}\}_1^{n_h} \Big\}  \;. 
\label{definition line bethe roots et nu roots}
\enq
In the following it will appear convenient to partition the set of
complex roots $\{z_a\}_1^n$ in two sub-sets: 
\begin{itemize}
\item  one built of close roots $\{z_a^{c} \}_{1}^{n_c}$, \textit{viz}. those satisfying  $ | \Im\big( z_a^{c} \big) | < \eta$;
\item one built out of the wide roots $\{z_a^{w} \}_{1}^{n_w}$, \textit{viz}. those satisfying $| \Im\big( z_a^{w} \big) | > \eta$.
\end{itemize}

In order to state the large-$L$ behaviour of the counting function, we
still need to introduce two auxiliary functions, namely, the dressed phase
$\phi$ and the dressed momentum $p$. They are defined as the analytic continuations, starting from
$\intff{-\tf{\pi}{2}}{\tf{\pi}{2}}$, of the unique solutions to the linear integral equations
\beq
\big(I+K\big)[\phi(*,z)](\om) \;  = \;  \th(\om-z) 
\label{definition eqn int phase habillee} 
\enq
and 
\beq
\big(I+K\big)[p](\om) \; = \;  \f{p_0(\om) }{ 2\pi } \; + \; 
\f{1}{2 \i \pi} \bigg\{ p\big(-\f{\pi}{2}\big) \cdot \th\big(\om+\f{\pi}{2} \big) \; - \; 
p\big(\f{\pi}{2}\big) \cdot \th\big(\om-\f{\pi}{2} \big) \bigg\} \;  .
\label{definition eqn int moment habille} 
\enq
Here we agree that $I$ is the identity operator while the operator $K$ acts as
\beq
K[f](\om) \, = \, \Int{ - \tf{\pi}{2} }{ \tf{\pi}{2} } K(\om-s) f(s) \cdot \dd s \; . 
\enq
Notice that in \eqref{definition eqn int phase habillee}, $*$ indicates
the running variable on which $I+K$ acts. 
The functions $\phi$ and $p$ can be explicitly represented in terms of
either their Fourier series or, respectively, in terms of ratios of
$q$-Gamma functions or Jacobi Theta functions (see Appendix
\ref{Appenix dressed momentum} and \ref{Appendix dressed phase} for
more details). It seems convenient in the following to use a homogenised
version of the dressed phase defined as
\beq
\vp(\om,z) \; = \; \phi(\om,z)\; - \; \f{1}{2}\cdot \bs{1}_{ |\Im(z)| < \eta }  \cdot \phi\big(\om, \f{\pi}{2} \big)
\enq
where we introduced the notation
\beq
\bs{1}_{\e{condition}} \; = \; \left\{  \ba{cc}   1 & \text{if `condition' is satisfied} \\ 
						   0  & \text{otherwise} \ea \right. 	\;\; . 
\enq
We call $\vp(\om,z)$ a homogenised version of the dressed phase since it solely depends on the difference $\om-z$, \textit{cf}. \eqref{ecriture homogenised dressed phase small z}, 
\eqref{ecriture propriete reduction dressed phase wide argument 1}-\eqref{ecriture propriete reduction dressed phase wide argument 2}.

In order to close the listing of solutions to linear integral equations that will be of use to our study, we introduce the dressed energy $\veps$
as the solution to the linear integral equation 
\beq
\big( I\,+ \, K \big)[ \veps ](\om) \; = \; \veps_0(\om)  \qquad \e{where} \quad  \veps_0(\la) \; = \; h- 2 J \sinh(\eta) p_0^{\prime}(\la) \;.
\label{definition dressed and bare energies}
\enq
The dressed energy enters in the description of the excitation energy above the ground state and can be expressed as 
\beq
\veps(\la) \; = \; \f{h}{2} \, - \; 4 \pi J \sinh(\eta) \cdot  p^{\prime}(\la) \;. 
\enq

We are  now in position to describe the large-$L$ asymptotic behaviour of the counting function for arguments lying 
uniformly away from the lines $\Im(z) \; = \; \pm \eta$. 
\begin{prop}
\label{Propositon large L asympt ctg fct}
Let the set of Bethe roots $\{\mu_a\}_1^N$ satisfy the above stated hypotheses and set 
\beq
\label{xiinfinity}
\xi_{\mu}^{(\infty)}(\om) \;= \; p( \om )
\; + \; \f{ 1 }{ 2 \i \pi L } \cdot \bigg\{ \sul{a=1}{n_h} \vp\big( \om , \nu_{h_a} \big) 
		\; - \; \sul{a=1}{n} \vp\big( \om , z_a \big) \bigg\} 
 %
% \; - \;  \f{   n_w  }{2 \i \pi L } \cdot \phi\Big(\om,\f{ \pi }{2} \Big) 
%
\; + \; \f{1-\a - \iota}{ 2 L } \;. 
\enq
Then, the counting function $\wh{\xi}_{\mu}$ admits the large-$L$ asymptotic expansion
\begin{itemize}
\item $\wh{\xi}_{\mu}(\om) \; = \; \xi_{\mu}^{(\infty)}(\om) \; + \; \e{O}\big( L^{-\infty} \big)$\, , \qquad for $|\Im(\om)|<\eta $;
\item $\wh{\xi}_{\mu}(\om) \; = \; \xi_{\mu}^{(\infty)}(\om) \,+ \, \xi_{\mu}^{(\infty)}(\om-\i\eta) \, - \, \f{1 - 2 \iota}{2L} \, + \, \e{O}\big( L^{-\infty} \big)$\, , \qquad for $\Im(\om)>\eta $;
\item $\wh{\xi}_{\mu}(\om) \; = \; \xi_{\mu}^{(\infty)}(\om) \,+ \, \xi_{\mu}^{(\infty)}(\om + \i\eta) \, - \, \f{1 - 2 \iota}{2L} \, + \, \e{O}\big( L^{-\infty} \big)$\, , \qquad for $\Im(\om) < -\eta $. 
\end{itemize}

In each of the three situations above, the remainder is uniform in $\om $ 
provided that it is located at a finite distance to the boundary of the
domain of interest.

\end{prop}
%
%
%
%Note that 

The asymptotic behaviour for $|\Im(\om)|>\eta$ is a typical effect of the so-called second determination introduced in \cite{DestriDeVegaNLIERatherCompleteAnalysisSineGordon} 
but already implicitly present in various earlier works. 

The above large-volume asymptotic expansions yield % allow one to write down
the large-$L$ asymptotic expansion of the parameter $x_{\mu}$ in (\ref{firstroot}). 
It is readily seen that, to leading order in $L$, one has 
\beq
\label{xplicit}
x_{\mu} \; = \; \f{ 1 }{ 2 \i \pi L p^{\prime}\big(-\tf{\pi}{2}\big) }
\cdot \bigg\{ \i \pi ( \a - \iota )
%
%\, + \, n_w \phi\Big( -\f{\pi}{2},\f{\pi}{2} \Big)
%
\; + \;  \sul{a=1}{n} \vp\Big( -\f{\pi}{2} , z_a \Big)
		\; - \; \sul{a=1}{n_h} \vp\Big( -\f{\pi}{2} , \nu_{h_a} \Big)  \bigg\}
 \; + \; \e{O}\Big( L^{-2} \Big) \; .
\enq
It is easy to check \textit{a posteriori}  that all of the hypotheses stated at the beginning of Sub-Section \ref{Subsection Asympt behaviour counting function} are  indeed fulfilled.

\Proof

Let $\Ga_{\mu}\, = \,  \Ga_{\mu}^{(\ua)} \cup \Ga_{\mu}^{(\da)} $ correspond to the loop around $\intff{ x_{\mu} - \tf{\pi}{2} }{ x_{\mu} +  \tf{\pi}{2} }$
depicted in Fig. \ref{Figure definition contour Gamma mu} where 
\beq
\left\{ \ba{ccc} 
\Ga_{\mu}^{(\ua)} & = & \intff{ x_{\mu} + \f{\pi}{2} }{  x_{\mu} + \f{\pi}{2} + \i \tau} \cup 
\intff{  x_{\mu} + \f{\pi}{2} + \i \tau}{  x_{\mu} -\f{\pi}{2} + \i \tau}
\cup  \intff{  x_{\mu} -\f{\pi}{2} + \i \tau}{  x_{\mu} -\f{\pi}{2} }   \vspace{2mm} \\
\Ga_{\mu}^{(\da)}  & = & \intff{  x_{\mu}  -\f{\pi}{2} }{  x_{\mu} -\f{\pi}{2} - \i \tau} \cup 
	\intff{  x_{\mu}  -\f{\pi}{2} - \i \tau}{  x_{\mu} + \f{\pi}{2} - \i \tau}
\cup  \intff{  x_{\mu} -\f{\pi}{2} -  \i \tau}{  x_{\mu} + \f{\pi}{2} }    \ea \right. \; \;  .
\label{definition du contour gamma mu}
\enq
\begin{figure}[ht]
\begin{center}

\begin{pspicture}(7,7)

\psline[linestyle=dashed, dash=3pt 2pt]{->}(1,4)(6.7,4)
\psdots(1.5,4)(6,4) 
\rput(0.7,3.5){$x_{\mu}-\f{\pi}{2}$}
\rput(6.6,3.7){$x_{\mu}+\f{\pi}{2}$}

\rput(0.7,5.9){$x_{\mu}-\f{\pi}{2}+\i\tau$}
\rput(6.8,6){$x_{\mu}+\f{\pi}{2}+\i\tau$}

\rput(0.5,1.8){$x_{\mu}-\f{\pi}{2}-\i\tau$}
\rput(6.8,1.8){$x_{\mu}+\f{\pi}{2}-\i\tau$}

\psline{-}(1.5,2.5)(1.5,5.5)
\pscurve{-}(1.5,5.5)(1.65,5.9)(2,6)

\psline{-}(2,6)(5.5,6)
\pscurve{-}(5.5,6)(5.85,5.9)(6,5.5)

\psline{-}(6,5.5)(6,2.5)
\pscurve{-}(6,2.5)(5.85,2.1)(5.5,2)

\psline{-}(5.5,2)(2,2)
\pscurve{-}(2,2)(1.65,2.1)(1.5,2.5)

%region p H_i

%\rput(14.5,4.5){$p\pa{H_{III}}$}

%region H_i

\rput(3.5,6.5){ $ \Ga_{\mu}^{(\ua)} $ }
\rput(4,1.5){ $ \Ga_{\mu}^{(\da)} $ }

\psline[linewidth=2pt]{->}(5,2)(5.1,2)
\psline[linewidth=2pt]{->}(4,6)(3.9,6)

\end{pspicture}
\caption{Contour $\Ga_{\mu}=\Gamma^{(\ua)}_{\mu} \cup \Gamma^{(\da)}_{\mu} $. \label{Figure definition contour Gamma mu} }
\end{center}
\end{figure}

\noindent By taking into account the absent roots $\nu_{h_1},\dots, \nu_{h_{n_h}}$, the 
representation \eqref{ecriture representation originale pour xi kappa} for $\wh{\xi}_{\mu}$ can be 
%recasted   
transformed as
\beq
\wh{\xi}_{\mu}(\om) \; = \; \f{ p_0(\om) }{ 2\pi }  \; + \; \f{1}{2 \i \pi L } 
\bigg\{ \sul{a=1}{n+n_{w}} \th(\om-\nu_{h_a}) \, - \, \sul{ a=1 }{ n } \th(\om-z_a) \bigg\}
\; + \; \f{ 1-2\a  }{ 2L } \; - \; 
\Oint{ \Ga_{\mu} }{} \f{\th(\om-s) \wh{\xi}_{\mu}^{\prime}(s) }{ \ex{2 \i \pi L \wh{\xi}_{\mu}(s)}-1 } \cdot \f{ \dd s }{ 2 \i \pi } \;. 
\label{reecriture NLIE xi hat step 1}
\enq
This  expression holds for any $\om$ such that $|\Im(\om)| \not= \eta$, as one can always pick $\tau$
small enough so that the cuts of $\th$ lie away of the integration domain. 
By hypothesis, the function $\wh{\xi}_{\mu}$ is real analytic in a neighbourhood of
$\R$ and strictly increasing along the real axis. Setting $x = \Re (\om)$, $y = \Im (\om)$
and using the Cauchy-Riemann equations we find that, for all $\la \in {\mathbb R}$, 
\beq
\frac{\Dp{} }{\Dp{} x} \Re\big(\wh{\xi}_{\mu} \big)   _{\mid  \om = \la} 
        \; =  \; \frac{\Dp{} }{\Dp{} y}  \Im\big(\wh{\xi}_{\mu} \big) _{\mid  \om = \la} \,  >  \, 0 \; . 
\enq
Thus, by $\pi$-periodicity of its derivative, $\wh{\xi}_{\mu}$ has a strictly
positive imaginary part in a finite strip above the real axis and a strictly
negative imaginary part in a finite strip below the real axis. This ensures that,
close to the real axis, $\wh{\xi}_{\mu}(\om)$ has no other zeros than the
$\nu_j$, $j = 1, \dots, N + n_w$, or these numbers shifted by integer multiples
of $\pi$.

Observe that the function
\beq
\wh{u}_{\mu}(s) \; = \; \left\{ \ba{cc}  
      -2 \i \pi L \wh{\xi}_{\mu}(s) \, + \, \wh{u}_{\mu}^{(+)}(s) & s \in \Ga_{\mu}^{(\ua)}  \vspace{2mm} \\
	  \wh{u}_{\mu}^{(-)}(s)   & s \in \Ga_{\mu}^{(\da)} \ea \right. 
\quad \e{where} \qquad \wh{u}_{\mu}^{(\eps)}(s) \; = \; \ln \Big(1-\ex{ 2 \i \pi \eps L \wh{\xi}_{\mu}(s)}   \Big)	  
\label{definition fonction hat u}
\enq
defines an anti-derivative of the counting function part of the integrand in \eqref{reecriture NLIE xi hat step 1}. 
Here, the logarithm is defined by its principal determination,
\textit{i}.\textit{e}. $\e{arg} \in \intoo{-\pi}{\pi}$. 
This ensures that the functions 
\beq
s \mapsto \ln \Big(1-\ex{ \pm 2 \i \pi L \wh{\xi}_{\mu}(s)}   \Big)
\enq
are $\pi$-periodic on $\mathbb{H}_{\pm}\cap V$, where $V$ is some sufficiently small 
open neighbourhood of $\intff{ - \tf{\pi}{2} }{ \tf{\pi}{2} }$. As a consequence, 
their $\pm$-boundary values on $V \cap \R$ are $\pi$-periodic as well. In particular, one has 
\beq
   \lim_{\vsg \tend 0^+} 
\bigg\{ \ln \Big(1-\ex{\pm 2 \i \pi L \wh{\xi}_{\mu}( x_{\mu}- \tf{\pi}{2} \pm \i \vsg)} \Big) \bigg\}
\; = \; \lim_{\vsg \tend 0^+}
 \bigg\{ \ln \Big(1-\ex{\pm 2 \i \pi L \wh{\xi}_{\mu}( x_{\mu} + \tf{\pi}{2} \pm \i \vsg)} \Big) \bigg\}
\; = \; \ln 2 \;. 
\label{ecriture identite valeur limites partie logarithmique}
\enq
  Then, upon an integration by parts, one obtains
\bem
-\Oint{ \Ga_{\mu} }{} \f{\th(\om-s) \wh{\xi}_{\mu}^{\prime}(s) }{ \ex{ 2 \i \pi L \wh{\xi}_{\mu}(s)}-1 } \cdot \f{ \dd s }{ 2 \i \pi }
\; =\; 
- \Oint{ \Ga_{\mu} }{} K(\om-s) \wh{u}_{\mu}(s) \cdot \f{ \dd s }{ 2 \i \pi  L } \\
\; + \; \f{1}{ 2 \i \pi } \bigg\{  \wh{\xi}_{\mu}\big( x_{\mu}-\f{\pi}{2} \big)   \cdot  \th\big(\om+\f{\pi}{2}-x_{\mu} \big) 
\, - \,  \wh{\xi}_{\mu}\big( x_{\mu} + \f{\pi}{2} \big)  \cdot  \th\big(\om-\f{\pi}{2}-x_{\mu} \big)\bigg\} \;. 
\label{ecriture resultat IPP}
\end{multline}
Note that in \eqref{ecriture resultat IPP} above, the boundary terms which involve the logarithmic part of $\wh{u}_{\mu}$  have canceled out due to 
\eqref{ecriture identite valeur limites partie logarithmique}. It solely remains to squeeze down to 
$\intff{ x_{\mu}-\tf{\pi}{2} }{ x_{\mu}+\tf{\pi}{2} }$ the part of the integral on the \textit{rhs} of \eqref{ecriture resultat IPP}
 that involves $-2 \i \pi L \wh{\xi}_{\mu}$ and then re-centre it on $\intff{ -\tf{\pi}{2} }{ \tf{\pi}{2} }$.
All in all, one arrives at 
the non-linear integral equation
\bem
\wh{\xi}_{\mu}(\om) + \Int{-\tf{\pi}{2} }{ \tf{\pi}{2} } \! K(\om-s) \cdot \wh{\xi}_{\mu}(s) \cdot \dd s  
\; = \; \f{ p_0(\om) }{ 2\pi } \; -\; \f{1}{ 4 \i \pi }\th\big( \om-\f{\pi}{2}\big) \; + \; \f{ 1 }{ 2\i\pi L } 
\bigg\{ \sul{a=1}{n+n_{w}} \th(\om-\nu_{h_a}) \, - \, \sul{ a=1 }{ n } \th(\om-z_a) \bigg\}  \\ 
\;+ \; \f{1 - 2 \iota}{2L} \bs{1}_{|\Im(\om)|<\eta} \; + \; \f{ 1-2\a  }{ 2L } \; - \; \f{ n_w }{ 2 \i \pi L }  \th\big(\om-\f{\pi}{2}\big) \; + \; \mf{r}_{\wh{\xi}_{\mu}}(\om)
\label{ecriture forme integeree premiere NLIE xi hat}
\end{multline}
where we agree upon
\beq
\mf{r}_{\wh{\xi}_{\mu}}(\om) \; = \; 
- \; \sul{\eps=\pm}{} \f{ 1 }{ 2 \i \pi L } 
\Int{ \Ga^{(\eps)} }{} K(\om-s) \cdot \wh{u}_{\mu}^{(\eps)}(s) \cdot  \dd s \; 
\qquad \e{with} \quad
\left\{ \ba{ccc} \Ga^{(+)} & = & \intff{ \tf{\pi}{2} }{ -\tf{\pi}{2} }  + \i \tau  \\   
 \Ga^{(-)} & = & \intff{ -\tf{\pi}{2} }{ \tf{\pi}{2} } - \i \tau  \ea \right. \; . 
\label{ecriture reste et defnition ctrs gama pm}
\enq
In order to obtain this representation, we
have invoked the $\pi$-periodicity of the integrand in
\eqref{ecriture reste et defnition ctrs gama pm}
so as to cancel out the contribution of the lines parallel to $\i\R$ and shift the 
integration along the upper/lower part of $\Ga_{\mu}$  up to $\Ga^{(\pm)}$. %Furthermore, 
%\textcolor{blue}{\sout{We have also used that}} 
%
%
%
%\beq
%%
%\textcolor{blue}{\xout{
%\Int{ \tf{\pi}{2} }{  x_{\mu}+\tf{\pi}{2}  } \hspace{-3mm} K(\om-s) \cdot \Big( \wh{\xi}_{\mu}\big(x_{\mu}+\tf{\pi}{2}\big)\, - \,  \wh{\xi}_{\mu}\big(s\big) \Big) \cdot \dd s
%
%\; + \;  \Int{ x_{\mu}-\tf{\pi}{2}  }{ -\tf{\pi}{2} }  \hspace{-3mm}  K(\om-s) \cdot \Big( \wh{\xi}_{\mu}\big(x_{\mu}-\tf{\pi}{2}\big)\, - \,  \wh{\xi}_{\mu}\big(s\big) \Big) \cdot \dd s 
%%
% \; = \; 0}}
%%
%\enq
%
%
%
%\textcolor{blue}{\sout{as can be seen by reducing the whole integration to the interval $\intff{0}{x_{\mu}}$ and using the periodicity or quasi-periodicity properties of the building blocks of the integrand.}
%[fg: I would like to delete this, since it is not important for
%the understanding. In fact, I think one can proceed easier without this formula.]}
Due to the remark that follows \eqref{reecriture NLIE xi hat step 1} and by its very construction  $\mf{r}_{\wh{\xi}_{\mu}}(\om) \; = \; \e{O}\big(L^{-\infty}\big)$. 
Therefore, when acting with $(I+K)^{-1}$ on \eqref{ecriture forme integeree premiere NLIE xi hat}
one obtains the asymptotic expansion in the region $|\Im(\om)|<\eta$. 

In order to obtain the asymptotic expansion of $\wh{\xi}_{\mu}(\om)$ outside of the strip $ |\Im(\om)|<\eta$, 
we introduce the functions
\beq
\left. \ba{lc} 
\e{for} \quad  \Im(\om) >\eta   & \kappa_{\ua}(\om )  \vspace{1mm}\\ 
 \e{for} \quad \Im(\om)  < \eta  &  \kappa_{c}(\om )    \vspace{1mm}  \\
\e{for} \quad \Im(\om) <-\eta  &  \kappa_{\da}(\om )   \ea \right\}  
\;  = \;   \Int{-\tf{\pi}{2} }{ \tf{\pi}{2} } \hspace{-2mm} K(\om-s) \cdot \xi_{\mu}^{(\infty)}(s) \cdot \dd s  \; . 
\enq
Then, it is %readily 
immediately seen that, for $|\Im(\om)| > \eta$,  $\wh{\xi}_{\mu}(\om)$  is given by %can be recast as
\bem
\wh{\xi}_{\mu}(\om)
\; = \; - \kappa_{\ua/\da}(\om) \; +\; \f{ p_0(\om) }{ 2\pi } \; -\; \f{ 1 }{ 4 \i \pi }\th\big(\om-\f{\pi}{2}\big) 
\; + \; \f{1}{2 \i \pi L } \bigg\{ \sul{a=1}{n+n_{w}} \th(\om-\nu_{h_a}) \, - \, \sul{ a=1 }{ n } \th(\om-z_a) \bigg\}  \\ 
\; + \; \f{ 1-2\a  }{ 2L } \; - \; \f{n_w}{ 2 \i \pi L }  \th\big(\om-\f{\pi}{2}\big) \; - \;
\Int{-\tf{\pi}{2} }{ \tf{\pi}{2} } K(\om-s) \cdot \underbrace{ \Big( I \,+ \, K\Big)^{-1}\big[ \mf{r}_{\wh{\xi}_{\mu}} \big](s)}_{= \e{O}\big( L^{-\infty} \big)} \cdot \dd s
\end{multline}
%
%
%
%for $\pm \Im(\om) > \eta$. 
The functions $\kappa_{\ua/\da}$ and $\kappa_{\e{c}}$ satisfy the system of jump conditions 
\beqa
\kappa_{\ua;+}(\om)-\kappa_{c;-}(\om) & = & 
-\xi_{\mu}^{(\infty)}(\om-\i\eta) \qquad \e{for} \qquad \om \in \intff{-\tf{\pi}{2}}{\tf{\pi}{2}}+\i\eta \; , \\
\kappa_{\da;-}(\om)-\kappa_{c;+}(\om) & = & -\xi_{\mu}^{(\infty)}(\om+\i\eta) \qquad \e{for} \qquad \om \in \intff{-\tf{\pi}{2}}{\tf{\pi}{2}}-\i\eta \;. 
\eeqa
Here and hereafter, we denote for an arbitrary function $g_{;\pm}(\om)=\lim_{\epsilon \searrow 0} g(\om \pm i \epsilon) $.
Furthermore, the integral equation satisfied by $\xi^{(\infty)}_{\mu}$ ensures that 
\bem
\kappa_{c}(\om)\; =  \; - \; \xi^{(\infty)}_{\mu}(\om) \; +\; \f{ p_0(\om) }{ 2\pi } \; -\; \f{ 1 }{ 4 \i \pi }\th\big(\om-\f{\pi}{2}\big) \\
\; + \; \f{1}{2 \i \pi L } \bigg\{ \sul{a=1}{n+n_{w}} \th(\om-\nu_{h_a}) \, - \, \sul{ a=1 }{ n } \th(\om-z_a) 
-n_w \th\big(\om-\f{\pi}{2}\big) \bigg\}  
\; + \; \f{ 1 - \a - \iota }{ L }   \;. 
\end{multline}
This provides us with the sought analytic continuation of $\kappa_{\ua/\da}$
and yields the claimed form of the large-$L$ asymptotics of $\wh{\xi}_{\mu}$
in the whole complex plane. \qed

\subsection{The higher-level Bethe Ansatz  equations}
\label{SousSection HLBAE}

We are now prepared to derive the so-called higher-level Bethe Ansatz equations that, in the limit of large system size, determine
the positions of the complex roots $\{z_a\}_1^n$ as functions of the holes $\{\nu_{h_a}\}_1^{n_h}$ which turn into free parameters in this limit. The concept of higher-level Bethe Ansatz equations emerged from the long 
struggle to characterise the structure of the complex solutions to the Bethe Ansatz equations describing the low-lying excited states.\footnote{In our terminology here, such
excited states have an energy above the ground state which stays bounded in the large-$L$ limit.}
Indeed, starting from Bethe's seminal work and until the early `80s it was widely accepted that such complex solutions organise into strings, with an exponential precision in $L$.
The counting of strings, which is an important consistency
test for the string-based thermodynamics, was even mistakenly
claimed to prove the completeness of the Bethe ansatz.
%The string hypothesis has even been used to provide wrong proofs
%of the completeness of the Bethe Ansatz.
The break-through came in 1982 with the pioneering analysis of Destri
and Lowenstein \cite{DestriLowensteinFirstIntroHKBAEAndArgumentForStringIsWrong} of the structure of the complex solutions to the Bethe equations 
describing the chiral invariant Gross-Neveu model. It was shown in \cite{DestriLowensteinFirstIntroHKBAEAndArgumentForStringIsWrong} that the complex solutions to the Bethe Ansatz equations describing
the low-lying excited states form two-strings, quartets and wide pairs, but do not form general, larger strings. The work of Destri and
Lowenstein was followed a few months later by two independent papers on the XXZ chain
\cite{BabelondeVegaVialletStringHypothesisWrongXXZ,WoynaorwiczHLBAEMAsslessXXZ0Delta1}. 
Woynarovich \cite{WoynaorwiczHLBAEMAsslessXXZ0Delta1} studied the massless regime
at anisotropy  $1>\De>0$, while the analysis by Babelon, de Vega and Viallet \cite{BabelondeVegaVialletStringHypothesisWrongXXZ} 
dealt with all real values of the anisotropy parameter. In 1984, the derivation of the higher-level Bethe Ansatz equations for the
massive regime of the XXZ chain was reconsidered by Virosztek and Woynarovich \cite{VirosztekWoynarovichStudyofExcitedStatesinXXZHigherLevelBAECalculations}
who confirmed again the picture of two-strings, quartets and wide pairs, but obtained a slightly different form of the higher-level
Bethe Ansatz equations. As we shall see, our analysis of the previous sub-section reconfirms the latter set of equations.

\begin{prop}
\label{hlbaes}

Let $\{\mu_a\}_1^N$ be a solution of the $\a$-twisted Bethe equations \eqref{ecriture BAE}
satisfying  the hypotheses stated in Sub-Section
\ref{Subsection Asympt behaviour counting function}. Then, in the large-$L$ limit, 
the real numbers $\{\nu_a \}_{1}^{N+n_w}$ densely fill the interval
$\intff{ - \tf{\pi}{2} }{ \tf{\pi}{2} }$.
Furthermore,  in the $L\tend +\infty$ limit, the complex roots $\{z_a\}_{1}^{n}$ organise into 
\begin{itemize}
\item wide pairs $\big\{ y_a,\ov{y}_a \big\}_{1}^{ \frac{n_w}{2} }$, $\Im(y_a)>0$; 
\item strings $\big\{ w_a, w_a - \i \eta \, + \, \de_a \big\}_{1}^{ \frac{n_c}{2} }$, $\Im(w_a)>0$ with $\de_a$, the string-deviation parameter, 
satisfying $\de_a=\e{O}(L^{-\infty})$.
\end{itemize}
The two types of limiting complex roots solve the following sets of equations
\beq
1\; = \; - \ex{-2 \i \pi \a} \pl{b=1}{n_h} \f{ \sin\big( w_a-\nu_{h_b} - \i \eta \big) }{ \sin\big( w_a-\nu_{h_b} \big) }
\pl{b=1}{\tf{n_c}{2}} \f{ \sin\big( w_a-w_b+ \i \eta \big) }{ \sin\big( w_a-w_b - \i \eta \big) }
\cdot \pl{b=1}{\tf{n_w}{2}} \f{ \sin\big( w_a-y_b + \i \eta \big) \sin\big( w_a-\ov{y}_b \big)}
			{  \sin\big( w_a-y_b - \i \eta \big) \sin\big( w_a-\ov{y}_b - 2 \i \eta \big) }
\label{ecriture HLBAE close roots}
\enq
in what concerns the close roots, and
\beqa
1 & = & -\ex{-2 \i \pi \a} \pl{b=1}{n+n_w} \f{ \sin\big( y_a-\nu_{h_b}-\i\eta \big) }{ \sin\big( y_a-\nu_{h_b} \big) }
\pl{b=1}{\tf{n_c}{2}} \f{ \sin\big( y_a-w_b + \i \eta \big) }{ \sin\big( y_a-w_b - \i \eta \big) }
\cdot \pl{b=1}{\tf{n_w}{2}} \f{ \sin\big( y_a-y_b + \i \eta \big) \sin\big( y_a-\ov{y}_b \big)}
			{  \sin\big( y_a-y_b - \i \eta \big) \sin\big( y_a-\ov{y}_b - 2 \i \eta \big) }  \\
1 & = & -\ex{-2 \i \pi \a} \pl{b=1}{n+n_w} \f{ \sin\big( \ov{y}_a-\nu_{h_b} \big) }{ \sin\big( \ov{y}_a-\nu_{h_b} + \i \eta  \big) }
\pl{b=1}{\tf{n_c}{2}} \f{ \sin\big( \ov{y}_a-w_b + 2 \i \eta \big) }{ \sin\big( \ov{y}_a-w_b \big) }
\cdot \pl{b=1}{\tf{n_w}{2}} \f{ \sin\big( \ov{y}_a-y_b + 2 \i \eta \big) \sin\big( \ov{y}_a-\ov{y}_b + \i \eta \big)}
			{  \sin\big( \ov{y}_a-y_b \big) \sin\big( \ov{y}_a-\ov{y}_b - \i \eta \big) } 			
\eeqa
in what concerns the wide roots.

\end{prop}

For  $\Im(w_a) \not= \i \tf{\eta}{2}$, the close roots organise in quartets
$\{w_a, w_a-\i \eta +\de_a, \ov{w}_a, \ov{w}_a + \i \eta + \ov{\de}_a \}$ 
or two-strings $\{w_a, w_a-\i \eta + \de_a\}$ if $w_a=\ov{w}_a + \i \eta + \ov{\de}_a $, 
since the complex roots always appear in complex conjugated pairs. 
We shall however not build on such a distinction in the present paper.
We also stress that the equation satisfied by $\ov{y}_a$ is indeed the complex conjugate of the one satisfied by 
$y_a$ since, due to our convention of defining the roots $w_a$, one has that 
$\big\{ w_a \big\}_1^{\frac{n_c}{2}} \, = \, \big\{ \ov{w}_a + \i \eta \big\}_1^{\frac{n_c}{2}} $ in the $L \tend + \infty$ limit.

We recall that one can represent the higher-level Bethe Ansatz equations in a more
symmetric form by reparametrising the roots as it has been suggested in \cite{BabelondeVegaVialletStringHypothesisWrongXXZ}:
\beq
\big\{ \chi_a \big\}_1^{n_{\chi}} \; = \; 
\Big\{ y_a-i\f{\eta}{2} \Big\}_{1}^{ \f{n_w}{2} } \cup \Big\{ \ov{y}_a+ \i \f{\eta}{2} \Big\}_{1}^{ \f{n_w}{2} }
\cup \Big\{ w_a - \i \f{\eta}{2} \Big\}_{1}^{ \f{n_c}{2} }  \quad \e{where} \qquad n_{\chi} \, = \, \f{n_c}{2}+n_w \;. 
\label{definition reparametrisation via chis}
\enq
The new parameters $\chi_a$, $a=1,\dots,n_{\chi}$, now satisfy
%
%
%It also guarantees that for λ belonging to some small neighbourhood of R in C and satisfying
%±I(λ) > 0, one has I(∓i ξ μ (λ)) > c > 0 for some L-independent constant c.
\beq
\ex{2 \i \pi \a} \; = \; - \pl{b=1}{ 2 n_{\chi} } \f{ \sin\big( \chi_a - \nu_{h_b} - \i \tf{\eta}{2} \big) }{ \sin\big( \chi_a - \nu_{h_b} + \i \tf{\eta}{2} \big) } 
\cdot \pl{ b=1 }{ n_{\chi} }  \f{ \sin\big( \chi_a - \chi_b + \i \eta \big) }{ \sin\big( \chi_a - \chi_b - \i \eta \big) }  \;. 
\label{ecriture HBAE in terms of chi's}
\enq
The higher-level Bethe Ansatz equations written in the above form look similar to the Bethe Ansatz equations associated with an inhomogeneous 
XXZ chain of length $2n_{\chi}$.

There is a discrepancy between the above  higher-level Bethe Ansatz equations and those obtained in \cite{BabelondeVegaVialletStringHypothesisWrongXXZ}.
The higher-level Bethe Ansatz equations in \cite{BabelondeVegaVialletStringHypothesisWrongXXZ}
contain an additional factor $\ex{\de}$ in the equations for the close roots. The additional constant $\ex{\de}$ takes the form 
\beq
\ex{\de} \; = \; \exp\Big\{ -2\i \sul{a=1}{ 2 n_{\chi} } \nu_{h_a}  \Big\} \cdot 
\pl{a=1}{ n_{\chi} } \bigg\{ \f{ \cos^2(\ov{\chi}_a-\i \tf{\eta}{2}) }{ \cos^2(\chi_a + \i \tf{\eta}{2}) }   \bigg\} \cdot 
\pl{a=1}{ 2 n_{\chi} } \pl{r=0}{+\infty}  \bigg\{ \f{ \cos^2(\nu_{h_a} + \i (2r+1) \eta) \cos^2(\nu_{h_a} - \i 2(r+1) \eta) }
{  \cos^2(\nu_{h_a} - \i (2r+1) \eta) \cos^2(\nu_{h_a} + \i 2(r+1) \eta) } \bigg\} \;. 
\label{ecriture formule exp delta}
\enq
The factor $\ex{\de}$ appeared in \cite{BabelondeVegaVialletStringHypothesisWrongXXZ}
since certain $\e{O}(\tf{1}{L})$ terms which would have produced counter-terms were disregarded. This fact becomes manifest within
our approach. By using the functional equation satisfied by the homogenised
dressed phase $\vp(\om,z)$, it is %readily seen that 
seen straightforwardly that 
\beq
\de = 2 \cdot 2\i \pi L \cdot \Big( \wh{\xi}_{\mu}(-\tf{\pi}{2}) \, - \,\wh{\xi}_{\mu}(x_{\mu} - \tf{\pi}{2}) \Big) \; + \; \e{O}\big( L^{-\infty} \big) \;. 
\enq
Thus, $\de$ is related to the small $\e{O}(L^{-1})$ shift $x_{\mu}$ of the
`Fermi-zone' where the roots $\{\nu_a\}_{1}^{N+n_w}$ condense. 

The absence of the $\ex{\de}$ term was already observed in
\cite{VirosztekWoynarovichStudyofExcitedStatesinXXZHigherLevelBAECalculations}, where the authors tried to explain it with the claim that $\ex{\de}=1$ when evaluated at solutions to the higher level Bethe equations.
This can be seen to be incorrect by solving explicitly the higher-level Bethe Ansatz equations with $n_{\chi}=1$ and
$\a=0$. In case of one complex root (and thus two holes) the  only
solutions, up to $\pi$-periodicity, read
\beq
\label{chi1andchi2}
\chi_{1}\, = \, \f{ \nu_{h_1}+ \nu_{h_2} }{ 2 } \quad , \qquad \chi_{2}\, = \, \f{ \nu_{h_1}+ \nu_{h_2} }{ 2 } + \frac \pi 2 \;. 
\enq
Inserting \textit{e.g.} $\chi_1$ into \eqref{ecriture formule exp delta} one concludes that $\ex{\de}\not=1$, unless $\nu_{h_1}=-\nu_{h_2}$ which
is clearly not  a generic distribution of hole parameters.

%\Proof 
{\it Proof of Proposition \ref{hlbaes} --- \ }

Up to corrections of the order $\e{O}\big( L^{-\infty} \big)$ the
counting function $\wh{\xi}_{\mu}$ is determined by $\xi_{\mu}^{(\infty)}$
(see Proposition~\ref{Propositon large L asympt ctg fct}).
Using Proposition~\ref{Propositon large L asympt ctg fct} in equation
(\ref{ecriture eqn Bethe ac fct cptge}), inserting $z_a$ and $\nu_{h_a}$
for $\mu_a$ and neglecting the $\e{O}\big( L^{-\infty} \big)$ corrections
we obtain a closed system of finitely many equations that determine
the holes and the close and wide roots. For the holes we obtain
\beq
\label{logbaeholes}
     \xi_{\mu}^{(\infty)} (\nu_{h_a})
        = \frac{h_a - \iota}{L} + \e{O}\big( L^{-\infty} \big) \;.
\enq
From this equation and the explicit form of the function $\xi_{\mu}^{(\infty)} (\nu_{h_a})$
(see~(\ref{xiinfinity})) we infer that the holes become free real
parameters for $L \rightarrow \infty$.

We now justify the organisation
of the close roots into strings. Expressing $\wh{\xi}_{\mu}$ by means 
of the non-linear integral equation it satisfies, one obtains the representation
\beq
\ex{2 \i \pi L \wh{\xi}_{\mu}(\om) } \; = \;
- \ex{ -  \i \pi ( \a + \iota ) } \ex{ 2 \i \pi L p(\om) } 
\cdot  \pl{b=1}{n}\f{ \sin\big( \om - z_b + \i \eta \big) }{ \sin\big( \om - z_b - \i \eta \big) } \cdot 
\pl{b=1}{n+n_w}\f{ \sin\big( \om - \nu_{h_b} - \i \eta \big) }{ \sin\big( \om - \nu_{h_b} + \i \eta \big) }
\cdot \ex{ \psi(\om) } \cdot \Big( 1+\e{O}\big( L^{-\infty} \big) \Big)
\enq
in which the $\e{O}\big( L^{-\infty} \big)$ remainder is regular in the strip $| \Im(\om) | <\eta$. 
The function 
\beq
\psi(\om) \; = \; \sul{a=1}{n}K\big[ \vp(*,z_a) \big](\om) \; - \;   \sul{a=1}{n+n_w}K\big[ \vp(*,\nu_{h_a}) \big](\om) 
\label{ecriture fonction psi}
\enq
arising above is already regular for $\om$ belonging to the strip $|\Im(\om)|<\eta$. 
%Note that, in \eqref{ecriture fonction psi}, the $*$ indicates the running variable on which $K$ acts. 

Let $z_b^{c}$, $0<\Im(z_b^{c})< \eta$, be a close root. Since, $\big| \ex{2i\pi p(\om)} \big| <1$ for 
$0<\Im(\om) < \eta$, \textit{cf}. \eqref{ecriture signe Re de dressed momentum}, it follows that, for 
$\ex{2\i \pi \wh{\xi}_{\mu}(z_a^{c})}=1$ to hold, $z_b^{c}$ has to approach,
up to some exponentially small correction in $L$, $z_{\ell}^{c}+ \i \eta$,
where $z_{\ell}^{c}$ is another close root such that $-\eta<\Im(z_{\ell}^{c})< 0$.
As a consequence, one can partition the roots as
\beq
\big\{ z_a \big\}_{1}^{n} \; = \; \Big\{ y_a, \ov{y}_a \Big\}_{1}^{\f{n_w}{2}} \bigcup 
  \Big\{ w_a, w_a - \i \eta +\de_a \Big\}_{1}^{\f{n_c}{2}} \; ,
\enq
where $\Im(y_a)>0$, $\Im(w_a)>0$ and $\de_a$ is some exponentially small correction in $L$, $\de_a = \e{O}\big( L^{-\infty} \big)$. 
In order to obtain an equation for $w_a$ that is free from divergent terms,
we multiply the equations satisfied by $w_a$ and $w_a- \i \eta+\de_a$. The 
evaluation of most terms is straightforward with the help of 
the identities \eqref{ecriture propriete reduction dressed phase close argument}, 
\eqref{ecriture propriete reduction dressed phase wide argument 1} and  \eqref{ecriture propriete reduction dressed phase wide argument 2} whereas, in what concerns the singular ones, 
it is enough to use the identity
\beq
\lim_{\de \tend 0^+} \bigg[
\exp\Big\{ \vp(\la,t)+\vp(\la-\i \eta+\de,t)+ \vp(\la,t-\i \eta+\de)+\vp(\la-\i \eta+\de,t- \i \eta+\de) \Big\}_{\mid \la=t} 
\bigg] \; = \; 1 \;. 
\enq
This leads precisely to equation \eqref{ecriture HLBAE close roots}. 

Finally, the form of the equations satisfied by the wide pairs follows straightforwardly from the large-$L$
asymptotic behaviour of $\wh{\xi}_{\mu}(\om)$ when $|\Im(\om)| >\eta$ provided that one 
uses the reduction properties of the dressed phase \eqref{ecriture propriete reduction dressed phase close argument},
\eqref{ecriture propriete reduction dressed phase wide argument 1} and  \eqref{ecriture propriete reduction dressed phase wide argument 2}. 

\qed

\subsection{Energy and momentum of excited states}

The energy and momentum of an excited state parametrised by the roots $\{\mu_a\}_1^N$ of
the $\a$-twisted Bethe equations \eqref{ecriture BAE} take the form 
\beq
\mc{E}_{\e{ex}} \; = \; \sul{a=1}{N} \Big( \veps_0(\mu_a) \, - \,  \veps_0(\la_a) \Big)
\qquad \text{and} \qquad 
\mc{P}_{\e{ex}} \; = \; \sul{a=1}{N} \Big( p_0(\mu_a) \, - \,  p_0(\la_a) \Big)
\; .
\label{definition excitation momentum}
\enq
Here $\{ \la_a \}_1^N$ denotes the Bethe Ansatz roots of the ground
state which solve \eqref{ecriture BAE} with $\alpha=0$.
%Similarly, the excitation energy reads
%
%
%
%\beq
%
%\mc{E}_{\e{ex}} \; = \; \sul{a=1}{N} \Big( \veps_0(\mu_a) \, - \,  \veps_0(\la_a) \Big) .
%
%\label{definition excitation energy}
%\enq
%
%
%
%where 
The bare excitation energy $\veps_0$ is defined in
\eqref{definition dressed and bare energies} while the momentum $p_0$
of bare excitations is given in \eqref{definition phase et moment nu}.
The large-$L$ behaviour of the counting functions associated with the
systems of Bethe roots $\{ \mu_a \}_1^N$ and $\{ \la_a \}_1^N$ 
allows one to compute both, the excitation energy and the momentum
of an $\a$-twisted excitation, up to $\e{O}\big( L^{-\infty} \big)$ corrections. 
\begin{prop}
\label{Proposition estimation excitation momentum and energy}
Under the hypothesis of Sub-Section \ref{Subsection Asympt behaviour counting function},
the quantities defined in \eqref{definition excitation momentum}
%-\eqref{definition excitation energy}
admit the large-$L$ asymptotic behaviour  
\beq
\mc{E}_{\e{ex}} \; =  \; - \; \sul{ a = 1 }{ 2 n_{\chi} } \veps^{(0)}(\nu_{h_a}) \; + \; \e{O}\big( L^{-\infty} \big)   \qquad 
and \qquad 
\mc{P}_{\e{ex}} \; = \; (\a + \iota) \pi  \, - \,  2\pi \sul{ a = 1 }{ 2 n_{\chi} } p(\nu_{h_a}) \; + \; \e{O}\big( L^{-\infty} \big) 
\label{expressions pour energie et impulsion excitation}
\enq
where $\veps^{(0)}$ corresponds to the dressed energy at zero magnetic field:
\beq
\veps^{(0)}(\la)  \; = \; - 4\pi J \sinh(\eta)p^{\prime}(\la) \;.
\enq
\end{prop}

The result stated above can be obtained by direct calculation of Fourier coefficients,
\textit{cf}.\ \cite{BabelondeVegaVialletStringHypothesisWrongXXZ},
or by the dressed function trick which relies on a direct handling of the
linear integral equations. It was first obtained by Johnson,
Krinsky and McCoy \cite{JohnsonKrinskyMcCoyCorrelationLength8VModelAndXYZ} based
on results for the transfer matrix of the eight-vertex model.

Here we propose a different proof, based solely on analytic continuation.
The technique we develop is useful for characterising many other
relations among solutions to linear integral equations in the massive
regime, in particular those involving the resolvent. 

\Proof 

We first estimate the excitation energy:
\beq
\mc{E}_{\e{ex}} \; = \; \sul{a=1}{n} \veps_0(z_a) \; - \;  \sul{a=1}{n+n_w} \veps_0(\nu_{h_a})  \; + \; \mc{E}^{(\nu)} \; - \; \mc{E}^{(\la)}
\qquad \e{with} \quad \left\{  \ba{c}  \mc{E}^{(\nu)} \, = \;  \sul{a=1}{N+n_w} \veps_0(\nu_{a})  \\ 
					\mc{E}^{(\la)} \, = \;  \sul{a=1}{N} \veps_0(\la_{a}) 				\ea \right. \;. 
\enq
 By applying an argument similar to the one in Section 1.2, the latter two terms have representations
\beq
\mc{E}^{(\nu/\la)} \, = \; L \Int{-\tf{\pi}{2} }{ \tf{\pi}{2} } \wh{\xi}_{\mu/\la}^{\prime}(s) \cdot \veps_0(s) \cdot \dd s  
	  \; - \; \sul{\eps=\pm }{} \Int{ \Ga^{(\eps)} }{} \veps_0^{\prime}(s) \cdot \wh{u}^{(\eps)}_{\mu/\la}(s) \cdot \f{ \dd s }{ 2\i \pi } \;,
\enq
where obvious notations are employed:   $\wh{\xi}_{\la}(s)$ denotes the counting function that is obtained from 
\eqref{ecriture representation originale pour xi kappa} upon replacing $\mu_k$ by $\la_k$ and setting $\alpha=0$,
while $\wh{u}^{(\eps)}_{\la}(s)$  is obtained from \eqref{definition fonction hat u} by replacing  $\wh{\xi}_{\mu}(s)$ by  $\wh{\xi}_{\la}(s)$.

Since $\veps_0$ is a periodic function, one obtains
\beq
\mc{E}_{\e{ex}} \; = \; \sul{a=1}{n} \veps_0(z_a) \; - \;  \sul{a=1}{n+n_w} \veps_0(\nu_{h_a})  \; - \; 
\Int{ - \tf{\pi}{2} }{ \tf{\pi}{2} } \wh{F}^{\prime}(s) \veps_0(s) \cdot \dd s \; + \; \e{O}\big( L^{-\infty} \big) \;,
\label{excitation energy exprimee en termes de F}
\enq
where
\beq
\label{shiftfunction}
\wh{F}(\om) \; = \; L\cdot \big( \wh{\xi}_{\la}(\om) \, - \, \wh{\xi}_{\mu}(\om) \big)
\enq
is the shift function associated with the states parametrised by
$\{\mu_a\}_1^N$ and $\{ \la_a \}_1^N$. Its large-$L$ expansion is readily 
deduced from Proposition \ref{Propositon large L asympt ctg fct}:
\beq
\wh{F}(s) \; = \; \f{ 1 }{ 2 \i \pi } \cdot \bigg\{  \sul{a=1}{n} \vp\big( s , z_a \big) - \sul{a=1}{n+n_w} \vp\big( s , \nu_{h_a} \big)  \bigg\}  
\; + \; \f{\a + \iota}{ 2 }  \; +  \; \e{O}\Big( L^{-\infty} \Big) \;. 
\label{F_aysmptotics}
\enq
Here we have anticipated that $\iota = 0$ for the ground state
(see section~\ref{explainiota}).
The large-$L$ asymptotics of the function $\wh{F}^{\prime}(s)$ can be
expressed in terms of the resolvent kernel, \textit{cf}.\
Appendix \ref{Subsection Resolvent kernel}. More precisely, it follows %readily
immediately  from the integral equation satisfied by $\phi(s,z)$ and the
quasi-periodicity properties of $\th$ that
\beq
\vp\big(s+\tf{\pi}{2},z\big) \, - \, \vp\big(s-\tf{\pi}{2},z\big) \, = \, \i \pi \cdot \bs{1}_{|\Im z |<\eta }
\enq
As a consequence, $\Dp{s}\vp(s,z)$ solves the linear integral equation 
\beq
\big(I+K\big)\big[ \Dp{*}\vp(*,z)\big](s) \; = \; 2\i \pi  K(s-z) \;. 
\enq
Applying the inverse $I-R$ to the operator $I+K$ and invoking
the integral equation (\ref{resolvent_kernel_ieqn}) satisfied by the
resolvent kernel $R(s)$ leads to 
\beq
\f{1}{2\i \pi }\Dp{s}\vp(s,z) \; = \; K(s-z) \; - \; \Int{-\f{\pi}{2} }{ \f{\pi}{2} }R(s-w)K(w-z) \dd w \;.
% + \; 
%
%\f{1}{2}\bs{1}_{|\Im z |<\eta } R\big(s-\f{\pi}{2} \big) \;. 
%
\enq
The expression for the integral term depends on the domain in the
complex plane where $z$ belongs to. The integral can, in fact, be
estimated by using a reasoning similar to that invoked when implementing
the analytic continuations of the functions $\kappa_{\ua/\da}$
and $\kappa_{\e{c} }$. One obtains
\beq
\f{1}{2\i \pi }\Dp{s}\vp(s,z) \; = \;
   \begin{cases}
      R(s-z) & | \Im(z) | < \eta \\ 
      R(s-z) \; + \; R\big(s-z \pm \i \eta  \big)  &   \pm  \Im(z) > \eta
   \end{cases}  \;. 
\enq
Now introduce the function 
\beq
\left. \ba{ccc} \mc{R}_{\ua}[\veps_0](z) & \e{for} & \Im(z) > \eta \vspace{2mm} \\  
		\mc{R}_{ \e{c} }[\veps_0](z) & \e{for} &  |\Im(z)| < \eta   \vspace{2mm}  \\
		\mc{R}_{\da}[\veps_0](z)  & \e{for} &  \Im(z) < -\eta     \ea \right\}
		 = \Int{-\tf{\pi}{2} }{ \tf{\pi}{2} } R(s-z) \veps_0(s) \cdot \dd s. 
\enq
It follows from 
\beq
R(s-z)  \underset{s\tend t}{\sim} \f{ \mp 1}{2\i \pi} \cot(s-t) \qquad \e{with}  \quad z=t\pm \i \eta
\enq
that the three functions above satisfy %to 
the jump relations
\beq
\mc{R}_{\ua;+}[\veps_0](x+\i\eta) \; - \; \mc{R}_{\e{c};-}[\veps_0](x+\i\eta) \; = \; -\veps_0(x) \quad \e{and} \quad
\mc{R}_{\da;-}[\veps_0](x-\i\eta) \; - \; \mc{R}_{\e{c};+}[\veps_0](x-\i\eta) \; = \; -\veps_0(x)\;, 
\enq
with $x \in \R$. 
Thanks to these, we are able to estimate $\mc{R}_{\ua/\da}[\veps_0](z)$ 
by analytic continuation based on $\mc{R}_{\e{c}}[\veps_0](z) \, = \, \veps_0(z) \, - \, \veps(z)$:
\beq
\mc{R}_{\ua}[\veps_0](z) \; = \; -\veps_0(z-\i \eta ) \, + \, \veps_0(z) \, - \, \veps(z)  \qquad \e{and} \qquad 
\mc{R}_{\da}[\veps_0](z) \; = \; -\veps_0(z+\i \eta ) \, + \, \veps_0(z) \, - \, \veps(z)  \;. 
\enq
In the above formulae, $\veps$ refers to the dressed energy %$\veps$ 
that has been defined in \eqref{definition dressed and bare energies}. 

\noindent We have just proven that 
\beqa
\veps_0(z) \; - \; \Int{-\f{\pi}{2} }{ \f{\pi}{2} } \Dp{s}\vp(s,z) \veps_0(s) \cdot \f{ \dd s}{2\i \pi } \; = \; \veps(z) \, +  \,\veps(z\mp \i \eta) 
 & \qquad \e{for} & \qquad \pm \Im(z) >\eta \\
\veps_0(z) \; - \; \Int{-\f{\pi}{2} }{ \f{\pi}{2} } \Dp{s}\vp(s,z) \veps_0(s) \cdot \f{ \dd s}{2\i \pi } \; = \; \veps(z)% \, +  \,
%
%    \f{1}{2}\big( \veps(\tf{\pi}{2}) - \veps_0(\tf{\pi}{2}) \big)
%
 & \qquad \e{for} & \qquad  |\Im(z)| <\eta  \; . 
\eeqa
By utilising  \eqref{excitation energy exprimee en termes de F} and the asymptotic form \eqref{F_aysmptotics} one obtains
\beq
\mc{E}_{\e{ex}} \; = \; \sul{ a=1 }{ \tf{n_w}{2} } \Big\{ \veps(y_a)+\veps(y_a-\i\eta) + \veps(\ov{y}_a)+\veps(\ov{y}_a+\i\eta)\Big\}
\; + \; \sul{ a=1 }{ \tf{n_c}{2} } \Big\{ \veps(w_a)+\veps(w_a-\i\eta+\de_a) \Big\} \; - \; \sul{ a=1 }{ n+n_w }  \veps(\nu_{h_a})  \; + \; \e{O}\big( L^{-\infty} \big) \;. 
\enq
Using the $\i \eta$ anti-periodicity of $p^{\prime}$, one arrives at the expression
\beq
\mc{E}_{\e{ex}} \; =  \;  n_{\chi} h \; - \; \sul{ a = 1 }{ 2 n_{\chi} } \veps(\nu_{h_a}) \;. 
\enq
It is then enough to use the explicit expression for $\veps$ so as to
see that the $h$-dependent terms cancel out, hence leading to  
\eqref{expressions pour energie et impulsion excitation}.

We now pass on to the estimation of the $\mc{P}_{\e{ex}}$.
A handling similar to the previous steps yields
\bem
\label{momentumthroughshiftfunction}
\sul{a=1}{N+n_w} p_0(\nu_a) \; = \; -L \Int{-\f{\pi}{2} }{ \f{\pi}{2} } \wh{\xi}_{\mu}(s)p_0^{\prime}(s) \cdot \dd s 
\, + \, p_0\big( \f{\pi}{2} \big) \cdot \big(N+n_w+\f{1}{2}\big) \; - \; 
\sul{\eps=\pm}{}  \Int{ \Ga^{(\eps)} }{} p_0^{\prime}(s) \wh{u}_{\mu}^{(\eps)} \cdot \f{ \dd s }{2\i\pi }  \\
\; + \; L \Int{0}{ x_{\mu} } p_0^{\prime}\big( s+\f{\pi}{2} \big)\cdot 
\underbrace{ \Big\{ \wh{\xi}_{\mu}\big( x_{\mu} - \f{\pi}{2} \big) \, - \, \wh{\xi}_{\mu}\big( x_{\mu} + \f{\pi}{2} \big) 
\, + \,  \wh{\xi}_{\mu}\big( s + \f{\pi}{2} \big) \, - \, \wh{\xi}_{\mu}\big( s - \f{\pi}{2} \big)  \Big\} }_{ =0 } \cdot \dd s \;. 
\end{multline}
Therefore,
\beq
\mc{P}_{ \e{e x}}\; = \; \f{\a + \iota}{2} p_0\big( \f{\pi}{2} \big) \; + \; \sul{a=1}{n} \wt{p}(z_a) \, - \, \sul{a=1}{n+n_w} \wt{p}(\nu_{h_a}) 
\; + \; n_w \wt{p}\big( \f{\pi}{2} \big) \; + \; \e{O}\big( L^{-\infty} \big) 
\enq
where we have set 
\beq
\wt{p}(z) \; = \; p_0(z) \; + \;  \Int{ -\f{\pi}{2} }{ \f{\pi}{2} } \phi(s,z)p_0^{\prime}(s) \cdot \f{ \dd s }{ 2\i \pi } \;. 
\enq

It follows from the integral equation satisfied by the dressed momentum and from the fact that $z\mapsto \phi(s,z)$ is analytic on $|\Im(z)|<\eta$
uniformly in $s\in \intff{-\tf{\pi}{2}}{\tf{\pi}{2}}$ that $\wt{p}(z)=2\pi p(z)$ for $|\Im(z)|<\eta$. Its value in other regions of the complex plane 
can be obtained by computing the jumps on the lines $\Im(z) = \pm \eta$ of the functions
\beq
\left. \ba{ccc} \Phi_{\ua}[p_0](z) & \e{for} &  \Im(z) > \eta \vspace{1mm} \\  
		\Phi_{ \e{c} }[p_0](z) & \e{for} &  |\Im(z)| < \eta   \vspace{1mm}  \\
		\Phi_{\da}[p_0](z)  & \e{for} &  \Im(z) < -\eta     \ea \right\}
		 = \Int{-\tf{\pi}{2} }{ \tf{\pi}{2} } \phi(s,z) p_0^{\prime}(s) \cdot \f{\dd s }{2\i\pi} \;. 
\enq
We start from the jump equations
\beq
\th\big(\la-x\pm \i \eta +\i 0^+ \big) \, - \,  \th\big( \la - x\pm \i \eta -\i 0^+\big)  \; =
 \; \mp 2\i \pi \bs{1}_{ \la  \in  \intff{x}{ \tf{\pi}{2} } }%(\la)
\enq
%
%
%
%where $\bs{1}_{A}$ is the indicator function of the set $A$ one gets that the function 
and introduce an convenient function $g_{\pm}$ defined by 
\beq
g_{\pm}(\la,x)\; = \; \phi\big( \la,x\pm\i\eta-\i0^+ \big) \, - \, \phi\big( \la,x\pm\i\eta+\i0^+ \big).
\enq
Immediately seen, it solves the integral equation
\beq
\big( I+K \big)[g(\cdot, x) ](\la)  \; = \; \mp 2\i \pi \bs{1}_{ \la  \in \intff{x}{\tf{\pi}{2}} } %(\la)
 \qquad viz. \qquad 
g_{\pm}(\la,x) \; = \; \mp 2\i \pi \bs{1}_{  \la  \in  \intff{x}{ \tf{\pi}{2} } }%(\la) 
\mp \phi\big(\la,\f{\pi}{2}\big) \pm \phi\big(\la,x\big) \;. 
\enq
This leads to the jump conditions 
\beq
\Phi_{\ua;+}[p_0](x+\i \eta) \,  - \, \Phi_{\e{c};-}[p_0](x+\i \eta)  \; = \; p(x) \, - \, p\big( \f{\pi}{2}\big)  \quad  \e{and}  \quad 
\Phi_{\da;-}[p_0](x+\i \eta) \,  - \, \Phi_{\e{c};+}[p_0](x+\i \eta)  \; = \; p(x)\, - \, p\big( \f{\pi}{2}\big)  \;. 
\nonumber
\enq
Thus, for $\pm \Im(z)>\eta$ but close to the line $\Im(z)= \pm \eta$ one has that 
\beq
\wt{p}(z) \; = \; p_0(z) \, + \, \Big( p(z)+p(z\mp\i \eta) -p_0(z) - p_0\big(\f{\pi}{2}\big)\Big) \; = \; - p_0\big(\f{\pi}{2}\big)
\enq
This formula holds, in fact, in the whole domain $|\Im(z)|> \eta$ by analytic continuation. 
All-in-all, this leads to \eqref{expressions pour energie et impulsion excitation}. \qed

\subsection{General parametrisation of the excited states for large $L$}
\label{explainiota}
To lay a firm ground for the calculation of form factors, we have
reconsidered the analysis of the excitations of the XXZ chain in the
massive regime in the large-$L$ limit, originally performed in
\cite{BabelondeVegaVialletStringHypothesisWrongXXZ,
VirosztekWoynarovichStudyofExcitedStatesinXXZHigherLevelBAECalculations}.
We have obtained, in particular, the large-$L$ limits of the counting
function and the shift function \eqref{shiftfunction} which will play a prominent role below.
It remains to summarize our results and to interpret the parameter
$\iota$ introduced in (\ref{firstroot}) and (\ref{nextroots}).
A similar parameter was first introduced in
\cite{VirosztekWoynarovichStudyofExcitedStatesinXXZHigherLevelBAECalculations}.
As we shall see, it distinguishes states that are degenerate in the
thermodynamic limit $L \rightarrow \infty$. It is clear that such a
degeneracy must exist for the ground state sector in the Ising limit $\Delta
\rightarrow \infty$, where the symmetric and antisymmetric combinations
of the two N\'eel states are the degenerate ground states of the model.
As was already observed by Orbach \cite{OrbachXXZCBASolution} such type
of degeneracy persists for any finite $\Delta > 1$ in the limit
$L \rightarrow \infty$. We shall see that the corresponding states
are distinguished by different values of~$\iota$.

Let us start with the discussion of the ground state.
It follows from the arguments brought up in \cite{Yang-YangXXZproofofBetheHypothesis}
that the Bethe roots $\{\la_a\}_1^N$ pertaining to the ground state
are all real and solve the logarithmic Bethe Ansatz equations, 
\beq
\f{ p_0(\la_a) }{ 2\pi }  \; - \; \sul{k=1}{N} \f{\th(\la_a-\la_k) }{ 2 \i \pi L } \; = \;  \f{a-\tf{1}{2}}{L}  \qquad \e{with} \quad a=1,\dots,N \;. 
\label{ecriture Log BAE GS}
\enq
Consequentially, they can be deduced from the counting function
(\ref{ecriture representation originale pour xi kappa}) with $\a=0$
and $n = n_w = n_h = 0$. 

%The XXZ chain in its massive regime possesses an excited state
%whose energy is exponentially in $L$ close to the one of the ground state.
%We shall refer to this eigenstate as the quasi ground state.
There is another state sharing these properties. The real Bethe roots
$\{\check{\la}_a\}_1^N$ describing this state correspond to the
solution to the logarithmic Bethe Ansatz equations
\beq
\f{ p_0(\check{\la}_a) }{ 2\pi }  \; - \;
\sul{k=1}{N} \f{\th(\check{\la}_a-\check{\la}_k) }{ 2 \i \pi L } \; = \; 
\f{a-\tf{3}{2}}{L}  \qquad \e{with} \quad a=1,\dots,N \;. 
\label{ecriture Log BAE QGS}
\enq
Again $n = n_w = n_h = 0$. Comparing
(\ref{ecriture representation originale pour xi kappa}),
(\ref{ecriture Log BAE GS}), (\ref{ecriture Log BAE QGS})
and (\ref{nextroots}) we infer that $\iota = 1$ for this state,
while $\iota = 0$ for the ground state. It follows from
Proposition \ref{Proposition estimation excitation momentum and energy}
that the excitation energy associated with this state is
$\; \mc{E}_{\e{ex     }} \, = \, \e{O}\big( L^{-\infty} \big)$.
Thus, in the thermodynamic limit it is degenerate with the
ground state. For this reason we shall call it the quasi
ground state. Again from Proposition
\ref{Proposition estimation excitation momentum and energy} it follows
that the momentum of the quasi ground state is
$\mc{P}_{\e{ex}} \, = \, \pi \, + \, \e{O}\big( L^{-\infty} \big)$.\footnote{%
In fact, it follows directly from (\ref{ecriture Log BAE GS}),
(\ref{ecriture Log BAE QGS}) that $\mc{P}_{\e{ex}} \, = \, \pi$ for arbitrary $L$.}
This situation is familiar from the Ising limit, where we have
two degenerate ground states with momenta $0$ and $\pi$.
%It follows %readily 
%from Proposition \ref{Proposition estimation excitation momentum and energy} that the excitation momentum and energy associated with these states
%is  $\mc{P}_{\e{ex}} \, = \, \pi \, + \, \e{O}\big( L^{-\infty} \big)$ and $\; \mc{E}_{\e{ex}} \, = \, \e{O}\big( L^{-\infty} \big)$. 

The logarithmic Bethe equations 
(\ref{ecriture Log BAE GS}), (\ref{ecriture Log BAE QGS}) are distinguished
by an overall shift of the parameters $a$ by one. Note that any other
overall shift would reduce the solution of the corresponding logarithmic
Bethe equations to one of the two solutions $\{\la_a\}_1^N$ or
$\{\check{\la}_a\}_1^N$ up to shifts by multiples of $\pi$ of the Bethe
roots. This can be seen as follows. Let $\{ \la_a \}_1^N$ be the solution
to \eqref{ecriture Log BAE GS} and define a set of roots 
\beq
\{v_a\}_1^N=\{v_1,\dots,v_N\}=\{\la_{k+1},\dots,\la_N,\la_1+\pi,\dots,\la_k+\pi \} \; .
\enq
It is then easy to see, by using the quasi-periodicity properties of
the bare phase and momentum, that 
\beq
\f{ p_0(v_a) }{ 2\pi }  \; - \; \sul{b=1}{N} \f{\th(v_a-v_b) }{ 2 \i \pi L } \; = \;  \f{a+2k-\tf{1}{2}}{L} \;. 
\enq
A similar statement holds with respect to the roots $\{ \check{\la}_a \}_1^N$,
replacing $2k \hookrightarrow 2k-1$.

The result of this section is that, in the large-$L$ limit, we can
associate with every solution $\{ \mu_a \}_1^N$ of the
$\a$-twisted Bethe equations (\ref{ecriture BAE}) a unique counting
function $\xi_\mu^{(\infty)}$ characterized by a set of `macroscopic data':
$\iota \in {\mathbb Z}$; $n, n_w \in {\mathbb Z}_+$, and $\{ \nu_{h_a} \}_1^{n+n_w}$,
$\{ \chi_a \}_1^{n_\chi}$ determined by (\ref{ecriture HBAE in terms of chi's}) and
(\ref{logbaeholes}). This counting function also determines the shift function
(\ref{shiftfunction}) and the energy and momentum of excitations through
(\ref{excitation energy exprimee en termes de F}) and
(\ref{momentumthroughshiftfunction}). Thus, all information about the
excited states at large $L$ is encoded in $\xi_\mu^{(\infty)}$ and its
parameters. They provide a general parametrisation of the excited states
which we will use below in order to describe the form factors of the
local operator $\sigma^z$.

It appears convenient for our further handlings to choose a slightly different parametrisation of the excited states. 
Originally, we fix $\iota \in \mathbb{Z}$ uniquely by demanding that $\{\nu_a\}_1^{N+n_w} \subset \intff{-\tf{\pi}{2} }{ \tf{\pi}{2} }$. 
However, changing $\iota$ by an even integer leaves 
${\rm e}^{2 \pi {\rm i} \wh{\xi}_\mu}$ unchanged while 
Bethe roots and hole parameters change at most by multiples of $\pi$. In such a 
situation they are not, in general, inside of the interval $\intfo{-\tf{\pi}{2}}{\tf{\pi}{2}}$
anymore. Still, such a shift does not change the expressions of normalised
Bethe states and form factors. 
Therefore, we will parametrise the excited states solely by the values 
$\iota \in \{0, 1\}$ while allowing some of the parameters $\{\nu_a\}_1^{N+n_w}$  to be located 
outside of $\intfo{-\tf{\pi}{2}}{\tf{\pi}{2}}$. In this situation, some of the Bethe roots $\{\mu_a\}_1^N$ may as well
move out of $\Re(\mu_a) \in \intfo{-\tf{\pi}{2}}{\tf{\pi}{2}}$. With this choice of parametrisation of the states, 
they are now defined through \eqref{definition line bethe roots et nu roots}.

As we have seen with the example of the ground state and the pseudo
ground state above, states with counting functions which differ by the
values of $\iota$ are generally inequivalent. Still, if we admit
$\iota = 0$ and $\iota = 1$ for any choice of $n, n_w$ we may encounter
the situation where two different sets of hole parameters
$\{\nu_a^{(\iota)}\}_1^{n_h}$, with $\iota = 0, 1$ are congruent
modulo $\pi$. Since $x_\mu^{(0)} - x_\mu^{(1)} = O (1/L)$, where we
denoted the corresponding boundary parameters by $x_\mu^{(\iota)}$,
we expect that only holes near the boundaries $\pm \pi/2$ may be
connected by shifts by $\pm \pi$. Consider a hole $\nu_{h_j}^{(1)}$
near the left boundary. Expanding (\ref{nextroots}) with $\nu_a = \nu_{h_j}$
around $x_\mu^{(1)} - \pi/2$ and using (\ref{firstroot}) and
(\ref{xiinfinity}) we obtain
\begin{equation}
     \nu_{h_j}^{(1)} + \pi = x_\mu^{(1)} + \frac \pi 2
                             + \frac{h_j^{(1)} - 1/2}{L p'(-\pi/2)}
			     + O(1/L^2) \; .
\end{equation}
If this is equal to a root $\nu_{h_k}^{(0)}$, then the right hand side
must be smaller than $x_\mu^{(0)} + \pi/2$. Assuming that there is
only a single such hole and using (\ref{xplicit}) we conclude that
\begin{equation}
     \frac{h_j^{(1)} - 1/2}{L p'(-\pi/2)} < x_\mu^{(0)} - x_\mu^{(1)} =
        \frac{1}{L p'(-\pi/2)} + O(1/L^2) \; .
\end{equation}
This can only be true if $h_j^{(1)} = 1$. It follows from (\ref{xiinfinity})
that the set of holes $\{\nu_1, \nu_{h_2}, \dots, \nu_{h_{n_h}}\}$ with
$\iota = 1$ and $\{\nu_{h_2}, \dots, \nu_{h_{n_h}}, \nu_1 + \pi \}$ with
$\iota = 0$ belong indeed to the same counting functions $\xi_\mu^{(1)} =
\xi_\mu^{(0)}$.

This means that if we take all solutions of (\ref{logbaeholes})
and (\ref{ecriture HBAE in terms of chi's}) with $\iota = 0, 1$
we double-count a certain number of equivalent solutions.
For a fixed number $n + n_w$ of holes there are
$\genfrac{(}{)}{0pt}{}{N + n_w - 1}{n + n_w - 1}$ sets of holes
with $h_1 = 1$. This implies that the relative number of
such double-counted states is
\begin{equation}
     \genfrac{(}{)}{0pt}{}{N + n_w - 1}{n + n_w - 1} \Bigg/
     \genfrac{(}{)}{0pt}{}{N + n_w}{n + n_w} = \frac{n + n_w}{N + n_w}
     = O(1/L) \;
\end{equation}
(compare \cite{VirosztekWoynarovichStudyofExcitedStatesinXXZHigherLevelBAECalculations}).
Using a similar argument as above one can easily show that equivalent
sets of holes with different values of $\iota = 0, 1$ can at most differ by
one element, in such a way that $\nu_1^{(1)} + \pi = \nu_{n_h}^{(0)}$.
Hence, the case considered above is already the most general
case of double-counting. Since the relative number of double-counted
states vanishes in the thermodynamic limit, we shall assume that $\iota$
can be arbitrarily chosen from $\{0, 1\}$.

To summarize, in the large-$L$ limit, the excited states having an excitation
energy with respect to the ground state which is finite in $L$
can be parametrised by 
\begin{itemize}
\item a choice of $\iota \in \{0,1\}$;
\item a choice of $n_h=n+n_w$ hole-integers $1<h_1<\dots<h_{n_h}<N+n_w$
which give rise to a set of $n_h$ hole parameters $\{\nu_{h_a}\}_1^{n_h}$
such that $\nu_{h_a}=\ga_{h_a}+\e{O}(L^{-1})$ with $\ga_k$
being the unique solution to $p(\ga_k)=\tf{k}{L}$, $k\in \mathbb{Z}$;
\item a set of complex roots $\{\chi_a\}_1^{n_{\chi}}$, $n_{\chi}= (n + n_w)/2$, which
corresponds to a solution to the higher-level Bethe Ansatz equations
\eqref{ecriture HBAE in terms of chi's}. 
\end{itemize}
%
%
%

%%%%%%%%%%%%%%%%%%%%%%%%%%%%%%%%%%%%%%%%%%%%%%%%%%%%%%%%%%%%%%%%%%%%%%%%%%%%%%%%%%%%%%%%%%%%%%%%%%%%%%%%%%%%%%%%%%%%%%%%%%%%%%%%%%%%%%%%%%%%%%%
%%%%%%%%%%%%%%%%%%%%%%%%%%%%%%%%%%%%%%%%%%%%%%%%%%%%%%%%%%%%%%%%%%%%%%%%%%%%%%%%%%%%%%%%%%%%%%%%%%%%%%%%%%%%%%%%%%%%%%%%%%%%%%%%%%%%%%%%%%%%%%%

%%%%%%%%%%%%%%%%%%%%%%%%%%%%%%%%%%%%%%%%%%%%%%%%%%%%%%%%%%%%%%%%%%%%%%%%%%%%%%%%%%%%%%%%%%%%%%%%%%%%%%%%%%%%%%%%%%%%%%%%%%%%%%%%%%%%%%%%%%%%%%%
%%%%%%%%%%%%%%%%%%%%%%%%%%%%%%%%%%%%%%%%%%%%%%%%%%%%%%%%%%%%%%%%%%%%%%%%%%%%%%%%%%%%%%%%%%%%%%%%%%%%%%%%%%%%%%%%%%%%%%%%%%%%%%%%%%%%%%%%%%%%%%%

%%%%%%%%%%%%%%%%%%%%%%%%%%%%%%%%%%%%%%%%%%%%%%%%%%%%%%%%%%%%%%%%%%%%%%%%%%%%%%%%%%%%%%%%%%%%%%%%%%%%%%%%%%%%%%%%%%%%%%%%%%%%%%%%%%%%%%%%%%%%%%%
%%%%%%%%%%%%%%%%%%%%%%%%%%%%%%%%%%%%%%%%%%%%%%%%%%%%%%%%%%%%%%%%%%%%%%%%%%%%%%%%%%%%%%%%%%%%%%%%%%%%%%%%%%%%%%%%%%%%%%%%%%%%%%%%%%%%%%%%%%%%%%%

%%%%%%%%%%%%%%%%%%%%%%%%%%%%%%%%%%%%%%%%%%%%%%%%%%%%%%%%%%%%%%%%%%%%%%%%%%%%%%%%%%%%%%%%%%%%%%%%%%%%%%%%%%%%%%%%%%%%%%%%%%%%%%%%%%%%%%%%%%%%%%%
%%%%%%%%%%%%%%%%%%%%%%%%%%%%%%%%%%%%%%%%%%%%%%%%%%%%%%%%%%%%%%%%%%%%%%%%%%%%%%%%%%%%%%%%%%%%%%%%%%%%%%%%%%%%%%%%%%%%%%%%%%%%%%%%%%%%%%%%%%%%%%%

\section{\boldmath The form factors of the spin operator $\sg^z$}
\label{Section Large L behaviour of Form Factors}
In the present section we shall estimate the large-$L$ behaviour of the quantity 
\beq
\mc{F}^{(z)}_m\Big( \{ \wt{\mu}_a\}_1^N ; \{\la_a\}_1^N\Big) \; \equiv  \; 
\f{ \bra{ \psi\big( \{\la_a\}_1^N \big) } \sg_1^z \ket{ \psi\big( \{ \wt{\mu}_a\}_1^N \big) }  
\bra{ \psi\big( \{ \wt{\mu}_a\}_1^N \big) } \sg_{m+1}^z \ket{ \psi\big( \{\la_a\}_1^N \big) }}
{ \norm{  \psi\big( \{\la_a\}_1^N \big) }^{2} \cdot  \norm{  \psi\big( \{ \wt{\mu}_a\}_1^N \big) }^{2}  } 
\enq
where $\{\wt{\mu}_a\}_1^N$ is a set of Bethe roots solving the Bethe
equations \eqref{ecriture BAE} at $\a = 0$ and satisfying the hypotheses
of Sub-Section \ref{Subsection Asympt behaviour counting function},
while $\{\la_a\}_1^N$ is the set of Bethe roots which characterize the
ground state. Note that $\mc{F}^{(z)}_m\Big( \{\la_a\}_1^N ; \{\la_a\}_1^N \Big)$
is the square of the ground state magnetization and therefore vanishes. For this
reason and in order to avoid to distinguish cases we assume in the following
that $\{\wt{\mu}_a\}_1^N \ne \{\la_a\}_1^N$.

\subsection{\boldmath Formulae at finite $L$}

The form factor $\mc{F}^{(z)}_m\Big( \{\wt{\mu}_a\}_1^N ; \{\la_a\}_1^N \Big) $
is most easily accessed by means of its generating function (see \textit{e}.\textit{g}.
\cite{KozKitMailSlaTerEffectiveFormFactorsForXXZ} and references therein).
More precisely, one has 
\beq
\mc{F}^{(z)}_m\Big( \{\wt{\mu}_a\}_1^N ; \{\la_a\}_1^N  \Big) \; = \; 
\f{2}{\pi^2}
%\sin^2\Big( \f{ \mc{P}_{\e{ex}}^{(\iota)} }{2 } \Big)\cdot 
%\ex{\i m \mc{P}_{\e{ex}}^{(\iota)} } \cdot
\sin^2\Big( \f{ \mc{P}_{\e{ex}} }{2 } \Big)\cdot 
\ex{\i m \mc{P}_{\e{ex}} } \cdot
\f{ \Dp{}^2 }{ \Dp{} \a^2 }
%\bigg\{
%
\mc{S}^{(\a)}_N \Big( \{ \mu_a \}_1^N \, ; \, \{ \la_a \}_1^N \Big)
%\bigg\}
\Biggr|_{\a=0} \;.
\label{ecriture relation FF et partial alpha of S alpha}
\enq
In this formula, the momentum $\mc{P}_{\e{ex}}$
and the $\a$-twisted scalar product $ \mc{S}^{(\a)}_N $
on the right-hand side are both evaluated at a solution $\{\mu_a \}$ to the
$\a$-twisted Bethe equation \eqref{ecriture BAE} such that
%${\mu_{a}}_{\mid \a= \iota} = \wt{\mu}_a$
${\mu_{a}}_{\mid \a= 0} = \wt{\mu}_a$.
%The excitation momentum is defined by
%
%
%
%\beq
%
% \mc{P}_{\e{ex}}^{(\iota)} \; = \; \sul{a=1}{N}\Big(p_0(\mu_a) \, - \, p_0(\la_a)  \Big) \; - \; \pi (\a-\iota) \;. 
%%
%\enq
%
%
%
%Its large-$L$ asymptotic behaviour is readily deduced from Proposition
%\ref{Proposition estimation excitation momentum and energy}. 

In order to present the expression for $\mc{S}^{(\a)}_N$, we first need to
introduce  some notation. 
Let $I+\wh{U}_{\th}$ be an operator acting on the space of functions supported
on the loop\footnote{By choosing this particular loop we have made the
assumption that $\om \mapsto 1-\ex{2 \i \pi \wh{F}(\om)}$
has no zeroes inside. This, however, is not a restriction. Indeed,
the assumption always holds for $\Im(\a)>0$ and large enough. The
expressions that are given below should then be understood as analytic
continuations from the domain $\Im(\a)>0$ and large enough to $\Im(\a)=0$.
See \cite{KozFFConjFieldNLSELatticeSpacingGoes0} for more details.}
$\Ga=\Ga^{(+)} \cup \Ga^{(-)}$, \textit{cf}.\ 
\eqref{ecriture reste et defnition ctrs gama pm}, whose integral
kernel $\wh{U}_{\th}\big( \om, \om^{\prime} \big)$ takes the form 
\beq
\wh{U}_{\th}\big( \om, \om^{\prime} \big) \; = \; 
\pl{a=1}{N}\bigg\{ \f{ \sin(\om - \la_a + \i  \eta) \sin(\om - \mu_a )  }{ \sin(\om - \mu_a + \i  \eta) \sin(\om - \la_a )}  \bigg\}
\cdot \f{ K_{\a}\big( \om - \om^{\prime}\big)  \, - \, K_{\a}\big( \th - \om^{\prime}\big)  }
{ 1-\ex{2 \i \pi \wh{F}(\om)}  } \;. 
\enq
We have set in the above
\beq
K_{\a}(\om) \; = \; \f{-1}{2\i \pi} \Big\{ \cot(\om + \i \eta) \; - \; \ex{2\i\pi \a} \cot(\om-\i\eta) \Big\} \;
\qquad \e{implying}\;\;  K_0(\om) \, = \, K(\om)\;.  
\enq
Finally, we need to introduce the shorthand notation for a double product of interest:
\beq
\mc{W}\Big( \{x_a\}_1^m;\{y_a\}_1^n\Big) \; = \; \f{ \prod_{a=1}^{m}\prod_{b=1}^{n} \big\{ \sin(x_a-y_b- \i \eta)\sin(y_b-x_a- \i \eta)  \big\}  }
{ \prod_{a,b=1 }^{m} \big\{ \sin(x_a-x_b - \i \eta) \big\} \cdot \prod_{a,b=1 }^{n} \big\{\sin(y_a-y_b- \i \eta)\big\}  } \;. 
\enq
With all these definitions in hand, the expression for the $\a$-twisted
form factor reads \cite{KozKitMailSlaTerEffectiveFormFactorsForXXZ,KMTFormfactorsperiodicXXZ}
\bem
\mc{S}^{(\a)}_N \Big( \{ \mu_a \}_1^N \, ; \, \{ \la_a \}_1^N \Big) \; = \; \mc{W}\Big( \{\la_a\}_1^N;\{\mu_a\}_1^N \Big)
%
%\pl{a,b=1}{N} \bigg\{ \f{ \sin\big( \mu_a-\la_b-i\eta \big)  \sin\big( \la_a-\mu_b-i\eta \big) }
%
%{ \sin\big( \mu_a-\mu_b-i\eta \big)  \sin\big( \la_a-\la_b-i\eta \big) } \bigg\} 
%
\cdot \det^2_N\bigg[ \f{ 1 }{ \sin\big( \la_a-\mu_b\big) }\bigg] \cdot 
\pl{a=1}{N} \bigg\{ \f{ \big|\ex{2\i\pi \wh{F}(\la_a)}-1  \big|^2 }
{   (2\pi L)^2 \wh{\xi}^{\prime}_{\mu}(\mu_a) \wh{\xi}^{\prime}_{\la}(\la_a) } \bigg\}   \\
\times   
\f{ \big(1- \ex{2i\pi \a} \big)^2 }{ \det_N \big[\Xi^{(\mu)}\big] \cdot \det_N \big[\Xi^{(\la)}\big] }  \cdot 
\pl{p=1}{2} \Bigg\{ \f{  \det_{\Ga}\big[ I + \wh{U}_{\th_p} \big] }
{1-\ex{2  \i \pi \wh{F}(\th_p)}   } 
 \pl{a=1}{N} \bigg\{ \f{ \sin(\th_p - \la_a + \i  \eta) }{ \sin(\th_p - \mu_a + \i  \eta) }  \bigg\} 
\;  \Bigg\}
\label{definition facteur S alpha gen spin spin}
\end{multline}
and involves determinants of the matrices 
\beq
\Xi^{(\la)}_{ab} \; = \; \de_{ab} \; + \; \f{ K(\la_a-\la_b) }{ L \wh{\xi}^{\prime}_{\la}(\la_b) }
\qquad \e{and} \qquad
\Xi^{(\mu)}_{ab} \; = \; \de_{ab} \; + \; \f{ K(\mu_a-\mu_b) }{ L \wh{\xi}^{\prime}_{\mu}(\mu_b) } \;. 
\enq

Note that the difference between the representation \eqref{definition facteur S alpha gen spin spin} for $\mc{S}^{(\a)}$ and the one that appeared in
\cite{KozKitMailSlaTerEffectiveFormFactorsForXXZ} stems from the fact that we have already implemented the rotation of Bethe roots by $\ex{\i\f{\pi}{2}}$ which is 
appropriate for treating the massive regime of the chain. The representation \eqref{definition facteur S alpha gen spin spin} involves two parameters $\th_1$ and $\th_2$. 
They can be chosen arbitrarily in that the representation \eqref{definition facteur S alpha gen spin spin}, taken as a whole, does not depend on the $\th$'s, 
see \cite{KozKitMailSlaTerXXZsgZsgZAsymptotics} for more details. 

The initial expression for $\mc{S}^{(\a)}_N$ is not convenient
for taking the large-$L$ limit. After some algebra, however, one can recast
$\mc{S}^{(\a)}_N$ into a form which is more convenient for the large-$L$ analysis: 
\beq
\mc{S}^{(\a)}_N \Big( \{ \mu_a \}_1^N \, ; \, \{ \la_a \}_1^N \Big) \; = \; \wh{\mc{D}}_1 \cdot \wh{\mc{D}}_2  \cdot \wh{\mc{A}}_{\e{reg}} \cdot \wh{\mc{A}}_{\e{sing}} \;. 
\label{product_of_four}
\enq
The coefficients $\wh{\mc{D}}_{k}$, $k=1,2$ are built out of the auxiliary function
\beq
\wt{\mc{D}}\Big( \{x_a\}_1^m;\{y_a\}_1^n\Big) \; = \; \f{ \pl{a\not=b }{m} \sin(x_a-x_b) \cdot \pl{a\not=b }{n} \sin(y_a-y_b) }
{ \pl{a=1}{m}\pl{b=1}{n} \sin(x_a-y_b)\sin(y_b-x_a)  } 
\enq
as well as of the products
\beq
V(\om) \; = \; \f{ \pl{a=1}{N} \sin(\om-\la_a) }{ \pl{a=1}{N+n_w} \sin(\om-\nu_a) } \qquad \e{and} \qquad
V_s(\om) \; = \; \f{ \pl{a=1}{N} \sin(\om-\la_a) }{ \pl{ \substack{ a=1 \\ \not= s} }{N+n_w} \sin(\om-\nu_a) } \;. 
\enq
Indeed, the coefficient $\wh{\mc{D}}_1$ is expressed as 
\beq
\wh{\mc{D}}_1 \; = \; (-1)^{N} (\i)^{n_w} ( 2 \i )^{n_w^2} \pl{a=1}{N} \bigg\{ \f{ \ex{2 \i \pi  \wh{F}(\la_a)}-1 }{ 2 \pi L \wh{\xi}^{\prime}_{\la}(\la_a) }  \bigg\}
\cdot \pl{a=1}{N+n_w} \bigg\{ \f{ \ex{-2 \i \pi \wh{F}(\nu_a)}-1 }{ 2 \pi L \wh{\xi}^{\prime}_{\mu}(\nu_a) }  \bigg\}
\cdot \wt{\mc{D}}\Big( \{\nu_a\}_1^{N+n_w};\{\la_a\}_1^N\Big) \;. 
\label{definition coefficient D1}
\enq
On this occasion, we recall that the $\{\nu_{a}\}_1^{N+n_w}$ correspond to the roots of $\ex{2\i\pi\wh{\xi}_{\mu}(\om)}-1$
located in the interval $\intff{ x_{\mu}-\tf{\pi}{2} }{ x_{\mu} + \tf{\pi}{2} }$.
In its turn, the coefficient $\wh{\mc{D}}_2$ reads 
\beq
\wh{\mc{D}}_2 \; = \;  \wt{\mc{D}}\Big( \{ \nu_{h_a} \}_1^{n_h};\{z_a\}_1^n\Big) \cdot 
 \pl{a=1}{n_h} \Big\{ 2 \pi L \wh{\xi}^{\prime}_{\mu}(\nu_{h_a}) \Big\}
\cdot \f{\pl{a=1}{n_h} \Big\{ V_{h_a}^2\big( \nu_{h_a} \big) \cdot V^{-1}\big( \nu_{h_a} + \i \eta \big)
			\cdot V^{-1}\big( \nu_{h_a} - \i \eta \big) \Big\}  }
			{ \pl{a=1}{n} \Big\{ V^2\big( z_a \big) \cdot V^{-1}\big( z_a+ \i \eta \big)
			\cdot V^{-1}\big( z_a - \i \eta \big) \Big\}     } \;. 
\label{definition coefficient D2}
\enq
The factor $\wh{\mc{A}}_{\e{reg}}$ contains all the terms that have
a regular behaviour in the $L \tend +\infty$ limit:
\bem
\wh{\mc{A}}_{\e{reg}} \; = \; 
\mc{W}\Big( \{\nu_a\}_1^{N+n_w};\{\la_a\}_1^N\Big) \cdot \mc{W}_{\e{reg}}\Big( \{ \nu_{h_a} \}_1^{n_h};\{z_a\}_1^n\Big) \cdot 
\f{ (\i)^{n_w}\cdot  (2\i)^{- n_w^{2}} \cdot (-1)^n \cdot \big(1-\ex{2\i\pi \a} \big)^2 }
	  { \det_{N+n_w}\big[ \Xi^{(\nu) }  \big] \cdot \det_N \big[ \Xi^{(\la)} \big] }  \\
\times \f{ \prod_{a=1}^{N}\Big(\ex{-2\i\pi \wh{F}(\la_a)}-1\Big) }{ \prod_{a=1}^{N+n_w}\Big(\ex{-2\i\pi \wh{F}(\nu_a)}-1\Big) }  \cdot 
\pl{p=1}{2} \Bigg\{ \f{  \det_{\Ga}\big[ I + \wh{U}_{\th_p} \big] }{1-\ex{2\i\pi \wh{F}(\th_p)}   } V\big( \th_p+\i\eta \big) 
\f{ \pl{a=1}{n_h}  \sin(\th_p - \nu_{h_a} + \i  \eta) }{ \pl{a=1}{n} \sin(\th_p - z_a + \i  \eta) }  
\;  \Bigg\} \;. 
\label{definition coefficient A}
\end{multline}
Above, we agree upon 
\beq
\mc{W}_{\e{reg}}\Big( \{ \nu_{h_a} \}_1^{n_h};\{z_a\}_1^n\Big) \; = \; 
\pl{a=1}{\tf{n_c}{2}}\big\{ \sin(-\de_a) \big\} \cdot \mc{W}\Big( \{ \nu_{h_a} \}_1^{n_h};\{z_a\}_1^n\Big) \;. 
\enq
In other words, $\mc{W}_{\e{reg}}\Big( \{ \nu_{h_a} \}_1^{n_h};\{z_a\}_1^n\Big) $ corresponds to the regular part of $\mc{W}\Big( \{ \nu_{h_a} \}_1^{n_h};\{z_a\}_1^n\Big) $, \textit{i}.\textit{e}. 
the one which admits a finite value for its large-$L$ asymptotics. We have also introduced the matrix
\beq
\Xi^{(\nu)}_{ab} \; = \; \de_{ab} \; + \; \f{ K(\nu_a-\nu_b) }{ L \wh{\xi}^{\prime}_{\mu}(\nu_b) }  \;. 
\enq
Finally, it remains to describe $\mc{A}_{\e{sing}}$. The latter term contains products of factors which, taken individually, exhibit strong singularities due to the formation of strings:
\beq
\wh{\mc{A}}_{\e{sing}}  \; = \; \pl{a=1}{ \tf{n_c}{2}} \bigg\{  \f{-1}{\sin(\de_a) } \bigg\} \cdot 
\pl{a=1}{n} \bigg\{ \f{ 1 }{ 2\pi L \wh{\xi}^{\prime}_{\mu}\big(z_a\big) } \bigg\}
\cdot \f{ \det_{N+n_w}\big[ \Xi^{(\nu) }  \big]   } { \det_N \big[ \Xi^{(\mu)} \big] } \;. 
\label{definition coefficient A sing}
\enq
%
%
%
%
%===========================================================================================================
%

\subsection{\boldmath The large-$L$ behaviour of the form factor: the result}

Each term in (\ref{product_of_four}) needs a careful analysis
to estimate the large-$L$ behaviour.
We therefore first present the final expression, then describe
the details of the analysis of each component in the subsequent sections.
We need some notation for the statement.

The thermodynamic limit of the shift function reads
\beq
F_{\iota}\Big(s \, | \,   \{\nu_a\}_1^{n_h} ;  \{\chi_a\}_1^{n_{\chi}}  \Big) \; = \; F_{\iota}(s)  
\; \equiv \; \f{\iota}{2} \; + \;  \f{1}{2\i \pi} \Big\{ \sul{a=1}{n} \vp(s,z_a) \, - \, \sul{a=1}{n_h} \vp(s , \nu_a)  \Big\} 
\label{definition fct comptage limite thermo}
\enq
and, in what concerns its periodised counterpart, one has  
\beq
 F_{\iota;\e{per}}(s) \; = \;  F_{\iota}(s)+\f{n_w}{\pi}(s+\f{\pi}{2}) \; - \; \f{ F_{\iota}(-\tf{\pi}{2}) -\iota }{2 \pi p^{\prime}(-\tf{\pi}{2}) } - \iota \;. 
\enq
Let $\{ \mf{z}_{a}^{(\iota)} \}_1^{n_\mathfrak{z}}$ be the set of zeroes
on $\intof{ -\tf{\pi}{2} }{ \tf{\pi}{2} }$ of 
\beq
s \mapsto \ex{-2 \i \pi F_{\iota}(s) } -1 \qquad \e{while} \qquad
G_{\iota}(s) \; = \; \ln \Bigg[  \f{ \ex{-2 \i \pi F_{\iota}(s) } -1 }{ \prod_{a=1}^{n_\mathfrak{z}} \sin(s-\mf{z}_a^{(\iota)}) } \Bigg] \; . 
\enq
For our convenience we introduce a $\pi$-periodic Cauchy transform by
\beq
\mc{C}\big[ F \big](\om) \; = \; \Int{-\tf{\pi}{2} }{ \tf{\pi}{2} } \f{ F(s) }{ \tan(s-\om) } \cdot \f{ \dd s }{2i\pi}.
\label{definition transfo Cauchy}
\enq
A particular combination of  $\pi$-periodic Cauchy transforms will be denoted by $\mc{L}\big[ F \big](\om) $.
\beq
\mc{L}\big[ F \big](\om) \; = \; 2\i \pi \Big\{ \mc{C}\big[ F \big](\om+i\eta)  \; + \; \mc{C}\big[ F \big](\om-i\eta) \; - \; 
 \mc{C}_+\big[ F \big](\om) \, - \, \mc{C}_-\big[ F \big](\om)  \Big\} \;. 
\label{definition transformee L}
\enq
%
%
%
%Note that 
The $\pm$-boundary values of the Cauchy transform only matter in the case when $\om \in \R$. We denote by  $\det\big[ I+K\big] $ and by $\det_{\Ga}\Big[ I + U_{\th_p}[ F_{\iota} ] \Big] $ the Fredholm determinants associated to 
the thermodynamic limit of   $\det_N\big[ \Xi^{(\la)}] $ and  $\det_{\Ga}\Big[ I+\wh{U}_{\th}\Big]$. For explicit forms see Proposition \ref{Areg_prop}.

Recall the reparametrisation of the complex roots $\{z_a\}_1^n$ in terms
of the roots $\{\chi_a\}_1^{n_{\chi}}$ as introduced in
\eqref{definition reparametrisation via chis}. We then introduce 
\beq
\mc{Y}_{\a}\Big( s_a \mid \{ s_b \}_1^{ n_{\chi} } ;  \{v_a\}_1^{ 2 n_{\chi} } \Big) \; = \; 1 \; + \; 
 \ex{- 2 \i \pi \a} \pl{b=1}{2 n_{\chi} } \ex{ \i p_0\big(s_a-v_b\big) }
\cdot \pl{b=1}{ n_{\chi} } \ex{-  \th\big(s_a-s_b \big)}  \;,
\enq
so that  the higher-level Bethe Ansatz  equations take the form
$\mc{Y}_{\a}\Big( \chi_a \mid  \{ \chi_b \}_1^{ n_{\chi} }; \{\nu_{h_s}\}_1^{ 2 n_{\chi} }\Big) \; = \; 0 $. 
We recall that $n_{\chi} =n_w+\tf{n_c}{2} = (n + n_w)/2$.

\begin{theorem}
\label{Theorem FF large L asymptotics}
Let $\{\wt{\mu}_a\}_1^N \ne \{\la_a\}_1^N$ be a
solution of the Bethe Ansatz equations such that
its associated counting function satisfies the hypotheses 
stated in Sub-Section \ref{Subsection Asympt behaviour counting function}.
Let $\{z_a\}_1^n$ be the complex roots for this state and
$\{\nu_{h_a}\}_1^{n_h}$ the positions of holes characterising this state. 
Then, the 2$n_\chi$-particle form factor
$\mc{F}_m^{(z)}\big( \{\wt{\mu}_a\}_1^N; \{\la_a\}_1^N\big)$ exhibits the
large-volume asymptotic behaviour  
\beq
\mc{F}_m^{(z)}\big( \{\wt{\mu}_a\}_1^N; \{\la_a\}_1^N\big) \; = \;  \ex{ \i \iota m \pi } \cdot 
\pl{a=1}{n_h } \bigg\{ \f{  \ex{-2\i \pi m p(\nu_{h_a})}  }{2\pi L p^{\prime}(\nu_{h_a})  } \bigg\} \cdot 
\f{ \Big(
\mc{F}^{(z)}_{\iota}\big( \{\nu_{h_a}\}_1^{n_h} ; \{\chi_a\}_1^{n_{\chi}} \big)\, \Big)^2 }
{ \det_{n_{\chi}} \Big[  \f{ \Dp{} }{ \Dp{} u_b } \mc{Y}_{0}\big( u_a\mid  \{u_c\}_1^{n_{\chi}} ;  \{\nu_{h_a}\}_1^{n_h} \big)  \Big]_{\mid u_a=\chi_a}  }
\cdot \Big(  1 + \e{O}(L^{-1}) \Big) \;. 
\enq
Here the normalized squared form factor of the spin operator takes the form 
\bem
\Big( \mc{F}^{(z)}_{\iota}\big( \{\nu_{h_a}\}_1^{n_h} ;  \{\chi_a\}_1^{n_{\chi}} \big)\, \Big)^2 \; = \; 
- 16  \sin^2\bigg( \f{\iota \pi}{2}+ \sul{a=1}{n_h}\pi p(\nu_{h_a}) \bigg) \times
\mc{W}_{\e{reg}}\Big( \{ \nu_{h_a} \}_1^{n_h};\{z_a\}_1^n\Big) \cdot \wt{\mc{D}}\Big( \{ \nu_{h_a} \}_1^{n_h};\{z_a\}_1^n\Big) \\
\times \exp\Bigg\{ \Int{-\tf{\pi}{2} }{ \tf{\pi}{2} }\f{ F_{\iota; \e{per}}(s)F^{\prime}_{\iota}(w)\, - \, F_{\iota; \e{per}}(w)F^{\prime}_{\iota}(s)  }
      {2 \tan\big(w \, -\, s \big) } \dd s \dd w   \; - \; 
\Int{ -\tf{\pi}{2} }{ \tf{\pi}{2} } \f{ F_{\iota; \e{per}}(s) F_{\iota; \e{per}}^{\,\prime}(w)  }{ \tan\big(w - s - \i\eta \big)  } \dd s \dd w  
\, - \! \Fint{-\tf{ \pi }{2} }{ \tf{ \pi }{2} }  F_{\iota; \e{per}}(s)
\ln^{\prime}\Big[  \ex{-2\i \pi F_{\iota}( s ) }-1 \Big] \cdot \dd s 	\Bigg\}  \\
\times \pl{a=1}{n_h} \Bigg\{ 4 \sin^2\big[ \pi F_{\iota}(\nu_{h_a}) \big]  \cdot \ex{\mc{L}[ F_{\iota; \e{per}} ](\nu_{h_a}) } \Bigg\} \cdot 
			 \pl{a=1}{n} \Big\{   \ex{-\mc{L}[ F_{\iota; \e{per}} ](z_a) }   \Big\}
\cdot \exp \bigg\{ -\f{n_w}{\pi} \Int{-\tf{\pi}{2} }{ \tf{\pi}{2} } G_{\iota}(s) \cdot \dd s +\i \pi \sul{\ell=1}{n_\mathfrak{z}}F_{\iota}(\mf{z}_{\ell}^{(\iota)}) \bigg\}    \\
\times  \f{ (\i )^{n_h}    2^{n_\mathfrak{z} n_w}  }{ \det^2\big[ I+K\big] } \cdot \ex{\eta n_w  (2n_h) } 
\cdot \pl{p=1}{2} \bigg\{ \f{  \det_{\Ga}\Big[ I + U_{\th_p}[ F_{\iota} ] \Big] }
{1-\ex{2 \i \pi F_{\iota}(\th_p)}   }  \cdot \f{ \ex{-2 \i\pi \mc{C}[ F_{\iota} ](\th_p+i\eta)} }{\big[ \cos(\th_p+\i\eta) \big]^{n_w} }
\f{ \pl{a=1}{n_h}  \sin(\th_p - \nu_{h_a} + \i  \eta) }{ \pl{a=1}{n} \sin(\th_p - z_a + \i  \eta) }  \;  \bigg\}  \;. 
\label{ecriture explicite FF GS to Excited squared}
\end{multline}
\end{theorem}
%
%
%

%\textcolor{blue}{\sout{Note that the ground state expectation value of
%$\sg^z$ vanishes, \textit{viz}. $\bra{  \{ \la_a \}_1^N } \sg^z_1
%\ket{  \{\la_a\}_1^N } =0$, for arbitrary volume $L$ due to the $\sin^2$
%factor in the expression for the squared amplitude.}}

The above theorem is the main result of this paper.   
We will present supplemental arguments  in Section 3, which show the usefulness
of  the above formula.  Especially  the comparison against the vertex-operator approach for the two-spinon case 
will be discussed in detail. In the rest of this section,  as promised above, 
the individual treatment of the four components in  (\ref{product_of_four})  is described in detail.

The large-$L$ behaviour of the coefficients $\wh{\mc{D}}_1$, $\wh{\mc{D}}_2$ is obtained in Subsection \eqref{SousSection Large L of D hat k}, Propositions
\ref{Proposition Asymp Beh D1}-\ref{Proposition Asymp Beh D2}. The one of the coefficients $\wh{\mc{A}}_{\e{reg}}$, $\wh{\mc{A}}_{\e{sing}}$ is obtained in Subsection \eqref{SousSection large L of A hat coeffs}, 
Propositions \ref{Areg_prop}-\ref{Proposition A sing large L}. All these asymptotic behaviours are uniform in $\a$ bounded. Gathering together these results 
and then applying \eqref{ecriture relation FF et partial alpha of S alpha} leads to Theorem \ref{Theorem FF large L asymptotics}.

%
%===========================================================================================================
%
\subsection{\boldmath Large-$L$ behaviour of the coefficients $ \wh{\mc{D}}_k$}
\label{SousSection Large L of D hat k}

In order to transform the expressions for the coefficients $ \wh{\mc{D}}_k$ into a form which allows one to take the large-$L$ limit easily,
we first need to establish a technical lemma. This lemma will involve
functions defined with the help of the determination `$\mf{ln}$' of the logarithm defined below:
\beq
\mf{ln} \big[ \sin(z ) \big] \; \equiv \; -\i \cdot \e{sgn}\big(\Im(z)\big) \cdot \big( z-\f{\pi}{2} \big) \, - \, \ln 2 \; + \; 
\ln\big[ 1\, - \, \ex{ \i 2 z\e{sgn}(\Im(z)) } \big] \qquad \e{for} \quad z \in \Cx \setminus \R. 
\label{definition du logarithme gothique}
\enq
We  stress  that the function `$\ln$' appearing  in the \textit{rhs}
of \eqref{definition du logarithme gothique} corresponds to
the principal branch of the logarithm. 

\begin{lemme}
\label{Lemme technique pour calcul sommes}
Let $\wh{\xi}_{\la/\mu}$ be a strictly increasing counting function
\begin{itemize}
\item  having $N_{\la/\mu}$ roots $\{\la_a\}/\{\nu_a\}$ in the
interval\footnote{We recall on this occasion that $x_\la, x_\mu$ are
the unique solutions on $\R$ to: $L \cdot \wh{\xi}_\la \big(x_\la - \tf{\pi}{2})
\,=\, \tf{1}{2}$,  $L \cdot \wh{\xi}_\mu \big(x_\mu - \tf{\pi}{2}) \,=\, \tf{1}{2}
- \iota$.} 
$x_{\la/\mu} \, + \, \intof{  - \tf{\pi}{2} }{ \tf{\pi}{2} }$;
\item  satisfying the quasi-periodicity requirement 
\beq
\wh{\xi}_{\la/\mu}\big(\om+\pi ) \,=\, N_{\la/\mu} \cdot L^{-1} \, + \, \wh{\xi}_{\la/\mu}\big(\om) \;; 
\enq
\item  being analytic in some open neighbourhood of the real axis. 
\end{itemize}
Let $\wh{\tau}_{\la/\mu}$ be the periodised form of $\wh{\xi}_{\la/\mu}$ that vanishes on the boundary of 
$ x_{\la/\mu} \, + \, \intff{ - \tf{\pi}{2} }{  \tf{\pi}{2} }$, %\textit{viz}.
\beq
\wh{\tau}_{\la/\mu}(\om) \; = \;  \wh{\xi}_{\la/\mu}(\om) \, - \,
\f{N_{\la/\mu}}{\pi L}\Big( \om - x_{\la/\mu}+\f{\pi}{2}  \Big) \; + \; \wh{C}_{\la/\mu} \quad \e{with} \quad
\left\{ \ba{ccc}  C_{\mu} & =& \; - \; \f{1 - 2 \iota}{2L}  \vspace{2mm} \\
		C_{\la} & =& \; - \; \f{1 }{2L}    \ea  \right.  \;. 
\enq
Then, the functions 
\beq
f_{\vsg}\big( \{\nu_a\} , \{\la_a\} \big) \; = \; \sul{ a=1 }{ N_{\mu } } \sul{ b = 1 }{ N_{\la} } \mf{ln} \Big[ \sin(\nu_a - \la_b - \i \vsg) \Big] 
\qquad \e{and} \qquad 
f\big( \om \mid \{\la_a\} \big) \; = \;  \sul{ b = 1 }{ N_{\la} } \mf{ln} \Big[ \sin(\om - \la_b ) \Big] \;,
\enq
where $\vsg>0$, can be recast as
\bem
f_{\vsg}\big( \{\nu_a\} , \{\la_a\} \big)  \; = \hspace{-2mm} \Int{-\tf{\pi}{2} }{ \tf{\pi}{2} } 
 \f{  L^2\,  \wh{\tau}_{\la}(s) \cdot \wh{\xi}^{\prime}_{\mu}(w) }{  \tan\big(w - s -\i \vsg \big) } \cdot \dd s \, \dd w
\, - \, \i L N_{\la} \Int{-\tf{\pi}{2} }{ \tf{\pi}{2} }  \wh{\tau}_{\mu}(s) \cdot \dd s  
\; + \; \bs{1}_{0<\vsg<\tau} \sul{ a=1 }{ N_{\mu} } \wh{u}_{\la}^{(-)}\big(\nu_{a}-i\vsg \big) \\
\; + \; \Big(  \vsg - \ln 2 - \i \big( \f{\pi}{2} + x_{\la} - x_{\mu} \big) \Big) \cdot N_{\la} N_{\mu}
 \; + \; \mf{r}_{\vsg}\big( \{\nu_a \} , \{ \la_a\} \big)
\label{expression asympt pour f vsg}
\end{multline}
and
\beq
f(\om\mid \{ \la_a \}) \; = \;  \Int{-\tf{\pi}{2} }{ \tf{\pi}{2} } 
 \f{ L \, \wh{\tau}_{\la}(s)  }{  \tan\big(\om - s \big) } \cdot \dd s
\; + \; \Big(  -\i \e{sgn}\big(\Im(\om)\big) \cdot \big( \om - x_{\la} - \f{\pi}{2} \big) - \ln 2\Big) N_{\la} 
\; + \; \sul{\eps= \pm }{} \bs{1}_{0< \eps \Im(\om) < \tau } \, \wh{u}_{\la}^{(\eps)}\big(\om \big) \; + \; \mf{r}\big( \om \mid \{\la_a\} \big) \;.
\enq
The indicator function $\bs{1}_{0<x<\tau}$ is non-zero only if $0<x<\tau$
where $\tau$ corresponds to the distance of the integration contours
$\Ga^{(\pm)}$ to $\R$, \textit{cf}.\ \eqref{ecriture reste et defnition ctrs gama pm},
$\mf{r}_{\vsg}\big( \{\nu_a \} , \{ \la_a\} \big)$ and
$\mf{r}\big( \om \mid \{\la_a\} \big)$ are  remainder terms. One has
\bem
\mf{r}_{\vsg}\big( \{\nu_a \} , \{ \la_a\} \big)\; = \; \sul{ \eps = \pm }{ }\; \Int{ \Ga^{(\eps)} }{} \f{ \dd s }{ 2 \i \pi }
\hspace{-3mm} \Int{ \Ga_{\e{in}}^{(+)}\cup \Ga_{\e{in}}^{(-)} }{}\hspace{-3mm}  \f{ \dd w }{2\i\pi}  
 \f{ \wh{u}_{\la}^{(\eps)}(s) \cdot \wh{u}_{\mu}^{\prime}(w) }{ \tan\big(w - s -\i \vsg \big) } \\
\; + \; L \sul{ \eps=\pm }{ } \Int{ -\tf{\pi}{2} + \tf{3 \i \tau  }{2} }{ \tf{\pi}{2} + \tf{3 \i \tau  }{2} } \hspace{-3mm} \f{ \dd s }{2 \i \pi} \Int{ \Ga^{(\eps)} }{}  \dd w 
 \f{ \wh{\tau}_{\la}(s) \cdot \big( \wh{u}_{\mu}^{(\eps)} \big)^{\prime}(w) }{ \tan\big(w - s -\i \vsg \big) }  
\; - \;  \i \; N_{\la} \sul{ \eps = \pm }{} \Int{ \Ga^{(\eps)} }{}  \wh{u}_{\mu}^{(\eps)} (s) \cdot \f{ \dd s }{ 2 \i\pi } \;. 
\end{multline}
Here the subscript `$\e{in}$' occurring in 
$\Ga_{\e{in}}^{(+)}\cup \Ga_{\e{in}}^{(-)}$ indicates that the latter contour is contained
inside of the outer contour $\Ga^{(\pm)}$, \textit{i}.\textit{e}. $\e{max}\Big\{ |\Im(s)| \; : \; s \in  \Ga_{\e{in}}^{(+)}\cup \Ga_{\e{in}}^{(-)} \Big\}< \tau$ and does not surround
the pole at $w=s-\i \vsg$. Analogously,
\beq
\mf{r}\big( \om \mid \{\la_a\} \big) \; = \; \sul{ \eps = \pm }{ } \, \Int{ \Ga^{(\eps)} }{} 
 \f{ \wh{u}_{\la}^{(\eps)}(s)  }{ \tan\big(\om - s  \big) } \cdot \f{ \dd s }{2 \i\pi}  \;. 
\enq
Finally, the functions $\wh{u}_{\la/\mu}$ and $\wh{u}_{\la/\mu}^{(\eps)}$ occurring above are as defined in \eqref{definition fonction hat u}. 
\end{lemme}
\Proof 

It is easily %readily 
seen by integrating by parts that 
\bem
f_{\vsg}\big( \{\nu_a \} , \{ \la_a\} \big)  \; = \; \sul{a=1}{N_{\mu}} \Oint{ \Ga_{\la} }{ }
\f{ \wh{u}_{\la}(s) }{ \tan\big( \nu_{a} - s - \i \vsg  \big) } \cdot \f{ \dd s}{ 2 \i \pi }
\; + \; \bs{1}_{0<\vsg<\tau} \sul{ a=1 }{ N_{\mu} } \wh{u}_{\la}\big(\wh{\nu}_{a}- \i \vsg \big)  \\
 \; - \; \sul{ \eps = \pm }{}L \eps \wh{\xi}_{\la}\big( x_{\la} - \eps \tf{\pi}{2} \big)
\sul{ a = 1 }{ N_{\mu} } \mf{ln} \big[\sin\big(\nu_a - x_{\la}+\eps \tf{\pi}{2} - \i \vsg \big) \big]
\end{multline}
where the contour $\Ga_{\la}$ is as defined by \eqref{definition du contour gamma mu}. 
The roots $\{ \wh{\nu}_a \}$ are translates by $\pi$ of the roots $\{ \nu_{a} \}$ so that 
the condition $\wh{\nu}_a \in \intff{x_{\la} -\tf{\pi}{2} }{ x_{\la} + \tf{\pi}{2}}$ holds. In other words, $\wh{\nu}_{a} = \nu_a + p \pi$, 
with $p\in \{1,0,-1\}$ chosen in such a way that the condition holds.

At this point, we split the integration \textit{vs}.  $\wh{u}_{\la}$. Due to $\pi$-periodicity, 
the parts involving the logarithms can be reduced to integrations along the lines  $\Ga^{(\pm)}$. 
Further, in the part involving $\wh{\xi}_{\la}$, we reconstruct the function $\wh{\tau}_{\la}$. 
This induces additional terms which cancel out the boundary contributions obtained above. 
Finally, observe that since $\wh{u}_{\la}^{(-)}$ is $\pi$-periodic in some open neighbourhood of $\R$ 
lying in the lower half-plane, one has $\wh{u}_{\la}^{(-)}\big(\wh{\nu}_{a}- \i \vsg \big) = \wh{u}_{\la}^{(-)}\big(\nu_{a}- \i \vsg \big) $.  
%All-in-all, this leads to 
Thus, the following expression holds, 
\bem
f_{\vsg}\big( \{\nu_a \} , \{ \la_a\} \big)  \; = \;  \Int{ x_{\la}-\f{\pi}{2} }{ x_{\la}+\f{\pi}{2} } \hspace{-2mm} \dd s \Int{-\f{\pi}{2} }{ \f{\pi}{2} } \hspace{-1mm} \dd w
\f{ L^2 \cdot \wh{\tau}_{\la}(s) \cdot \wh{\xi}_{\mu}^{\prime} (w) }{ \tan\big(w - s - \i \vsg  \big) } 
\, + \, \f{ N_{\la} }{ \pi } \sul{a=1}{N_{\mu} } \Int{ x_{\la}-\f{\pi}{2} }{ x_{\la}+\f{\pi}{2} } \mf{ln}\Big[ \sin\big(\nu_{a} - s - \i \vsg  \big) \Big] \cdot \dd s \\
\; + \; \bs{1}_{0<\vsg<\tau} \sul{ a=1 }{ N_{\mu} } \wh{u}_{\la}^{(-)}\big(\nu_{a}- \i \vsg \big)   
\; + \;  \sul{\eps=\pm}{} \Int{ x_{\la}-\f{\pi}{2} }{ x_{\la}+\f{\pi}{2} } \hspace{-2mm} \dd s \Int{ \check{\Ga}^{(\eps)} }{  } \f{ \dd w }{ 2 \i \pi}
\f{ L \cdot \wh{\tau}_{\la}(s) \cdot \big(\wh{u}_{\mu}^{(\eps)}\big)^{\prime} (w) }{ \tan\big(w - s - \i \vsg  \big) } 
\; + \; \sul{ \eps=\pm }{ } \Int{ \Ga^{(\eps)} }{} \f{ \dd s }{2 \i \pi} 
 \hspace{-3mm} \Int{ \Ga_{\e{in}}^{(+)}\cup \Ga_{\e{in}}^{(-)} }{} \hspace{-3mm} \f{ \dd w }{2 \i \pi} \f{ \wh{u}_{\la}^{(\eps)}(s) \cdot \wh{u}_{\mu}^{\, \prime}(w) }{ \tan\big(w - s - \i \vsg \big) } \;. 
\label{ecriture de la reecriture de f vsg}
\end{multline}
Above, the contour $ \check{\Ga}^{(\eps)}$ stands for a variant of the contour $ \Ga^{(\eps)}$ where the parameter $\tau$ is chosen such that $0<\tau<\vsg$. Since $\wh{\tau}_{\la}$
is $\pi$-periodic and holomorphic in an open neighbourhood of $\intff{ -\tf{\pi}{2}  }{ \tf{\pi}{2}}$ one can deform the $s$-contours in the $4^{\e{th}}$ term in \eqref{ecriture de la reecriture de f vsg}
to  $\intff{ -\tf{\pi}{2}  }{ \tf{\pi}{2}} + \tf{ 3 \i \tau }{2} $ and then deform the $w$-contour from $ \check{\Ga}^{(\eps)}$ to $ \Ga^{(\eps)}$. 

To conclude, it still remains to estimate 
\bem
\f{ N_{\la} }{ \pi }  \sul{a=1}{N_{\mu} } \Int{ x_{\la}-\f{\pi}{2} }{ x_{\la}+\f{\pi}{2} }
\mf{ln}\Big[ \sin\big(\nu_{a} - s - \i \vsg  \big) \Big] \cdot \dd s \; = \; 
N_{\la}N_{\mu} \Big( \vsg - \ln 2 - \i \big( x_{\la}+\tf{\pi}{2}\big) \Big) \; + \; 
\i N_{\la} \Oint{ \Ga_{\mu} }{} s \wh{u}_{\mu}^{\,\prime}(s) \cdot \f{ \dd s }{ 2 \i \pi } \\
\; = \; N_{\la}N_{\mu} \Big( \vsg - \ln 2 -\i \big( x_{\la}-x_{\mu} + \f{\pi}{2} \big)  \Big)  
\, + \, \i L N_{\la} \Int{ -\tf{\pi}{2} +x_{\mu} }{ \tf{\pi}{2} + x_{\mu} } \hspace{-3mm} s \, \wh{\tau}_{\mu}^{\, \prime}(s) \cdot \dd s
\; - \;\i N_{\la} \sul{ \eps = \pm }{} \Int{ \Ga^{(\eps)} }{}  \wh{u}_{\mu}^{(\eps)}(s) \cdot \f{ \dd s }{ 2 \i \pi } \;. 
\end{multline}
Then, a set of straightforward manipulations leads to \eqref{expression asympt pour f vsg}. 
The rewriting of $f\big(\om\mid \{\la_a\} \big)$ follows the same steps, 
so that we leave the details to the reader. \qed

The above lemma already  %allows one to %recast
yields the 
representation of   $\wh{\mc{D}}_1$ in a compact form on the level of which,
moreover, the $L\tend+\infty$ limit is transparent. Namely, we have the 
\begin{prop}
\label{Proposition Asymp Beh D1}
The coefficient $\wh{\mc{D}}_1$ introduced in \eqref{definition coefficient D1}
admits the representation
\beq
\wh{\mc{D}}_1 \; = \;  \exp\Bigg\{ \Int{-\tf{\pi}{2} }{ \tf{\pi}{2} }\f{ \wh{F}_{\e{per}}(s)\wh{F}^{\prime}(w)\, - \, \wh{F}_{\e{per}}(w)\wh{F}^{\prime}(s)  }
      {2 \tan\big(w \, -\, s \big) } \dd s \dd w  \Bigg\} \cdot \ex{ \mf{r}_{\mc{D}_1} }
\label{ecriture rep int pour D hat 1}
\enq
where $\mf{r}_{\mc{D}_1}$ is a remainder such that $\mf{r}_{\mc{D}_1} \, = \, \e{O}\big( L^{-\infty} \big)$ whereas $\wh{F}_{\e{per}}$
is the periodised shift function:
\beq
\wh{F}_{\e{per}}(\la) \; = \; \wh{F}(\la)+\f{n_w}{\pi}\cdot \big( \la  + \f{ \pi }{ 2 } \big) + \f{x_{\la} N_{\la}  - x_{\mu} N_{\mu} }{ \pi } - \iota \;. 
\enq
\end{prop}

\Proof 

The singular factor $\wt{\mc{D}}$ admits the representation
\beq
\wt{\mc{D}}\Big( \{\nu_a\}_1^{N+n_w};\{\la_a\}_1^N\Big) \; = \; \lim_{\vsg\tend 0^+}  \Bigg\{ \big(-\i \vsg \big)^{-2N - n_w}
\f{ \pl{a,b=1 }{N+n_w} \sin(\nu_a-\nu_b-\i\vsg) \cdot \pl{a,b=1 }{N} \sin(\la_a-\la_b-\i\vsg) }
{ \pl{a=1}{N+n_w}\pl{b=1}{N} \sin(\la_b-\nu_a- \i\vsg)\sin(\nu_a-\la_b- \i \vsg)  } \Bigg\} \;. 
\enq
One can use the first part of Lemma \ref{Lemme technique pour calcul sommes} to
compute the different products. Note that, in the intermediate calculations,
one is free to choose any determination of the logarithm, and $\mf{ln}$ in particular. 
This yields 
\bem
\wt{\mc{D}}\Big( \{\nu_a\}_1^{N+n_w};\{\la_a\}_1^N\Big) \; = \; \lim_{\vsg\tend 0^+}  \Bigg\{ \big(-\i \vsg \big)^{-2N - n_w}
\pl{a=1}{N+n_w} \bigg\{\f{1-\ex{-2 \i \pi L \wh{\xi}_{\mu}(\nu_a - \i \vsg) } }{1-\ex{-2 \i \pi L \wh{\xi}_{\la}(\nu_a- \i \vsg) } }  \bigg\} \cdot 
\pl{a=1}{N} \bigg\{ \f{1-\ex{-2 \i \pi L \wh{\xi}_{\la}(\la_a - \i \vsg) } }{1-\ex{-2i\pi L \wh{\xi}_{\mu}(\la_a- \i\vsg) } }  \bigg\}  \Bigg\} \\
\times \big( 2 \i \big)^{-n_w^2}   \exp\Bigg\{ \Fint{-\tf{\pi}{2} }{ \tf{\pi}{2} }
\f{ \wh{F}_{\e{per}}(\la)\wh{F}^{\prime}(\nu)  }{\tan\big(\nu \, -\, \la \big) } \dd \la \dd \nu  
\; + \; \i\pi \Int{-\tf{\pi}{2} }{ \tf{\pi}{2} } \hspace{-2mm} \wh{F}_{\e{per}}(\la)\wh{F}^{\prime}(\la)   \dd \la  
+ \i n_w \Int{-\tf{\pi}{2} }{ \tf{\pi}{2} } \hspace{-2mm} \wh{F}_{\e{per}}(\la)  \dd \la   \Bigg\} \cdot \ex{ \mf{r}_{\mc{D}_1} } \;. 
\end{multline}
%
%
%
%Note that
We have made use of the Plemelj formula at an intermediate stage. 
Also, $ \mf{r}_{\mc{D}_1} $ is the remainder whose explicit expression can be
obtained by combining the ones provided in Lemma
\ref{Lemme technique pour calcul sommes}. It is clear that
$\mf{r}_{\mc{D}_1}=\e{O}\big(L^{-\infty}\big)$,
so we shall not dwell any longer on this quantity. 
It remains to observe that
\beq
\Int{-\tf{\pi}{2} }{ \tf{\pi}{2} } \hspace{-2mm} \wh{F}_{\e{per}}(\la)\wh{F}^{\prime}(\la)   \dd \la   \; = \; 
\f{1}{2} \Big[ \wh{F}_{\e{per}}^{2}\big(\tf{\pi}{2}\big) \, - \, \wh{F}_{\e{per}}^{2}\big(-\tf{\pi}{2}\big) \Big]
\;- \; \f{n_w}{\pi}\Int{-\tf{\pi}{2} }{ \tf{\pi}{2} } \hspace{-2mm} \wh{F}_{\e{per}}(\la) \dd \la \; . 
\enq
Finally, the integral representation \eqref{ecriture rep int pour D hat 1} follows upon symmetrising the principal-value integral. \qed

In order to obtain a similar type of expression for $\wh{\mc{D}}_2$ some more work is necessary with the additional 
factors. Still, one obtains the
\begin{prop}
\label{Proposition Asymp Beh D2}
The coefficient $\wh{\mc{D}}_2$ introduced in \eqref{definition coefficient D2}
admits the representation 
\beq
 \wh{\mc{D}}_2 \; = \;   \wt{\mc{D}}\Big( \{ \nu_{h_a} \}_1^{n_h};\{z_a\}_1^n\Big) \cdot \ex{2\eta n_w(n_h)}\cdot 
 \pl{a=1}{n_h} \Bigg\{ \f{ 4 \sin^2\big[ \pi \wh{F}(\nu_{h_a}) \big] }{ 2 \pi L \wh{\xi}^{\prime}_{\mu}(\nu_{h_a})  } 
 \cdot \ex{\mc{L}\big[ \wh{F}_{\e{per}} \big](\nu_{h_a}) } \Bigg\} \cdot 
			 \pl{a=1}{n} \Big\{   \ex{-\mc{L}\big[ \wh{F}_{\e{per}} \big](z_a) }   \Big\} \cdot \ex{\mf{r}_{\mc{D}_2}}
\enq
where $\mf{r}_{\mc{D}_2}=\e{O}\big( L^{-\infty} \big)$ is a 
remainder and $\mc{L}$ is defined in \eqref{definition transformee L}.
\end{prop}
%
%
%
%Note that the $\pm$-boundary values of the Cauchy transform only matter in the case when $\om \in \R$. 

\Proof

On the basis of  arguments similar to the ones invoked in the course of the proof of Proposition
\ref{Proposition Asymp Beh D1}, one establishes that, for any $\om$ away from the
real axis (\textit{viz}. one can choose $\tau$ such that $|\Im(\om)|>\tau$), 
\beq
V(\om) \; = \; 2^{n_w} \exp\bigg\{  \i \e{sgn}(\Im(\om)) \Big( n_w(\om-\f{\pi}{2}) +  N x_{\la} - (N + n_w) x_{\mu} \Big)  \bigg\} \cdot 
\ex{ -2 \i\pi \mc{C}\big[ \wh{F}_{\e{per}}\big](\om) } \cdot \ex{ \mf{r}_{V} (\om) } 
\enq
whereas, one has,
\beq
V_{h_b}\big( \nu_{h_b} \big) \; = \; (-2)^{n_w}    \f{ -2 \sin \big[ \pi \wh{F}(\nu_{h_b} ) \big]   }{  2\pi L \wh{\xi}_{\mu}^{\prime}\big( \nu_{h_b}\big)  }
\exp\bigg\{ - \i\pi  \Big(  \mc{C}_-\big[ \wh{F}_{\e{per}}\big]\big( \nu_{h_b}\big) \, + \, \mc{C}_+\big[ \wh{F}_{\e{per}}\big]\big( \nu_{h_b}\big) \Big)  \bigg\}
\cdot \ex{ \mf{r}_{V} (\nu_{h_b}) } \;. 
\enq
In the two cases above, the remainder reads
\beq
\mf{r}_V(\om) \; = \; \sul{\eps = \pm}{} \;  \Int{ \Ga^{(\eps)}}{} \f{ \wh{u}^{(\eps)}_{\la} (s)\, - \, \wh{u}^{(\eps)}_{\mu}(s)}{\tan(\om - s)} \cdot \f{ \dd s }{ 2i\pi}  \;. 
\enq
Upon putting the various bits together, one obtains the desired representation for $\wh{\mc{D}}_2$. In particular, the remainder
$\mf{r}_{\mc{D}_2}$ is then expressed in terms of $\mf{r}_V(\om)$  evaluated at the hole $\nu_{h_a}$ or complex root $z_a$. \qed

\subsection{\boldmath The $\wh{\mc{A}}$-coefficients}
\label{SousSection large L of A hat coeffs}

\subsubsection{\boldmath The regular factor $\wh{\mc{A}}_{\e{reg}}$}

Observe that the shift function satisfies the quasi-periodicity property
\beq
\wh{F}\big(x+\tf{\pi}{2} \big) - \wh{F}\big(x-\tf{\pi}{2} \big)  \, = \, -n_w \;. 
\enq
Since it is real valued on $\intof{ - \tf{\pi}{2} }{ \tf{\pi}{2} }$ and continuous, the function 
\beq
s \mapsto \ex{-2i\pi \wh{F}(s)}-1
\enq
has at least $n_w$ zeroes on $\intof{-\tf{\pi}{2} }{ \tf{\pi}{2} }$.
We denote these zeroes by $\mf{z}_1, \dots , \mf{z}_{n_\mathfrak{z}}$. As a
consequence, the function 
\beq
G(s) \; = \; \ln \Bigg[ \f{ \ex{-2i\pi \wh{F}( s ) }-1 }{ \prod_{ a=1 }^{n_\mathfrak{z}} \sin\big( s-\mf{z}_{a} \big) } \Bigg] 
\enq
is holomorphic in some open neighbourhood of the real axis.

\begin{prop}\label{Areg_prop}
The regular coefficient $\wh{\mc{A}}_{\e{reg}}$ defined in \eqref{definition coefficient A}  admits the representation 
\bem
\wh{\mc{A}}_{\e{reg}} \; = \;  \mc{W}_{\e{reg}}\Big( \{ \nu_{h_a} \}_1^{n_h};\{z_a\}_1^n\Big)
\cdot \exp\Bigg\{  -\Int{ -\tf{\pi}{2} }{ \tf{\pi}{2} } 
\f{ \wh{F}_{\e{per}}(s) \wh{F}_{\e{per}}^{\,\prime}(w)  }{ \tan\big(w - s - \i\eta \big)  } \dd s \dd w  
\, - \! \Fint{-\tf{ \pi }{2} }{ \tf{ \pi }{2} }  \wh{F}_{\e{per}}(s)
\ln^{\prime}\Big[  \ex{-2i\pi \wh{F}( s ) }-1 \Big] \cdot \dd s
				\Bigg\}  \\
\times (\i )^{n_w} (-1)^n 2^{n_\mathfrak{z} n_w} \ex{-n_w^2\eta} \cdot
\exp \bigg\{ -\f{n_w}{\pi} \Int{-\tf{\pi}{2} }{ \tf{\pi}{2} } G(s) \cdot \dd s +\i \pi \sul{\ell=1}{n_\mathfrak{z}}F(\mf{z}_{\ell}) \bigg\} 
\cdot  \f{ \big(1-\ex{2\i \pi \a} \big)^2 }{ \det^2\big[ I+K\big] }  \\
\times \pl{p=1}{2} \bigg\{ \f{  \det_{\Ga}\Big[ I + U_{\th_p}\big[ \wh{F} \big] \Big] }
{1-\ex{2 \i \pi \wh{F}(\th_p)}   }  \cdot \f{ \ex{-2 \i\pi \mc{C}\big[ \wh{F}\big](\th_p+i\eta)} }{\big[ \cos(\th_p+\i\eta) \big]^{n_w} }
\f{ \pl{a=1}{n_h}  \sin(\th_p - \nu_{h_a} + \i  \eta) }{ \pl{a=1}{n} \sin(\th_p - z_a + \i  \eta) }  
\;  \bigg\} \cdot \ex{ \mf{r}_{ \mc{A}_{\e{reg}} } }
\end{multline}
Above, $\det\big[ I+K\big]$ is the Fredholm determinant of the integral operator $I+K$ acting on $L^2\big( \intff{-\tf{\pi}{2}}{\tf{\pi}{2}}\big)$
with the integral kernel $K(\la-\mu)$. The Cauchy transform $\mc{C}$ is as defined in \eqref{definition transfo Cauchy}.
Finally, the integral kernel $U_{\th}\big[F\big]$ reads
\bem
U_{\th}\big[\wh{F}\big]\big( \om, \om^{\prime} \big) \; = \;  \ex{-n_w\eta}  \cdot 
\exp\bigg\{ 2i\pi \cdot \Big( \mc{C}\big[\wh{F}_{\e{per}} \big](\om) \, - \,  \mc{C}\big[\wh{F}_{\e{per}} \big](\om+\i\eta) \Big) \bigg\} 
\cdot \f{ K_{\a}\big( \om - \om^{\prime}\big)  \, - \, K_{\a}\big( \th - \om^{\prime}\big)  }
{ 1-\ex{2 \i\pi \wh{F}(\om)}  }\\
\times \pl{a=1}{n_h}\bigg\{ \f{ \sin(\om - \nu_{h_a} + \i  \eta) }{ \sin(\om - \nu_{h_a} )  } \bigg\}
\cdot \pl{a=1}{n} \bigg\{ \f{ \sin(\om - z_a ) }{ \sin(\om - z_a + \i  \eta) }  \bigg\} \\
\times \exp\bigg\{ \i(1-\e{sgn}\big( \Im(\om) \big) \big(n_{w}(\om-\f{\pi}{2})+N x_{\la} - (N + n_w) x_{\mu}  \big) \bigg\} \;. 
\label{expression pour U theta}
\end{multline}
\end{prop}

Thanks to the identity
\beq
\ex{ 2i\pi \mc{C}\big[\wh{F}_{\e{per}} \big](\om) }\cdot 
\ex{ -\i \e{sgn}\big( \Im(\om) \big) \big(n_{w}(\om-\f{\pi}{2})+N x_{\la} - (N + n_w) x_{\mu}  \big) }
\; = \; \big[ 2\cos(\om)  \big]^{n_w} \cdot \ex{ 2i\pi \mc{C}\big[\wh{F} \big](\om) } \;,
\enq
one derives the following expression for $\wh{U}_{\th}\big[\wh{F}\big]\big( \om, \om^{\prime} \big) $
\bem
U_{\th}\big[\wh{F}\big]\big( \om, \om^{\prime} \big) \; = \; 
\exp\bigg\{ 2i\pi \cdot \Big( \mc{C}\big[\wh{F} \big](\om) \, - \,  \mc{C}\big[\wh{F} \big](\om  + \i\eta) \Big) \bigg\}
	\cdot \Big( \f{ \cos(\om) }{ \cos(\om +\i \eta) } \Big)^{n_w} \\
\times \pl{a=1}{n_h}\bigg\{ \f{ \sin(\om - \nu_{h_a} + \i  \eta) }{ \sin(\om - \nu_{h_a} )  } \bigg\}
\cdot \pl{a=1}{n} \bigg\{ \f{ \sin(\om - z_a ) }{ \sin(\om - z_a + \i  \eta) }  \bigg\}
\cdot \f{ K_{\a}\big( \om - \om^{\prime}\big)  \, - \, K_{\a}\big( \th - \om^{\prime}\big)  }
{ 1-\ex{2\i\pi \wh{F}(\om)}  } \;. 
\end{multline}

\Proof

The double product $\mc{W}$ depending on $N$ as well as parts of the integral kernel $\wh{U}_{\th}$ that involve $N$-dependent products are 
 recast  by means of Lemma \ref{Lemme technique pour calcul sommes}, leading to:
\beq
 \mc{W}\Big( \{ \nu_{a} \}_1^{N+n_w};\{\la_a\}_1^N\Big) \; = \; \big( 2\i \ex{-\eta}  \big)^{n_w^2} \cdot \exp\Bigg\{
 -\Int{ -\tf{\pi}{2} }{ \tf{\pi}{2} } 
\f{ \wh{F}_{\e{per}}(s) \wh{F}_{\e{per}}^{\,\prime}(w)  }{ \tan\big(w - s - i\eta \big)  } \dd s \dd w   \Bigg\}\cdot \ex{ \mf{r}_{\mc{W}} }
\enq
and the representation \eqref{expression pour U theta}
for $\wh{U}_{\th}\big[\wh{F}\big]$, up to $1+\e{O}\big( L^{-\infty} \big)$ corrections. 
Further, the two determinants appearing in the denominator of $\mc{A}_{\e{reg}}$
have representations in terms of Fredholm determinants, 
\beq
\det_{N+n_w}\Big[ \Xi^{(\nu)} \Big] \; = \; \det_{\Ga} \Big[ I\,+\, \mc{K}^{(\nu)} \Big]
\qquad \e{and} \qquad 
\det_{N}\Big[ \Xi^{(\la)} \Big] \; = \; \det_{\Ga} \Big[ I\,+\, \mc{K}^{(\la)} \Big] \;, 
\enq
where $I\,+\, \mc{K}^{(\nu)}$, resp.\ $I\,+\, \mc{K}^{(\la)}$, is an integral
operator acting on functions supported on the loop $\Ga$ whose integral kernel reads 
\beq
\mc{K}^{(\nu)}\big( \om, \om^{\prime} \big) \; = \; 
\f{ K(\om - \om^{\prime} )  }{ \ex{2 \i\pi L \wh{\xi}_{\mu}(\om^{\prime}) } -1 }
\qquad \e{and} \; \e{resp}. \qquad \mc{K}^{(\la)}\big( \om, \om^{\prime} \big) \; = \; 
\f{ K\big(\om - \om^{\prime} \big)  }{ \ex{2 \i\pi L \wh{\xi}_{\la}(\om^{\prime}) } -1 } \;. 
\enq
It is then enough to decompose the kernels as, \textit{e}.\textit{g}., 
\beq
\mc{K}^{(\nu/\la)}\big( \om, \om^{\prime} \big) \; = \; K(\om - \om^{\prime} ) \cdot \bigg\{ 
- \bs{1}_{ \Im(\om^{\prime})>0 } \; + \; \sul{ \eps=\pm }{}   \f{ - \eps \cdot \bs{1}_{ \eps \Im(\om^{\prime})>0 } }
{ \ex{-2 \i \pi \eps L \wh{\xi}_{\mu/\la}(\om^{\prime}) } -1 }   \bigg\} \;, 
\enq
deform the contour of the action of the first term up to
$\intff{\tf{\pi}{2}}{-\tf{\pi}{2}}$ and finally 
apply standard continuity theorems with respect to the kernel for Fredholm
determinants \cite{GohbergGoldbargKrupnikTracesAndDeterminants}
so as to drop the exponentially small in $L$ terms. This yields
\beq
\det_{N+n_w}\Big[ \Xi^{(\nu)} \Big] \cdot \det_{N}\Big[ \Xi^{(\la)} \Big]\; = \; 
\det^2\big[ I+K\big] \cdot  \Big( 1+\e{O}\big(L^{-\infty} \big) \Big) \;. 
\label{ecriture cptmt asymptotique des determinants}
\enq
Hence, the only factor left to evaluate is the product
\beq
\mc{P} \; = \; \f{ \pl{a=1}{N} \ex{-2 \i \pi \wh{F}( \la_{a}) }-1  }{ \pl{a=1}{N+n_w} \ex{-2 \i \pi \wh{F}( \nu_{a}) }-1  } \;. 
\enq
In order to %recast 
rewrite the product in a form that is convenient for taking the $L\tend  +\infty$ limit, we extract explicitly the product over the 
real zeroes of $ \ex{-2 \i\pi \wh{F}( \om) }-1 $ and treat these separately:
\beq
\ex{-2 \i \pi \wh{F}(\om)}-1 \; = \; \ \ex{G(\om) }\cdot  \pl{\ell=1}{n_\mathfrak{z}}\sin(\om-\mf{z}_{\ell}) \;. 
\enq
The function $G$ is already holomorphic on some open neighbourhood of $\R$, hence leading to 
\beq
\sul{a=1}{N}G(\la_a) \; = \; -L\Int{-\tf{\pi}{2} }{ \tf{\pi}{2} } G^{\prime}(s) \wh{\tau}_{\la}(s) \cdot \dd s 
	    \; + \; \f{ N_{\la} }{ \pi } \Int{-\tf{\pi}{2} }{ \tf{\pi}{2} } G(s) \cdot \dd s   
-\sul{\eps=\pm}{} \Int{ \Ga^{(\eps)} }{} G^{\prime}(s)\wh{u}_{\la}^{(\eps)}\cdot \f{ \dd s  }{ 2 \i \pi } \;. 
\enq
The product involving the real roots 
$\mf{z}_{\ell}$ is readily estimated by using the results gathered in Lemma \ref{Lemme technique pour calcul sommes}. 
%All-in-all, one gets
Altogether, one obtains
\beq
\mc{P} \; = \; 
2^{n_\mathfrak{z} n_w} \ex{-\f{n_w}{\pi} \Int{-\tf{\pi}{2} }{ \tf{\pi}{2} } G(s) \cdot \dd s + \i \pi \sul{\ell=1}{n_\mathfrak{z}} \wh{F}(\mf{z}_{\ell}) } \cdot
\exp\bigg\{ -\Fint{-\tf{ \pi }{2} }{ \tf{ \pi }{2} }  \wh{F}_{\e{per}}(s)
\ln^{\prime}\Big[  \ex{-2 \i \pi \wh{F}( s ) }-1 \Big] \cdot \dd s \bigg\} \cdot \ex{ \mf{r}_{\mc{P} } }
\enq
in which the remainder term reads
\beq
\mf{r}_{\mc{P} } \; =  \; - \sul{ \eps = \pm }{} \; 
\Int{ \Ga^{(\eps)} }{} \big[ \wh{u}^{(\eps)}_{\la} (s)- \wh{u}^{(\eps)}_{\mu}(s)\big] \cdot
\ln^{\prime}\Big[  \ex{-2 \i \pi \wh{F}( s ) }-1 \Big] \cdot \f{ \dd s }{ 2 \i \pi } \;. 
\enq
It now only remains to add up all of the results together. \qed

\subsubsection{\boldmath The singular factor $\wh{\mc{A}}_{\e{sing}}$}

\begin{prop}
\label{Proposition A sing large L}

The singular coefficient $\wh{\mc{A}}_{\e{sing}}$ defined in \eqref{definition coefficient A sing} admits the 
large-$L$ asymptotic expansion
\beq
\wh{\mc{A}}_{\e{sing}} \; = \; \f{ (- \i)^{n}  \cdot  \ex{ \mf{r}_{\mc{A}_{\e{sing}} } } }
{\det_{n_{\chi}} 
\Big[ \f{ \Dp{} }{ \Dp{} u_b } \mc{Y}_{\a}\Big( u_a  \mid \{ u_b \} _1^{n_{\chi}}  ;  \{\nu_{h_s}\}_1^{ 2 n_{\chi} } \Big) \Big]_{\mid u_a=\chi_a} } \;, 
\label{ecriture Jacobien des HLBAE}
\enq
where $\mf{r}_{\mc{A}_{\e{sing}} }  \, = \, \e{O}\Big( L^{-1}\Big)$ is a remainder. 

\end{prop}

The Jacobian in \eqref{ecriture Jacobien des HLBAE} can be thought of
as the higher-level norm formula for an excited state. 
Indeed, it arises in a way similar to the higher-level Bethe Ansatz
equation; namely by factoring out from the norm formula for 
an excited state the contribution of the `bulk' roots
$\{\nu_a\}_1^{N+n_w}$that form a dense distribution on
$\intff{ - \tf{\pi}{2} }{ \tf{\pi}{2} }$.

\Proof 

We start by explicitly dividing out the determinant of the matrix $\Xi^{(\nu)}$ out of the matrix $\Xi^{(\mu)}$. 
We introduce the integral kernel
\beq
\mc{K}^{(\mu)} (s,w) \; = \; \f{ K(s-w) }{  \ex{2 \i \pi L \wh{\xi}_{\mu}(w)} -1  }\;. 
\enq
Let $V=\{\nu_{h_1},\dots,\nu_{h_{n_h}}\}$, $Z=\{z_1,\dots,z_n\}$ and $\msc{C}_{\e{tot}}=
\Big\{\Ga_\mu \setminus\cup_{a=1}^{n_h}\Dp{}\mc{D}_{\nu_{h_a},\eps}\Big\} \bigcup \Bigl\{  \cup_{a=1}^{n}\Dp{}\mc{D}_{z_a,\eps}\,\Bigr\}$. 
Here, $\Dp{}\mc{D}_{z,\eps}$ stands for the boundary (oriented counterclockwise) of the disk of radius $\eps$
centred at $z$. Then, one has  
\beq
\det_{  \msc{C}_{\e{tot}}  }\big[ I + \mc{K}^{(\mu)} \big] \; = \; \sul{n \geq 0}{} \f{1}{n!}
\pl{a=1}{n} \Bigg\{ \Oint{ \Ga_\mu  }{} \f{ \dd \la_a }{ \ex{2 \i\pi L \wh{\xi}_{\mu}(\la_a)}-1  } 
\; - \; \sul{\la_a\in V }{} \f{1}{L\wh{\xi}_{\mu}^{\prime}(\la_a) }
\; + \;  \sul{\la_a\in Z }{} \f{1}{L\wh{\xi}_{\mu}^{\prime}(\la_a) }  \Bigg\}
\cdot \det_n\Big[ K(\la_a-\la_b)  \Big] \;. 
\enq
One can thus interpret the determinant of the operator $I+\mc{K}^{(\mu)}$
understood as acting on functions supported on $\msc{C}_{\e{tot}}$ as
the one of an operator $I+\wh{\mc{K}}^{(\mu)}$ acting on the space of functions 
supported on $\mf{X}=\Ga_\mu  \cup V \cup Z$. The operator $\wh{\mc{K}}^{(\mu)}$
has the matrix decomposition with respect to such a partition of the space $\mf{X}$:
\beq
\wh{\mc{K}}^{(\mu)} \; = \; \left( \ba{ccc}  \mc{K}^{(\mu)}(s,w)   &  -K(s-\nu_{h_b}) \cdot \big\{ L\, \wh{\xi}_{\mu}^{\prime}(\nu_{h_b}) \big\}^{-1}    
 												&  K(s-z_b) \cdot \big\{ L\, \wh{\xi}_{\mu}^{\prime}(z_b) \big\}^{-1}  \vspace{2mm} \\
	 \mc{K}^{(\mu)}(\nu_{h_a},w)  & -K(\nu_{h_a}-\nu_{h_b}) \cdot \big\{ L\, \wh{\xi}_{\mu}^{\prime}(\nu_{h_b}) \big\}^{-1}    
 												&  K(\nu_{h_a}-z_b) \cdot \big\{ L\, \wh{\xi}_{\mu}^{\prime}(z_b) \big\}^{-1}   \vspace{2mm} \\
	\mc{K}^{(\mu)}(z_a,w) & 	-K(z_a-\nu_{h_b}) \cdot \big\{ L\, \wh{\xi}_{\mu}^{\prime}(\nu_{h_b}) \big\}^{-1}    
 												&  K(z_a-z_b) \cdot \big\{ L\, \wh{\xi}_{\mu}^{\prime}(z_b) \big\}^{-1} 													\ea \right)   \;. 
\enq
Then, evaluating explicitly the contour integrals corresponding to the support $\Ga_\mu $, one obtains that 
\beq
\det_N\big[  \Xi^{(\mu)} \big] \;= \;  \det_{N+2(n+n_w)}\big[  \wh{\Xi}^{(\mu)} \big] \; ,
\enq
where the $[N+2(n+n_w)]\times [N+2(n+n_w)]$ matrix $\wh{\Xi}^{(\mu)}$ reads 
\beq
\wh{\Xi}^{(\mu)}  \; = \; \left( \ba{ccc}  \Xi^{(\nu)}_{ab}  &  -K(\nu_a-\nu_{h_b}) \cdot \big\{ L\, \wh{\xi}_{\mu}^{\prime}(\nu_{h_b}) \big\}^{-1}    
 												&  K(\nu_a-z_b) \cdot \big\{ L\, \wh{\xi}_{\mu}^{\prime}(z_b) \big\}^{-1}  \vspace{2mm} \\
	 K(\nu_{h_a}-\nu_b) \cdot \big\{ L\, \wh{\xi}_{\mu}^{\prime}(\nu_{b}) \big\}^{-1}      & 
				\de_{h_a h_b}-K(\nu_{h_a}-\nu_{h_b}) \cdot \big\{ L\, \wh{\xi}_{\mu}^{\prime}(\nu_{h_b}) \big\}^{-1}    
 												&  K(\nu_{h_a}-z_b) \cdot \big\{ L\, \wh{\xi}_{\mu}^{\prime}(z_b) \big\}^{-1}   \vspace{2mm} \\
	K(z_a-\nu_b) \cdot \big\{ L\, \wh{\xi}_{\mu}^{\prime}(\nu_{b}) \big\}^{-1} & 	-K(z_a-\nu_{h_b}) \cdot \big\{ L\, \wh{\xi}_{\mu}^{\prime}(\nu_{h_b}) \big\}^{-1}    
 												&  \de_{ab} + K(z_a-z_b) \cdot \big\{ L\, \wh{\xi}_{\mu}^{\prime}(z_b) \big\}^{-1} 													\ea \right)   \;. 
\enq
The inverse matrix to $\Xi^{(\nu)}_{ab}$ can be represented with the help of the so-called  discrete resolvent $\wh{R}$. 
The latter is defined as the unique solution to the equation 
\beq
\wh{R}\big(\nu_{j},z\big)  \;=\; K(\nu_j-z) 
			\; - \; \sul{\ell=1}{N+n_w} \wh{R}\big(\nu_{j}, \nu_{\ell}\big) \f{ K(\nu_{\ell}-z) }{ L \wh{\xi}_{\mu}^{\prime}(\nu_{\ell}) }
\enq
in which $z \in \Cx$. The discrete resolvent evaluated at two arbitrary complex numbers $(z,z^{\prime})$ is then defined through
the formula 
\beq
\wh{R}\big(z,z^{\prime}\big)  \;=\; K(z-z^{\prime}) 
			\; - \; \sul{\ell=1}{N+n_w}  \f{ K(z-\nu_{\ell} )\wh{R}\big(\nu_{\ell},z^{\prime}\big)  }{ L \wh{\xi}_{\mu}^{\prime}(\nu_{\ell}) } \, . 
\enq
With this object at hand, one readily checks that\footnote{It follows from the large-$L$ behaviour given in \eqref{ecriture cptmt asymptotique des determinants}
and from $ \det[I+K]>0$ that $ \Xi^{(\nu)}$ is indeed invertible.} 
\beq
\Big[ \big( \Xi^{(\nu)} \big)^{-1}  \Big]_{ab} \; = \; \de_{ab} \, - \,  \f{ \wh{R}(\nu_a,\nu_b) }{ L \wh{\xi}_{\mu}^{\prime}(\nu_b)} \;. 
\enq
By using the factorisation of determinants of block matrices 
\beq
\det\left( \ba{cc} A & B \\ C & D \ea \right) \; = \; \det[A] \cdot \det\big[D \, - \, C\cdot A^{-1} \cdot B  \big]
\enq
with $A = \Xi^{(\nu)}$, we are able to recast the determinant $\det_N\big[ \Xi^{(\mu)} \big]$ as
\beq
\det_{  N  }\big[ \Xi^{(\mu)}\big] \; = \; \det_{N+n_w}\big[\Xi^{(\nu)}\ \big] \cdot 
\pl{a=1}{n} \bigg\{ \f{1}{ L\, \wh{\xi}^{\prime}_{\mu}(z_a) } \bigg\} \cdot \det_{2n+n_w}\big[  \Ups \big] \;  . 
\enq
The matrix $\Ups$ appearing above takes the form  
\beq
\Ups \; = \;  \left( \ba{cc} 
\de_{ab} - \wh{R}\big(\nu_{h_a},\nu_{h_b}\big)\cdot \big\{ L\, \wh{\xi}_{\mu}^{\prime}(\nu_{h_b}) \big\}^{-1}   
																				 & \wh{R}\big(\nu_{h_a},z_b\big) \vspace{2mm} \\
 - \wh{R}\big(z_a,\nu_{h_b}\big)\cdot \big\{ L\, \wh{\xi}_{\mu}^{\prime}(\nu_{h_b}) \big\}^{-1}   
 		& \de_{ab} L\, \wh{\xi}_{\mu}^{\prime}(z_b) + \wh{R}\big(z_a,z_b\big)   \ea \right)  \; .  
 \label{ecriture forme blocks matrice Upsilon}
\enq
%
%
%
%All-in-all, we have proven that 
%
%
%
%\beq
%
%\det_{  N  }\big[ \Xi^{(\mu)}\big] \; = \; \det_{N+n_w}\big[\Xi^{(\nu)}\ \big] \cdot 
%
%\pl{a=1}{n} \bigg\{ \f{1}{ L\wh{\xi}^{\prime}_{\mu}(z_a) } \bigg\} \cdot \det_{2n+n_w}\big[  \Ups \big]  
%
%\cdot \Big( 1+ \e{O}\big(L^{-\infty} \big) \Big) \; . 
%
%\enq
%
%
%

In order to proceed further, we need to recall a technical result established in \cite{KirillovKorepinNormsStateswithStrings}. 
Let $\bs{P}_M =(p_1,\dots,p_M)$, $X$ be an $M \times M$ symmetric matrix and set
\beq
\De_M\Big( \bs{P}_M \; ; \; \{ X_{ab} \}_1^M \Big) \;  = \; \det_M\Big[ \de_{ab}\cdot \big( p_a - \sul{k=1}{M} X_{a,k}\big)  \,  + \, X_{a,b} \Big] \;. 
\label{definition determinant DeltaN}
\enq
Then, in the $X_{M-2(s-1),M-2s+1} \tend \infty$ limit, $s=1,\dots,\tf{n_c}{2}$, $\De_M$ admits the large-$M$ asymptotic behaviour
\bem
\De_M\Big( \bs{P}_M \; ; \; \{ X_{ab} \}_1^M \Big) \;  = \; \pl{s=1}{\tf{n_c}{2}} \Big\{- X_{M-2(s-1),M-2s+1} \Big\}
\cdot \De_{M-\tf{n_c}{2}}\Big( \bs{P}_{M-\tf{n_c}{2}}^{(\tf{n_c}{2})} \; ; \; \{ X_{a,b}^{(\tf{n_c}{2})} \}_1^{M-\tf{n_c}{2}} \Big) \\
\times \Big( 1+ \e{O}\Big( \e{max}_{s} \big| X_{M-2(s-1),M-2s+1}\big|^{-1} \Big) \, \Big)
\end{multline}
where
\beq
\bs{P}_{M-\tf{n_c}{2}}^{(\tf{n_c}{2})} \; = \; \Big(p_1,\dots ,p_{M-n_c}, p_{M-n_c+1}+p_{M-n_c+2},\dots,p_{M-1}+p_M \Big) 
\enq
and, for $1 \leq a,b \leq M-n_c$ and $1\leq p,\ell \leq \tf{n_c}{2}$
\beq
X_{a,b}^{(\tf{n_c}{2})} \; = \; X_{a,b}  \quad , \quad \; 
X_{a , M-n_c+p}^{(\tf{n_c}{2})} \; = \;    X_{M-n_c+p,a}^{(\tf{n_c}{2})} \; = \; X_{a,M-n_c+2(p-1)} + X_{a,M-n_c+2p-1}
\enq
whereas 
\bem
X_{M-n_c+\ell , M-n_c+p}^{(\tf{n_c}{2})} \; = \;  X_{ M-n_c+2(\ell-1) ,M-n_c+2(p-1)} \, + \,  X_{M-n_c+2(\ell-1),M-n_c+2p-1} \\
 + \, X_{ M-n_c+2\ell-1 ,M-n_c+2(p-1)} \, + \, X_{M-n_c+2\ell-1 ,M-n_c+2p-1}\;. 
\end{multline}
%
%
%
%It is readily seen that 
The determinant of $\Ups$ is indeed of the type \eqref{definition determinant DeltaN}. 
Thus, one obtains the reduction
\beq
\det_{2n+n_w}\big[ \Ups \big] \; = \; \pl{s=1}{\tf{n_c}{2}} \Big\{ -K\big(\de_s-i\eta\big) \Big\}
\cdot \det_{n+n_w+n_{\chi}}\big[  \wh{Y} \big] \cdot \Big( 1+ \e{O}\big(L^{-\infty} \big) \Big)
\enq
where 
\beq
\wh{Y}\;= \; \left( \ba{ccc} \de_{ab} - \wh{R}\big(\nu_{h_a},\nu_{h_b}\big)\cdot \big\{ L\, \wh{\xi}_{\mu}^{\prime}(\nu_{h_b}) \big\}^{-1}   & 
									    \wh{R}(\nu_{h_a},y_b)  \;\; \;   \wh{R}(\nu_{h_a}, \ov{y}_b)  & 
							  \wh{R}(\nu_{h_a},w_b) \, + \, \wh{R}(\nu_{h_a},w_b-\i \eta +\de_b )\\ 
 -\wh{R}\big(y_a ,\nu_{h_b}\big)\cdot \big\{ L\, \wh{\xi}_{\mu}^{\prime}(\nu_{h_b}) \big\}^{-1}  & Y^{(11)}  & Y^{(12)}    \\ 
 -\wh{R}\big(\ov{y}_a ,\nu_{h_b}\big)\cdot \big\{ L\, \wh{\xi}_{\mu}^{\prime}(\nu_{h_b}) \big\}^{-1} &             &            \vspace{1mm}\\
- \f{ \wh{R}\big(w_a ,\nu_{h_b}\big) \, + \, \wh{R}\big(w_a -\i \eta +\de_a,\nu_{h_b}\big) }{  L\, \wh{\xi}_{\mu}^{\prime}(\nu_{h_b}) }  &  Y^{(21)}  & Y^{(22)} 
\ea\right) \;. 
\enq
The entries $Y^{(ab)}$ of the block matrix $Y$ read
\beq
Y^{(11)} \; = \; \left( \ba{cc} L\wh{\xi}^{\prime}_{\mu}(y_a) \de_{ab} \, + \, \wh{R}(y_a,y_b)   &   \wh{R}(y_a,\ov{y}_b)  \\
							\wh{R}(\ov{y}_a,y_b)   &  L\wh{\xi}^{\prime}_{\mu}(\ov{y}_a) \de_{ab} \, + \, \wh{R}(\ov{y}_a,\ov{y}_b)  \ea \right)
\enq
and
\beq
Y^{(12)} \; = \; \left( \ba{c}  \wh{R}(y_a,w_{\ell}) \;+ \; \wh{R}(y_a,w_{\ell}-i\eta + \de_{\ell} )   \\
							 \wh{R}(\ov{y}_a,w_{\ell}) \;+ \; \wh{R}(\ov{y}_a,w_{\ell}-i\eta + \de_{\ell} )  \ea \right)				 
\enq
and
\beq
Y^{(21)} \; = \; \left( \ba{c c }  \wh{R}(w_{p},y_b) \;+ \; \wh{R}(w_{p}-i\eta + \de_{p},y_b )   &
							  \wh{R}(w_{p},\ov{y}_b) \;+ \; \wh{R}(w_{p}-i\eta + \de_{p},\ov{y}_b )  \ea \right)		
\enq
and, finally,
\bem
Y^{(22)}_{p \ell} \; = \; \de_{p\ell} \cdot L \Big(  \wh{\xi}^{\prime}_{\mu}(w_{\ell}) \; + \;   \wh{\xi}^{\prime}_{\mu}(w_{\ell}-i\eta+ \de_{\ell})  \Big) 
\; + \; \wh{R}(w_{p},w_{\ell}) \;+ \; \wh{R}(w_{p}-i\eta + \de_{p},w_{\ell} )   \\
		 \; + \; \wh{R}(w_{p}, w_{\ell}-i\eta+ \de_{\ell} ) \;+ \; \wh{R}(w_{p}-i\eta + \de_{p}, w_{\ell}-i\eta+ \de_{\ell} )  \;. 
\end{multline}
The expression for the entries of the various blocks defining the matrix $Y$ can be further simplified. 
The counting function can be expressed, in the large-$L$ limit, by means of Proposition \ref{Propositon large L asympt ctg fct}
and use of the functional equation satisfied by the homogenised dressed phase. 
Furthermore, the large-$L$ behaviour of the discrete resolvent $\wh{R}(z,z^{\prime})$
can be characterised by means of the representation
\beq
\wh{R}(z,z^{\prime}) \;= \; K(z-z^{\prime}) \; - \; \Int{-\tf{\pi}{2} }{ \tf{\pi}{2} } K(z-s) \cdot \wh{R}(s,z^{\prime}) \cdot \dd s  \; + \, \e{O}(L^{-\infty}) \;. 
\enq
For $|\Im(z)|< \eta$ and $|\Im(z^{\prime})|<\eta$, the equation can be %readily 
solved elementarily. 
The leading in $L$ expression for 
$\wh{R}(z,z^{\prime})$ when  $|\Im(z)|> \eta$ or $|\Im(z^{\prime})|>\eta$ is then obtained by analytic continuation
on the basis of the method that has been used in the proof of Proposition \ref{Proposition estimation excitation momentum and energy}. 
All-in-all, one obtains 
\beq
\ba{lcl} 
\e{if} \; |\Im(z)|,|\Im(z^{\prime})|<\eta   & \e{then} &  \wh{R}\big(z,z^{\prime}\big) = R\big(z-z^{\prime}\big)\; + \, \e{O}(L^{-\infty}) \vspace{1mm} \\
 \e{if} \;  |\Im(z)| < \eta \; \e{and}\;  \pm \Im(z^{\prime})>\eta  & \e{then} & 
\wh{R}\big(z,z^{\prime}\big) = R\big(z-z^{\prime}\big) \, + \, R\big(z-z^{\prime} \pm \i \eta\big)\; + \, \e{O}(L^{-\infty}) \; \vspace{1mm} \\
 \e{if} \; \Im(z) >\eta \; \e{and} \;  \Im(z^{\prime})>\eta & \e{then} &
\wh{R}\big(z,z^{\prime}\big) = 2 R\big(z-z^{\prime}\big) \, + \, R\big(z-z^{\prime}- \i \eta\big)+R\big(z-z^{\prime} + \i \eta\big)\; + \, \e{O}(L^{-\infty}) \vspace{1mm} \\
 \e{if} \;  \Im(z) >\eta \; \e{and} \; \Im(z^{\prime})<-\eta  &  \e{then} & 
\wh{R}\big(z,z^{\prime}\big) = 2 R\big(z-z^{\prime}-\i\eta\big) \, + \, R\big(z-z^{\prime}- 2 \i \eta\big)+R\big(z-z^{\prime} \big)\; + \, \e{O}(L^{-\infty}) \ea 
\enq
where $R$ is the resolvent operator to $I+K$ understood as
 acting on functions supported on $\intff{ - \tf{\pi}{2} }{ \tf{\pi}{2} }$, see 
Appendix \ref{Subsection Resolvent kernel}. 
All other instances of the parameters $z,z^{\prime}$ are obtained from the symmetry 
$\wh{R}(z,z^{\prime})=\wh{R}(z^{\prime},z)$ and the reflection property 
$\ov{ \wh{R}(z,z^{\prime}) } =\wh{R}(\ov{z} , \ov{z}^{\prime}) $.

By using the above asymptotic expression for $\wh{R}$, the shift recurrence relation satisfied by the resolvent 
\eqref{ecriture relation shift resolvent R} and the $\chi$-reparametrisation of the complex roots $\{z_a\}_1^n$, one obtains that 
\beq
Y_{ab} \; = \; \de_{ab}\Big( \f{1}{2\pi}\sul{s=1}{ 2 n_{\chi} } p_0^{\prime}\big( \chi_a-\nu_{h_s} \big)  \, - \, \sul{s=1}{n_{\chi} } K(\chi_a-\chi_s) \Big)
\; + \; K\big(\chi_a-\chi_b\big) \; + \; \e{O}\Big( L^{-\infty} \Big) \;. 
\enq
Now observe that the first block column of the matrix $\wh{Y}$ is of the form $\de_{ab} \, + \, \e{O}\big(L^{-1}\big)$.
Therefore, since $\det_{n_{\chi}}[Y]\not=0$, up to $\e{O}\big(L^{-1}\big)$ corrections, the determinant of the matrix $\wh{Y}$ reduces to the one of the matrix 
$Y$. As a consequence, one finds that 
\beq
\det_{n_{\chi}+n+n_w} \Big[ \wh{Y} \Big] \; = \; \big(-2\i \pi \big)^{-n_{\chi}} \cdot 
\det_{n_{\chi}} \Big[ \f{ \Dp{} }{ \Dp{} u_b } \mc{Y}_{\a}\Big( u_a \mid \{ u_c \} _1^{n_{\chi}} ; \mid  \{\nu_{h_s}\}_1^{ 2 n_{\chi} } \Big) \Big]_{\mid u_a=\chi_a} 
\cdot  \bigg( 1+\e{O}\Big(L^{-1}\Big) \bigg)  \;. 
\enq
The claim then follows. \qed

\section{The form-factor series}
\label{Section FF series expansion}

\subsection{The form-factor series}

In this section we build on the large-volume behaviour of individual form factors so as to 
write down the form-factor series expansion for the spin-spin correlation function
in the massive regime. The first term in this series corresponds to the staggered
magnetisation. In the large-distance limit, the $\sg^z$-$\sg^z$ correlator
approaches the staggered magnetisation exponentially fast.

Recall that the temporal evolution of an operator takes the form
\beq
\sg_k^{z}(t) \; = \; \ex{ \i t H }\sg_{k}^z \ex{ - \i t H} \;. 
\enq
Within such a convention for the temporal evolution, the space- and
time-dependent spin-spin correlation function in the limit $L\tend + \infty$
admits the  form-factor expansion
\bem
\moy{ \sg_1^z(0) \cdot \sg_{m+1}^z(t) } \; = \; (-1)^m \pl{n \geq 1}{} \bigg( \f{ 1- \ex{-2n\eta} }{ 1+\ex{-2n\eta} } \bigg)^4  \\
\; + \; \sul{\iota =0}{1} \sul{n_h \in 2 {\mathbb N} }{ } 
\f{ (-1)^{ \iota m }  }{ (n_h)! }
\Int{-\tf{\pi}{2} }{ \tf{\pi}{2} } \f{ \dd^{n_h} \nu }{ (2\pi)^{n_h} }
\pl{a=1}{n_h} \Big\{ \ex{\i t \veps^{(0)}(\nu_a) - 2\i \pi m p(\nu_{a})}   \Big\} \cdot \msc{F}^{(z)}_{\iota}\big(   \{\nu_{a}\}_1^{n_h}  \big) \;.
\label{ecriture series FF}
\end{multline}
The function $\veps^{(0)}$ stands for the dressed energy at zero magnetic
field (see Proposition \ref{Proposition estimation excitation momentum and energy}).
The non-trivial part of the integrand is defined in terms of a multi-dimensional residue
\beq
\msc{F}^{(z)}_{\iota}\big(   \{\nu_{a}\}_1^{n_h}  \big) \; = \; \f{ 1 }{ n_{\chi}! }
\Oint{ \Ga_{\eps}(\{  \nu_a \}) }{} \f{  \Big( \mc{F}^{(z)}_{\iota}\big(   \{\nu_{c}\}_1^{n_h} ; \{\psi_{c}\}_1^{n_{\chi}} \big)\, \Big)^2 }
{ \pl{a=1}{n_{\chi} }\mc{Y}_{0}\big( \psi_a\mid  \{\psi_c\}_1^{n_{\chi}} ;  \{\nu_{c}\}_1^{n_h} \big)   }
\cdot \f{ \dd^{n_{\chi}} \psi }{ (2\i \pi)^{n_{\chi}}  } \;.
\label{ecriture integrale multidim residus}
\enq
More precisely, the $n$-dimensional integral runs through the skeleton
associated with the higher-level Bethe Ansatz equations subordinate to the 
choice of holes $\{ \nu_a \}_1^{n_h}$:
\beq
\Ga_{\eps}\big( \{ \nu_a \} \big) \; = \;
\Big\{ \bs{\psi} \in \Cx^{n_{\chi}} \; : \; \big| \mc{Y}_{0}\big( \psi_a\mid  \{\psi_c\}_1^{n_{\chi}} ;  \{\nu_{a}\}_1^{n_h} \big) \big| \;= \; \eps \quad a=1,\dots, n_{\chi}   \Big\}
\enq
with $\eps>0$ but small enough. Upon computing the integral, it produces a summation over all solutions $\{\chi_a\}_1^{n_{\chi}}$  to the higher-level Bethe Ansatz equations 
\beq
\mc{Y}_{0}\big( \chi_a\mid  \{\chi_c\}_1^{n_{\chi}} ;  \{\nu_{a}\}_1^{n_h} \big) \;= \; 0 \quad a=1,\dots, n_{\chi} \;, 
\enq
(see \textit{e.g.}\ \cite{AizenbergYuzhakovIntRepAndMultidimensionalResidues} for more
details). 

\Proof 

The above series expansion for the spin-spin two-point function can be
obtained as follows. We first focus on excited states (\textit{i}.\textit{e}. those containing a non-zero
number of holes). In this case, the contribution to the form-factor expansion originating from the sector with
$n_h=n+n_{w}$ hole excitations takes the form 
\beq
\moy{ \sg_1^z(0) \cdot \sg_{m+1}^z(t) }_{n_h} \; = \; \sul{\iota=0}{1} \sul{  h_1<\dots  <h_{n_h}  }{} 
\sul{ \substack{ \{\chi_a\} \\ HBAE } }{}
\f{ \ex{ \i \iota m \pi} \cdot \prod_{a=1}^{n_h } \Big\{  \ex{\i t \veps^{(0)}(\nu_{h_a}) -2\i \pi m p(\nu_{h_a})}   \Big\} \cdot 
\Big( \mc{F}^{(z)}_{\iota}\big(   \{\nu_{h_a}\}_1^{n_h} ; \{\chi_a\}_1^{n_{\chi}} \big)\, \Big)^2   }
{\pl{a=1}{n_h} \Big\{  2\pi L p^{\prime}(\nu_{h_a}) \Big\} \cdot 
	  \det_{n_{\chi}} \Big[  \f{ \Dp{} }{ \Dp{} u_b } \mc{Y}_{0}\big( u_a\mid  \{u_c\}_1^{n_{\chi}} ;  \{\nu_{h_a}\}_1^{n_h} \big)  \Big]_{\mid u_a=\chi_a}  }
\cdot \Big(  1 + \e{O}(L^{-\infty}) \Big)  \;. 
\enq
The first sum goes though the two possible choices of the $\iota$ parameter
which allows us to distinguish between the excitations above the ground state or
the quasi-ground state. The second sum runs
over all the possible choices of integers $1\leq h_1 < \dots < h_{n_h} \leq N+n_{w}$
which determine the configuration $\{ \nu_{h_a} \}_{1}^{n_h}$ of hole positions
in the given excited state. Finally, the last summation symbol runs through
all solutions to the higher-level Bethe Ansatz equations subordinate to the choice 
$\{ \nu_{h_a} \}_{1}^{n_h}$ of the hole parameters, namely:
\beq
\mc{Y}_{0}\big( \chi_a\mid  \{\chi_c\}_1^{n_{\chi}} ;  \{\nu_{h_c}\}_1^{n_h} \big) \;= \; 0 \quad a=1,\dots, n_{\chi} \;.  
\enq

Clearly, one can drop the exponentially small in $L$ corrections. Furthermore, the sum over all solutions to the higher-level Bethe Ansatz 
equation can be recast as a multi-dimensional residue integral 
\beq
\sul{ \substack{ \{\chi_a\} \\ \e{sols} HBAE} }{} \f{ \Big( \mc{F}^{(z)}_{\iota}\big(   \{\nu_{h_a}\}_1^{n_h} ; \{\chi_a\}_1^{n_{\chi}} \big)\, \Big)^2   }
{ \det_{n_{\chi}} \Big[  \f{ \Dp{} }{ \Dp{} u_b } \mc{Y}_{0}\big( u_a\mid  \{u_c\}_1^{n_{\chi}} ;  \{\nu_{h_a}\}_1^{n_h} \big)  \Big]_{\mid u_a=\chi_a}  } 
\; = \; \msc{F}^{(z)}_{\iota}\big(   \{\nu_{h_a}\}_1^{n_h}  \big)
\enq
in which $\msc{F}^{(z)}_{\iota}\big(   \{\nu_{h_a}\}_1^{n_h}  \big)$ is as it has been defined in \eqref{ecriture integrale multidim residus}. 
Note that this function is analytic in some open neighbourhood of $\intff{ - \tf{\pi}{2} }{ \tf{\pi}{2} }$. 

Finally, it follows from Proposition \ref{Propositon large L asympt ctg fct} that
the hole parameters satisfy the equation
\beq
 p(\nu_{h_a}) \; = \; \f{h_a}{L} \; + \; \e{O}(L^{-1})
\enq
with a remainder that is uniform in respect to $h_a$. Thus,
in the $L\tend + \infty$ limit, the sum over the hole parameters turns into a integral over 
$\intff{-\tf{\pi}{2}}{ \tf{\pi}{2}}$ as a Riemann sum:
\bem
\lim_{L\tend + \infty}  %\bigg\{ 
\sul{ h_1<\cdots  < h_{n_h}  }{}  
\pl{a=1}{n_h } \bigg\{ \f{  \ex{\i t \veps^{(0)}(\nu_{h_a}) - 2\i \pi m p(\nu_{h_a})}  }{2\pi L p^{\prime}(\nu_{h_a})  } \bigg\}
\cdot  \msc{F}^{(z)}_{\iota}\big(   \{\nu_{h_a}\}_1^{n_h}  \big)  \\ 
\; = \; \Int{-\tf{\pi}{2} }{ \tf{\pi}{2} } 
\pl{a=1}{n_h} \Big\{ \ex{\i t \veps^{(0)}(\nu_{a}) -2\i \pi m p(\nu_{a})} \Big\} \cdot \f{  \msc{F}^{(z)}_{\iota}\big(   \{\nu_{a}\}_1^{n_h}  \big)  }{ (n_h)! } 
\cdot \f{ \dd^{n_h} \nu }{ (2\pi)^{n_h} }  \;. 
\end{multline}

We now focus on the case when there are no holes and $\iota \in \{1,0\}$.
When $\iota=0$, it follows from Theorem \ref{Theorem FF large L asymptotics} 
that the form factor is simply zero. Since we are in a situation when
$\iota=1$ and $\{\nu_{h_a} \}=\{z_a\}=\{\emptyset\}$ most of the terms in
\eqref{ecriture explicite FF GS to Excited squared} simplify. 
Denoting by $ \{\wt{\la}_a\}_1^N$ the solution of the Bethe Ansatz equations describing the quasi-ground state, one has
\beq
\mc{F}_m^{(z)}\big( \{\wt{\la}_a\}_1^N; \{\la_a\}_1^N\big) \; = \;  - 4 (-1)^{m} 
\exp\bigg\{ - \Int{ - \tf{\pi}{2} }{ \tf{\pi}{2} }  \f{ \dd s }{ \tan(s-\th-\i \eta) } \bigg\} 
\cdot \f{\det_{\Ga}^{2}\Big[ I + U_{\th}\big[ F_{1}\big(*\mid \{ \emptyset \};\{ \emptyset \}\big) \big] \Big] } { \det^2\big[ I+K\big] } \cdot \Big( 1+ \e{O}\big( L^{-\infty} \big) \Big)\;. 
\enq
Above, we have already specified the two arbitrary parameters $\th_1$ and $\th_2$ to take the same value $\th \in \intof{ - \tf{\pi}{2} }{ \tf{\pi}{2} }$
and made use of the fact that 
\beq
F_{1}\big(*\mid \{ \emptyset \};\{ \emptyset \}\big) \; = \; \f{1}{2} \;. 
\enq
The integral in the exponent can be computed explicitly:
\beq
\Int{ - \tf{\pi}{2} }{ \tf{\pi}{2} }  \f{ \dd s }{ \tan(s-\th-\i\eta) }  \; = \; \i \pi \;. 
\enq
Furthermore, the integral kernel of the operator $ U_{\th}[ \tf{1}{2} ]$ takes the simple form 
\beq
U_{\th}[ \tf{1}{2} ](\om,\om^{\prime} ) \; = \; \f{ K(\om-\om^{\prime}) - K(\th-\om^{\prime}) }{ 2 }
\cdot \exp\Big\{ \i \f{\pi}{2} \big(\e{sgn}(\Im(\om))-1 \big) \Big\} \;. 
\enq
Thus, squeezing the contour $\Ga$ to $\intff{ -\tf{\pi}{2} }{ \tf{\pi}{2} }$, one obtains
\beq
\det_{\Ga}\Big[ I + U_{\th}\big[ F_{1}\big(*\mid \{ \emptyset \};\{ \emptyset \}\big) \big] \Big]\; = \; \underset{ \intff{-\f{\pi}{2}}{\f{\pi}{2}} }{\det}[I+V] \qquad 
\e{with} \qquad 
V(\om,\om^{\prime}) \; = \; - \big[ K(\om-\om^{\prime}) - K(\th-\om^{\prime}) \big] \;. 
\enq
The determinants of $I+K$ and $I+V$ can be computed explicitly. Indeed, with respect to the orthonormal basis $\big\{ \tf{ \ex{2\i n \la} }{ \pi } \big\}_{n \in \mathbb{Z} } $, one has 
\beq
K\big[ \tf{ \ex{2\i n * } }{ \pi }  \big](\la) \; = \; \ex{-2|n|\eta}  \cdot \f{ \ex{2\i n \la} }{ \pi } \qquad \e{and} \qquad
-\Int{ -\tf{\pi}{2} }{ \tf{\pi}{2} } K(\th-\la) \ex{-2\i n \la} \cdot \f{ \dd \la }{ \pi } \; = \; 
\f{-1}{\pi} \ex{-2\i n \th-2|n|\eta} \;. 
\label{calcul action sur base fourier de K et V}
\enq
%
%
%
%We remind that the $*$ refers to the running variable on which $K$ acts.
 As a consequence of \eqref{calcul action sur base fourier de K et V} one has 
\beq
\det[I+K] \; = \; \pl{n \in \mathbb{Z}}{} \Big( 1+ \ex{-2|n|\eta} \Big) \qquad \e{and} \qquad
\det[I+V] \; = \; \pl{n \in \mathbb{Z}\setminus \{0\} }{} \Big( 1- \ex{-2|n|\eta} \Big)  \; . 
\enq
Thus, %, all-in-all, 
one arrives at  the representation
\beq
\mc{F}_m^{(z)}\big( \{\wt{\la}_a\}_1^N; \{\la_a\}_1^N\big) \; = \; 
(-1)^m \pl{n \geq 1}{} \bigg( \f{ 1- \ex{-2n\eta} }{ 1+\ex{-2n\eta} } \bigg)^4  \cdot \Big( 1+ \e{O}\big( L^{-\infty} \big) \Big)  \;. 
\enq

It then solely remains to put all the partial results together, which
leads to the form-factor series \eqref{ecriture series FF} upon the
hypothesis of its convergence. \qed

\subsection{Comparison with the vertex-operator approach}

Multiple-integral representations for the form factors of the spin
operators in the XXZ spin-$\frac{1}{2}$ chain in the massive regime 
were computed by means of the vertex-operator formalism in
\cite{JimboMiwaFormFactorsInMassiveXXZ}. The results of \cite{JimboMiwaFormFactorsInMassiveXXZ} allowed for an analysis of the density structure factor in the 
massive regime of the chain \cite{BougourziWestonZeroAndTwoParticleDSFInMassiveXXZ}.
% %
This work was later generalized by Lashkevich to the XYZ case
\cite{LashkevichElemBlocksEightVertex} whose result was then used to conjecture the
form factors of local fields in the sine-Gordon model \cite{LukyanovTerrasSpinSpinAmplitudesBetterTreatementXXZ}
and also to compute the longitudinal structure factor for the XXZ model
in the massless regime at zero magnetic field
\cite{CauxKonnoSorrelWestonFFofMasslessXXZfromXYZResults}. 
% % % % % % % %
For the moment, we are unable to make a direct and general connection
between our formulae and those obtained within the vertex-operator approach.
Still, in the subsequent analysis, we find agreement in the case of
two-hole excitations. 
% %

It is interesting to note that the formula by Jimbo and Miwa for the
two-particle form factor of $\sigma^z$ involves a non-trivial contour
integral, whereas the expression found by Lashkevich does not contain
any integrals. Therefore, we prefer to compare our result with
Lashkevich's result \cite{LashkevichElemBlocksEightVertex}. Taking the limit
to the massive XXZ model, we obtain for the two-particle amplitude
\begin{equation}
\label{voa}
\frac{1}{4} \, \left| \phantom{.}^{(0)} \langle vac | \sigma^z | \nu_1,\nu_2 \rangle^{(0)}_{\epsilon,-\epsilon} \pm  \phantom{.}^{(1)} \langle vac | \sigma^z | \nu_1,\nu_2 \rangle^{(1)}_{\epsilon,-\epsilon} \, \right|^2 = \left| f_\pm (\nu_1,\nu_2) \right|^2 
\end{equation} 
where the indices $(0)$ and $(1)$ label the ground states and $\epsilon \in \{ +,- \} $ is the spin index.
The functions $f_\pm$ are defined by
\begin{equation}
f_-(\nu_1,\nu_2) = f_+(\nu_1+\pi,\nu_2) = \frac{\pi}{\eta} \,  \frac{G(\nu_1 - \nu_2) \, \vartheta_1 \left(\frac{\pi(\nu_1+\nu_2)}{2\mathrm{i}\eta} \, |\, \mathrm{e}^{-\pi^2/\eta} \right)}{\vartheta_1 \left(\frac{\pi \nu_1}{2\mathrm{i}\eta}-\frac{\pi}{4} \, | \, \mathrm{e}^{-\pi^2/2\eta} \right) \vartheta_1 \left(\frac{\pi \nu_2}{2\mathrm{i}\eta}-\frac{\pi}{4} \, | \, \mathrm{e}^{-\pi^2/2\eta} \right) } \frac{ \mathrm{i} \, \vartheta^\prime_1 \left(0 \, | \, \mathrm{e}^{-\pi^2/\eta} \right) }{\sin\left(\frac{\nu_1-\nu_2}{2} + \frac{\mathrm{i}\eta}{2} \right)}~,
\end{equation}
wherein
\begin{equation}
G(x) = \mathrm{e}^{\frac{\eta}{4}(\frac{x}{\eta}+\mathrm{i})^2} \left( \frac{\qtriprod{q^4}}{\qtriprod{q^2}} \right)^2 \frac{\qtriprod{\mathrm{e}^{-2\mathrm{i}x}} 
\qtriprod{q^4 \mathrm{e}^{2\mathrm{i}x}} }{\qtriprod{q^2 \mathrm{e}^{-2\mathrm{i}x}} \qtriprod{q^6 \mathrm{e}^{2\mathrm{i}x}} }~.
\end{equation}
Here we have introduced the notation
\begin{equation}
(x;p_1,p_2)_\infty = \prod_{k,l=0}^\infty (1- x \, p^k_1 \, p^l_2)~.
\end{equation} 
For our conventions for Jacobi-theta functions we refer to (\ref{JacobiThetaFunctions}).

Let us consider now our result, Theorem \ref{Theorem FF large L asymptotics}, in the case of an arbitrary two-hole
excitation. This implies that we have a single higher-level Bethe equation
(for one root $\chi$) whose solutions were given in (\ref{chi1andchi2}).
%read
%\begin{equation}
%\chi_1 = \frac{\nu_1+\nu_2}{2} ~, \qquad \chi_2 = \frac{\nu_1+\nu_2}{2} + \frac{\pi}{2}~.
%\end{equation} 
Inserting the solution $\chi_1$ in Theorem \ref{Theorem FF large L asymptotics} we obtain the amplitude 
\begin{align}
\label{twoparticle}
\mc{F}_m^{(z)}\big( \{\wt{\mu}_a\}_1^N; \{\la_a\}_1^N\big) \; &= \sin^2 \big( \iota\pi /2 + \pi p(\nu_1) + \pi p(\nu_2) \big) 
\left(  \frac{ \mathrm{e}^{-2\pi \mathrm{i} m (p(\nu_1)+ p(\nu_2))} }{ (2\pi L)^2 p^\prime(\nu_1) p^\prime(\nu_2) }   \right) \, (-1)^{\iota m} 
\notag \\[1em] &~\times  128~ \frac{\sin\big(\pi F_\iota(\nu_1) \big) \sin\big(\pi F_\iota(\nu_2)\big)}{ \sinh^2(\eta)}~\sin^2(\nu_{12})~ \frac{\mathcal{D}}{\prod_{n \in \mathbb{Z}} (1+ q^{2 |n|})^2 } \notag \\[1em]
&\times \qprod{q^2}^4  \left( \frac{\qtriprod{q^4}}{\qtriprod{q^2}}  \right)^8  \prod_{\epsilon=\pm} \qprod{q^2 e^{2 \i\epsilon \nu_{12}}}^2 \left( \frac{\qtriprod{q^4 e^{2 \i\epsilon \nu_{12}}}}{\qtriprod{q^2 e^{2 \i\epsilon \nu_{12}}}}  \right)^4~,
\end{align}
where $\iota \in \{0, 1\}$ and  $\nu_{12}=\nu_1-\nu_2$. The infinite product is defined by
\begin{equation}
(x;p)_\infty = \prod_{k=0}^\infty (1- x \, p^k)~.
\end{equation} 
The factor $\mathcal{D}$ is given by Fredholm determinants,
\begin{equation}
\label{determinants}
\mathcal{D} = \det_{[-\frac{\pi}{2} - \mathrm{i} \epsilon,\frac{\pi}{2} -\mathrm{i} \epsilon]} \{ I + \bar{U}_{\nu_1}  \} \cdot \mathcal{R}_{\nu_1} \cdot  \det_{[-\frac{\pi}{2} - \mathrm{i} \epsilon,\frac{\pi}{2} -\mathrm{i} \epsilon]} \{ I + \bar{U}_{\nu_2}  \} \cdot \mathcal{R}_{\nu_2} ~.
\end{equation}
The determinants involve operators acting on the interval $[-\pi/2 , \pi/2 ]$ shifted in the lower half plane by $0<\epsilon<\eta$.
The kernel is given by
\begin{equation}
\bar{U}_{\vartheta} (x,y) = e^{- \i \pi \iota } ~ \frac{\vartheta_4(x-\nu_1 \, | \, q^2)\vartheta_4(x-\nu_2  \, | \, q^2)}{\vartheta_1(x-\nu_1  \, | \, q^2)\vartheta_1(x-\nu_2  \, | \, q^2)} ~ [ K (x-y) - K (\vartheta -y) ]~,
\end{equation}
where $\iota \in \{ 0,1\}$.
Finally, the factor $\mathcal{R}_{\nu_{1}} $ is defined by
\begin{align}
\mathcal{R}_{\nu_1} &= \notag 1 + \frac{2 \pi \mathrm{i} ~ \mathrm{e}^{- \mathrm{i} \pi \iota} }{\mathrm{e}^{-2 \pi \mathrm{i} F_\iota(\nu_1)}-1} ~ \frac{\vartheta_4(0 \, | \, q^2)\vartheta_4(\nu_1-\nu_2 \, | \, q^2)}{\vartheta^\prime_1(0 \, | \, q^2)\vartheta_1(\nu_1-\nu_2 \, | \, q^2)}
\\  & \hspace{1cm} \times \left(  K(0) - K(\nu_2-\nu_1) - \int_{-\frac{\pi}{2} -\mathrm{i} \epsilon}^{\frac{\pi}{2} - \mathrm{i} \epsilon} \mathrm{d}z ~R_{\nu_2}(\nu_1,z)[K(\nu_1-z) -K(\nu_2-z)]  \right)~.
\end{align}
The corresponding equation for $\mathcal{R}_{\nu_2} $ is obtained by interchanging $\nu_1$ and $\nu_2$.
The function $R_\vartheta(x,y)$ is the resolvent of the operator $\bar{U}_\vartheta$ defined as the solution to the linear
integral equation
\begin{equation}
R_\vartheta(x,y) = \bar{U}_\vartheta(y,x) - \int_{-\frac{\pi}{2} -\mathrm{i} \epsilon}^{\frac{\pi}{2} -\mathrm{i} \epsilon} \mathrm{d}z ~R_{\vartheta}(z,y) \bar{U}_\vartheta(z,x)~.  
\end{equation}
% % % % % % % % %
Note that the amplitude for the second solution $\chi_2$ of the higher-level equations can be obtained by replacing $(\nu_1,\nu_2) \hookrightarrow (\nu_1+\pi,\nu_2) $ in \eqref{twoparticle}.

\begin{figure}
	\centering
	\includegraphics[width=0.9\linewidth]{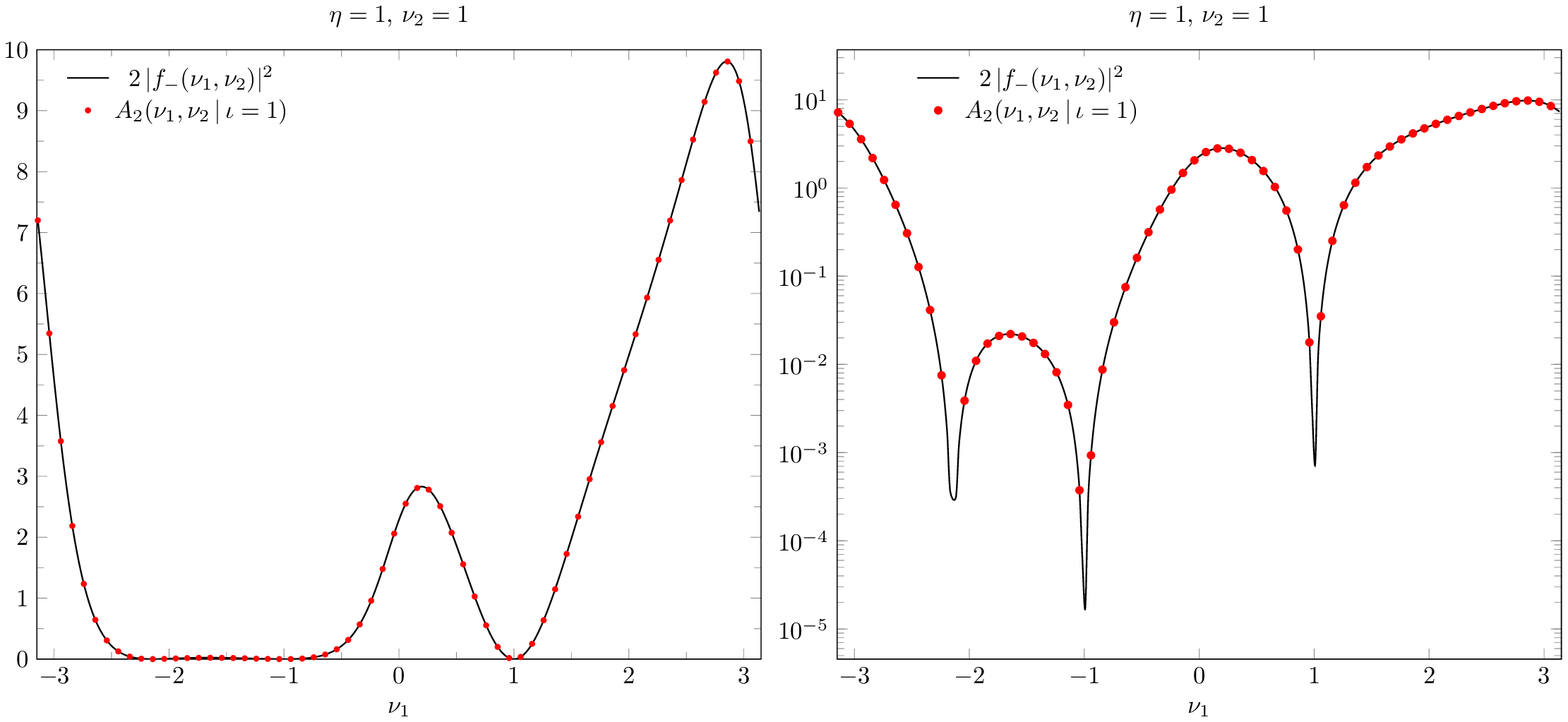}
	\caption{Comparison of our result $A_2(\nu_1,\nu_2 \, | \, \iota=1)$ (red dots) and the prediction $2 \, | f_-(\nu_1,\nu_2)   |^2$ from the vertex-operator approach (black line) as a function of $\nu_1$ 
	for a fixed value of $\nu_2$ (left panel). The right panel shows the same plot with logarithmic coordinate.}
	\label{fig:plots}
\end{figure}

The quantity we should compare with the vertex-operator formula \eqref{voa} is the form-factor density
\begin{equation}
A_2(\nu_1,\nu_2 \, | \, \iota) = \mc{F}_{m=0}^{(z)}\big( \{\wt{\mu}_a\}_1^N; \{\la_a\}_1^N\big) \cdot \left(  (2\pi L)^2  p^\prime(\nu_1) p^\prime(\nu_2)   \right)~.
\end{equation}
For the moment, we do not know how to calculate the Fredholm determinant part $\mathcal{D}$ analytically. However, we can compute it numerically and then compare with \eqref{voa}.
Remarkably, we find from a numerical calculation that $\mathcal{R}_{\nu_{1,2}}=0$
for the cases $\chi=\chi_1,~\iota=0$ and $\chi=\chi_2,~\iota=1$. The remaining two
non-zero amplitudes are connected by a shift $\nu_1 \hookrightarrow \nu_1 + \pi$.
Thus, our results agree with those from the vertex-operator approach if
\begin{equation}
\label{conjecture}
A_2(\nu_1,\nu_2 \, | \, \iota=1) = 2 \, | f_-(\nu_1,\nu_2)   |^2 
\end{equation}
holds.\footnote{The factor 2 is due to the summation over the spin index
$\epsilon$ in \eqref{voa}.} Our numerical calculation shows that our conjecture
\eqref{conjecture} is indeed correct (cf.\ Figure \ref{fig:plots}). Note
that \eqref{conjecture} implies a non-trivial identity for the Fredholm determinant
part $\mathcal{D}$ defined in \eqref{determinants}. It would be interesting to
have a direct proof for this identity.

We have shown the equivalence of our expressions to those from the
vertex-operator approach in the case of two-hole excitations.
We expect that the equality holds, in fact, in each $2n_\chi$-hole
excitation sector, hence leading to highly non-trivial identities between
multiple integrals. We plan to explore this question in a separate publication.

\subsection{\boldmath The large-$m$ asymptotic expansion}

\begin{prop}

 Assume that the form-factor series \eqref{ecriture series FF} is convergent. Then, one has the large-distance asymptotic expansion  
\beq
\moy{ \sg_1^z(0) \cdot \sg_{m+1}^z(0) } \; = \; (-1)^m \pl{n \geq 1}{} \bigg( \f{ 1- \ex{-2n\eta} }{ 1+\ex{-2n\eta} } \bigg)^4
 \; + \; \e{O}\big( \ex{-cm} \big)
 \label{asymptotics0}
\enq
for some $c>0$. 
\end{prop}

\Proof 

The sole complication in getting the result stems from the justification of the possibility to deform contours from $\intff{-\tf{\pi}{2}}{\tf{\pi}{2}}$
up to $\intff{-\tf{\pi}{2}}{\tf{\pi}{2}}-\i \tau$, with $\tau>0$ but small enough. In deforming the contours, one will, in principle
get the contributions of the boundaries $\intff{\tf{\pi}{2}}{\tf{\pi}{2}-\i \tau}\cup\intff{-\tf{\pi}{2}-\i \tau}{-\tf{\pi}{2}}$. These do not cancel out directly since the 
integrand, is \textit{not} $\pi$-periodic with respect to the parameters $\{\nu_a\}_1^{n_h}$.

Indeed, by using the explicit expression for the thermodynamic limit of the counting function, one obtains
\beq
 F_{\iota}\Big(s \, | \,   \{\nu_a +\pi \de_{ab} \}_1^{n_h} ;  \{\chi_a\}_1^{n_{\chi}}  \Big) 
 \; = \;  F_{\iota}\Big(s \, | \,   \{\nu_a\}_1^{n_h} ;  \{\chi_a\}_1^{n_{\chi}}  \Big) \; + \; \f{1}{2} \;. 
\enq
This property along with straightforward manipulations implies that 
\beq
\msc{F}^{(z)}_{0}\big(   \{\nu_{a} +\pi \de_{ab} \}_1^{n_h}  \big) \; = \; \msc{F}^{(z)}_{1}\big(   \{\nu_{a} \}_1^{n_h}  \big) 
\quad \e{and} \quad
\msc{F}^{(z)}_{1}\big(   \{\nu_{a} +\pi \de_{ab} \}_1^{n_h}  \big) \; = \; \msc{F}^{(z)}_{0}\big(   \{\nu_{a} \}_1^{n_h}  \big)   \;. 
\enq
Likewise, the quasi-periodicity of the dressed momentum $p(s+\pi)=p(s)+\tf{1}{2}$ ensures that 
\beq
(-1)^{ \iota m }  \pl{a=1}{n_h} \ex{-2\i \pi m p(\nu_{a})}  \quad \underset{ \nu_c \hookrightarrow \nu_c+\pi \de_{cb} }{ \hookrightarrow  } \quad
(-1)^{ (\iota+1) m }  \pl{a=1}{n_h} \ex{-2\i \pi m p(\nu_{a})} 
\enq
Hence, all-in-all, one has that the function 
\beq
\{\nu_{a}\}_1^{n_h} \; \mapsto \; \sul{\iota=0}{1}  (-1)^{ \iota m }  \pl{a=1}{n_h} \ex{-2\i \pi m p(\nu_{a})} \cdot \msc{F}^{(z)}_{\iota}\big(   \{\nu_{a}\}_1^{n_h}  \big)
\enq
is $\pi$-periodic in each of the variables $\nu_a$. Furthermore, the integrands in each term of the form-factor series are holomorphic
in some open neighbourhood of $\intff{-\tf{\pi}{2} }{ \tf{\pi}{2} }$. We thus deform the original contour to the lower-half plane. 
Note that, due to the $\pi$-periodicity, the contributions issuing from an integration on 
$\intff{\tf{\pi}{2}}{\tf{\pi}{2}-\i \tau}$ cancel out with those issuing from an integration on $\intff{-\tf{\pi}{2}-\i \tau}{-\tf{\pi}{2}}$. \qed

%Note that 
The lack of a better insight into the analytic structure of the integrand
\eqref{ecriture integrale multidim residus} does not allow us, for the moment,
to obtain a better estimation of the constant $c>0$ in \eqref{asymptotics0}
directly from Theorem 2.1. However, if we assume that 
\begin{itemize}
\item conjecture \eqref{conjecture} is correct,
\item the higher-spinon excitations in the form-factor expansion
give rise to a sub-dominant large-$m$ asymptotic behaviour
relative to the two-spinon excitations,
\end{itemize}
we can calculate the next term in the asymptotic expansion \eqref{asymptotics0}.
The second hypothesis might seem trivial on first thought. However, the
features of a large-parameter asymptotic behaviour of multi-dimensional
deformations of a Fredholm  determinant -- which is basically the case of 
all series of multiple integrals of representations for the correlation
functions in integrable models away from their free fermion point -- can
go quite far outside the scheme of classical asymptotic analysis. 
See, \textit{e}.\textit{g}. \cite{KozMailletSlaLongDistanceTemperatureNLSE},
where it was shown that the large-distance asymptotic behaviour of
the generating function of density-to-density correlation functions
in the non-linear Schr\"{o}dinger model gives rise to a tower of correlation
lengths that is quite different from the one that could be expected on
the basis of a `classical' term-by-term analysis of the individual integrals
building up the series.

\begin{prop}
Under the above assumptions, the asymptotic expansion below holds
\begin{equation}
\moy{ \sg_1^z(0) \cdot \sg_{m+1}^z(0) } \; = \; (-1)^m \pl{n \geq 1}{} \bigg( \f{ 1- \ex{-2n\eta} }{ 1+\ex{-2n\eta} } \bigg)^4
	\; + \; A \cdot \frac{\mathrm{k}\big(q^2 \big)^m}{m^2} \, \bigg( (-1)^m - \tanh^2\left(\frac{\eta}{2} \right)  \frac{(q;q^2)^4_\infty}{(-q;q^2)^4_\infty}         \bigg) \cdot \left( 1+ \e{O}\left( m^{-1} \right) \right)
	\label{asymptotics1}
\end{equation}	
where
\begin{equation}
\mathrm{k}\big(q^2 \big) = \frac{\vartheta_2^2(0 \, | \,  q^2)}{\vartheta_3^2(0 \, | \,  q^2)}~,\qquad A = 
\frac{1}{\pi \, \sinh^2 \big(\frac{\eta}{2} \big)} \, \frac{(-q;q^2)^4_\infty}{(q^2;q^2)^2_\infty} \, \left( \frac{\qtriprod{q^4}}{\qtriprod{q^2}} \right)^8~.
\end{equation}
\end{prop}

\Proof 

Suppose that \eqref{conjecture} is correct. Then the two-hole contribution to the
form-factor series \eqref{ecriture series FF} is given by 
\begin{equation}
I_2 (m) = \int_{-\pi}^{0} \frac{\mathrm{d}\nu_2}{2\pi} \,  \mathrm{e}^{2\pi \mathrm{i} m p(\nu_2) } \int_{-\pi}^{0} \frac{\mathrm{d}\nu_1}{2\pi} \,  \mathrm{e}^{2\pi \mathrm{i} m p(\nu_1) } \,  \left( (-1)^m \cdot |f_-(\nu_1,\nu_2)|^2  + |f_+(\nu_1,\nu_2)|^2 \right)
\label{twospinoncontribution}
\end{equation}
Since the integrand is $\pi$-periodic and holomorphic\footnote{Of course,
we have to evaluate the modulus for real arguments and then continue analytically
to the upper half-plane.} in the strip $0\leq\Im(\nu)\leq\eta/2$, we may
shift the contour by $\eta/2$ in the upper half-plane (note that
${\rm e}^{2 \pi {\rm i} p(\nu)}$ vanishes at $\{ -\pi+\i\eta/2, \i \eta/2 \}$
hence compensating the poles stemming from the form factor density).
The exponent becomes real and negative on this contour with a single maximum
at $\nu=-\pi/2 + \mathrm{i} \eta/2$. A saddle-point analysis then leads to
\eqref{asymptotics1}. Note that a similar analysis was performed in
\cite{JohnsonKrinskyMcCoyCorrelationLength8VModelAndXYZ} to determine
the correlation lengths in the eight-vertex model.

A similar method can be used to study the limit $m,t \rightarrow \infty$
(with fixed $m/t=\mathrm{v}$) of the dynamical correlation functions.
We plan to study this problem in future work.

\begin{figure}
\centering
\includegraphics[width=0.55\linewidth]{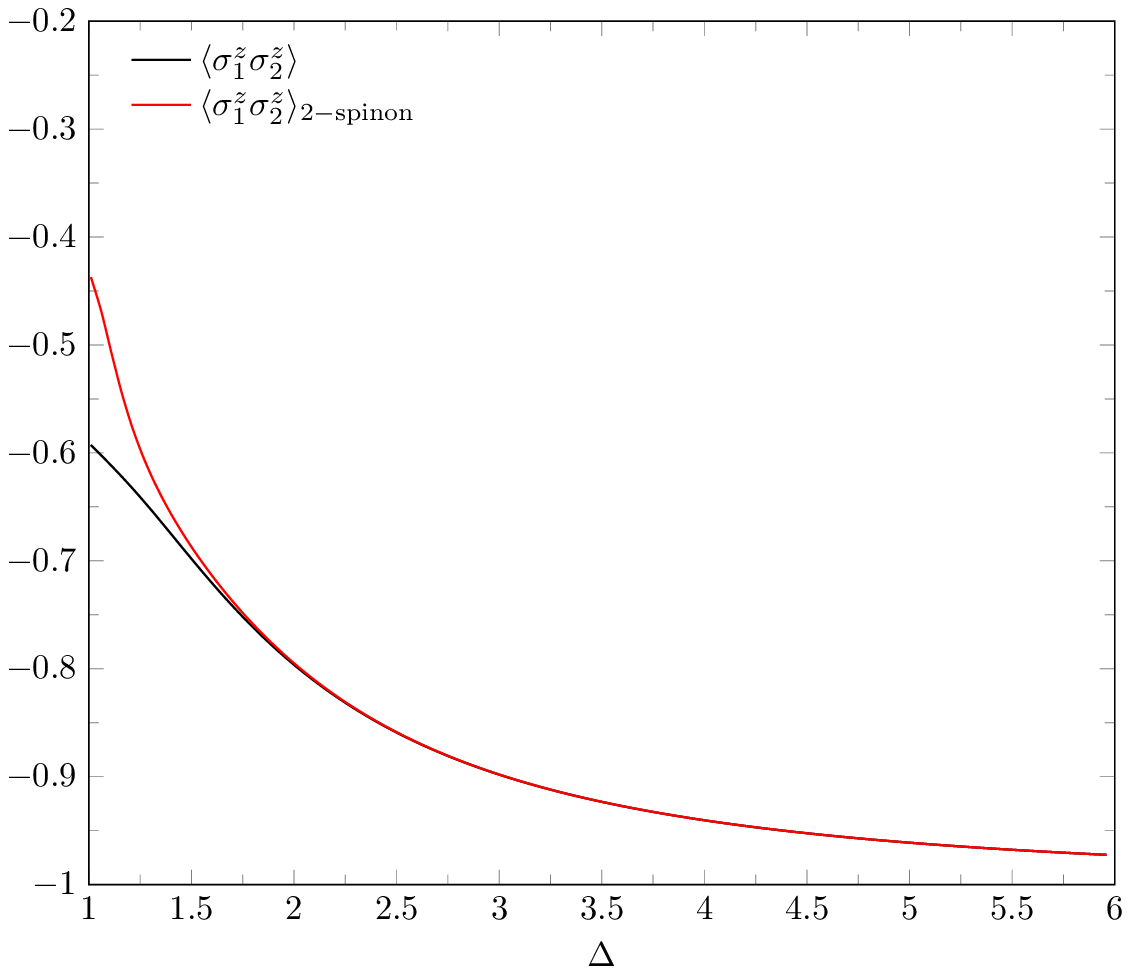}
\caption{Comparison of the exact nearest-neighbour correlator
\cite{KatoShiroishiTakahashiNextNearestCorrelationFctInMassiveXXZ} (black line)
and the 2-spinon approximation (red line). The relative difference between
both curves is of the order $10^{-3}$ for $\Delta>2$. Note that both
functions remain finite in the limit $\Delta\rightarrow 1$ with the known
ratio of ca. 73\% (\textit{cf}.\ \cite{BougourziFledderjohanKarbachMullerMutterDynamicTwoSpinonStructureFactor})}
\label{fig:Comparison_Exact_TwoSpinon}
\end{figure}

We would like to add that the integral $I_2 (m)$ can also be computed
numerically (using the shifted contour) for short distances $m$ where the
asymptotic expansion is not efficient. Remarkably, away from the isotropic
point $\Delta = 1$, the 2-spinon contribution
\begin{equation}
\moy{ \sg_1^z(0) \cdot \sg_{2}^z(0) }_{2-\text{spinon}} \; = \; - \pl{n \geq 1}{} \bigg( \f{ 1- \ex{-2n\eta} }{ 1+\ex{-2n\eta} } \bigg)^4 \, + \, I_2 (1)
\end{equation}
to the nearest-neighbour correlation functions approximates the exact result
\cite{KatoShiroishiTakahashiNextNearestCorrelationFctInMassiveXXZ} with high precision (\textit{cf}.\
Figure~\ref{fig:Comparison_Exact_TwoSpinon}). 
    
\section{Summary and outlook}

We revisited the problem of the evaluation of form factors of the spin-$\frac{1}{2}$ XXZ model in the massive regime from the algebraic Bethe Ansatz perspective. We
started from a determinant expression of the form factor of the operator $\sigma^z$ in the finite volume
\cite{KMTFormfactorsperiodicXXZ,KozKitMailSlaTerXXZsgZsgZAsymptotics} and performed a careful analysis of its large-$L$ behaviour. For this purpose we re-analysed the Bethe Ansatz
equations using a non-linear integral equation for the counting function.
This allowed us to resolve a certain controversial fine-point in the older literature
\cite{BabelondeVegaVialletStringHypothesisWrongXXZ,VirosztekWoynarovichStudyofExcitedStatesinXXZHigherLevelBAECalculations}. From the non-linear integral
equations we obtained the higher-level Bethe Ansatz equations \cite{VirosztekWoynarovichStudyofExcitedStatesinXXZHigherLevelBAECalculations}
that determine the complex Bethe roots pertaining to the low-lying excitations of the model in the thermodynamic limit. We managed to analyse form factors
parametrised by such complex roots in the large-$L$ limit. Our main result is their explicit characterization in Theorem \ref{Theorem FF large L asymptotics}.

We would like to emphasise that, at least for a small number of holes,
the formulae in Theorem \ref{Theorem FF large L asymptotics} are efficient for a
numerical calculation of form factors and amplitudes. We exemplified this
in Section \ref{Section FF series expansion}, where we considered the
form-factor expansion of the $\sigma^z$-$\sigma^z$ two-point function.
Assuming convergence of the form-factor series we could show that the
first term is given by the staggered magnetization of Baxter
\cite{BaxterStaggeredPolarisationInFModel,JimboMiwaFormFactorsInMassiveXXZ},
while the remainder decays exponentially fast with the distance. We further
compared our result for the two-spinon case with the formula
obtained by Lashkevich \cite{LashkevichElemBlocksEightVertex} in
the context of the XYZ model and found numerical agreement. We naturally
expect agreement in general, which implies non-trivial identities among
multiple-contour integrals and Fredholm determinants. We hope to clarify
this point in near future. Using the form of the two-spinon amplitude implied
by Lashkevich's form-factor expression, we obtained an explicit result
for the first decaying correction to the zeroth order formula for the
two-point function given by the staggered magnetization. This result
appears to be heretofore unknown and generalizes an old formula of Johnson,
Krinsky and McCoy \cite{JohnsonKrinskyMcCoyCorrelationLength8VModelAndXYZ}.

We have also started to analyse the temperature correlation functions of the XXZ model in the massive regime by means of the quantum transfer
matrix approach at low temperatures. This provides a different view on the two-point functions and again structurally different formulae
which we plan to publish separately. In separate work we shall also work out the asymptotics of the dynamical two-point functions that
follows from the form-factor series expansion. It would be interesting to compare, in this case, the results with those issuing from the analysis
of the density structure factor \cite{BougourziWestonZeroAndTwoParticleDSFInMassiveXXZ}.

%%%%%%%%%%%%%%%%%%%%%%%%%%%%%%%%%%%%%%%%%%%%%%%%%%%%%%%%%%%%%%%%%%%%%%%%%%%%%%%%%%%%%%%%%%%%%%%%%%%%%%%%%%%%%%%%%%%%%%%%%%%%%%%%%%
%%%%%%%%%%%%%%%%%%%%%%%%%%%%%%%%%%%%%%%%%%%%%%%%%%%%%%%%%%%%%%%%%%%%%%%%%%%%%%%%%%%%%%%%%%%%%%%%%%%%%%%%%%%%%%%%%%%%%%%%%%%%%%%%%%

%%%%%%%%%%%%%%%%%%%%%%%%%%%%%%%%%%%%%%%%%%%%%%%%%%%%%%%%%%%%%%%%%%%%%%%%%%%%%%%%%%%%%%%%%%%%%%%%%%%%%%%%%%%%%%%%%%%%%%%%%%%%%%%%%%
%%%%%%%%%%%%%%%%%%%%%%%%%%%%%%%%%%%%%%%%%%%%%%%%%%%%%%%%%%%%%%%%%%%%%%%%%%%%%%%%%%%%%%%%%%%%%%%%%%%%%%%%%%%%%%%%%%%%%%%%%%%%%%%%%%

%%%%%%%%%%%%%%%%%%%%%%%%%%%%%%%%%%%%%%%%%%%%%%%%%%%%%%%%%%%%%%%%%%%%%%%%%%%%%%%%%%%%%%%%%%%%%%%%%%%%%%%%%%%%%%%%%%%%%%%%%%%%%%%%%%
%%%%%%%%%%%%%%%%%%%%%%%%%%%%%%%%%%%%%%%%%%%%%%%%%%%%%%%%%%%%%%%%%%%%%%%%%%%%%%%%%%%%%%%%%%%%%%%%%%%%%%%%%%%%%%%%%%%%%%%%%%%%%%%%%%

\section*{Acknowledgments}
The authors would like to thank Andreas Kl\"umper and Nicolai Kitanine for fruitful discussions. MD and FG acknowledge financial support by the
Volkswagen Foundation and by the Deutsche Forschungsgemeinschaft
under grant number Go 825/7-1. KKK is supported by the CNRS. His work has
been partly financed by the Burgundy region PARI 2013 $\&$ 2014 FABER grants
`Structures et asymptotiques d'int\'egrales multiples'. KKK also enjoys
support from the ANR `DIADEMS' SIMI 1 2010-BLAN-0120-02. JS is supported by
a JSPS Grant-in-Aid for Scientific Research (C) No.\ 24540399. MD, FG and
JS are grateful for the warm hospitality extended to them at the IMB,
Universit\'e de Bourgogne. KKK and JS would like to express their gratitude
to the physics department of the University of Wuppertal for providing
a stimulating work environment during their visit. 

\appendix

\section{Solutions to linear integral equations}
\label{Appendix Explicit resolution LIE}

\subsection{The Fourier coefficients}

The linear integral equations driven by the integral operator $I+K$ can be solved by means
of Fourier series expansion of $\pi$-periodic functions 
\beq
f(\la) \; = \; \sul{ n \in \mathbb{Z} }{} c_n[f] \cdot \ex{2 \i n\la} \qquad \e{with} \qquad 
c_n[f] \; =  \; \Int{ -\tf{\pi}{2} }{ \tf{\pi}{2} } f(\la) \ex{-2 \i n \la} \cdot \f{ \dd \la }{ \pi } \; . 
\enq
It is %readily seen that 
easy to check that 
\beq
c_n[K]=\f{ \ex{-2 |n| \eta} }{\pi} \qquad \e{and} \qquad 
\Int{-\tf{\pi}{2} }{ \tf{ \pi }{ 2 } } f(\la-\mu) g(\mu) \cdot \dd \mu  \; = \; \pi \sul{n}{} c_n[f] \cdot c_n[g] \cdot \ex{2 \i n\la} \;. 
\enq
Further, defining  
\beq
b_n(t) \; = \; \Int{-\tf{\pi}{2} }{ \tf{\pi}{ 2}} \th(\la-t)  \ex{-2 \i n \la} \cdot  \f{ \dd \la }{ \pi }   
\enq
one obtains that
\begin{itemize}
\item when $|\Im(t)|>\eta$
\beq
b_0(t) \; = \;  2 \eta \e{sgn}\big( \Im(t) \big)  \qquad \e{and} \qquad 
b_n(t) \; = \; \f{2}{n} \sinh(2n \eta) \ex{-2 \i nt}  \cdot 
\Big( \de_{n < 0} \bs{1}_{\Im(t)> \eta} \; - \; \de_{n >0} \bs{1}_{\Im(t) < -\eta}  \Big)
\enq
for $n \not=0$ ; 
\item and, when $|\Im(t)|<\eta$, one has 
\beq
b_0(t) \; = \; \i (\pi \; - \; 2  t)  \qquad \e{and} \qquad 
b_n(t) \; = \;  \f{ (-1)^{n+1}  }{ n }  \; + \; \f{1}{n} \cdot \ex{-2\i nt - 2 |n| \eta } 
\enq
for $n \not=0$.
\end{itemize}

The coefficients $b_n(t)$, for $n\not=0$, are readily obtained through an integration by parts. 
In order to compute $b_0(t)$, it remains to observe that $t \mapsto b_0(t)$ is analytic in the regions $|\Im(t)|>\eta$ and 
$|\Im(t)|<\eta$, and that 
\beq
b_0^{\prime}(t) \; = \;  \left\{  \ba{cc} 0 &   |\Im(t)|>\eta  \\ 
                          	-2 \i  & |\Im(t)|<\eta  \ea \right.  \;. 
\enq
Further, the function $\th$ has  jumps on $\intff{-\tf{ \pi}{2} }{ \tf{\pi}{2}} \pm \i \eta$
\beq
\th\big( \la-x\pm \i \eta + \i 0^+ \big) \, - \,  \th\big( \la-x \pm \i \eta - \i 0^+ \big)  \; = \; 
\mp 2 \i \pi \bs{1}_{\la \in \intff{ x }{ \tf{\pi}{2} }} \;. 
\enq
The asymptotics
\beq
\lim_{y \tend \pm \infty} \th(\la-\i y) \; = \;  \pm 2\eta \quad \e{and}\; \e{the} \; \e{jump} \quad 
b_0\big( x \pm \i \eta - \i 0^+ \big) \, - \, b_0\big( x \pm \i \eta +\i 0^+ \big)\; = \; \pm 2 \i \big( \tf{\pi}{2} \, - \,  x \big)
\enq
%
%
%
%allow one to
 fix the values of $b_0(t)$ in each of the regions.

\subsection{The dressed momentum and energy}
\label{Appenix dressed momentum}

The integral equation \eqref{definition eqn int moment habille} %satisfied by the dressed momentum 
can be solved explicitly by means of Fourier transformations. One finds, for $\la \in \intff{-\tf{\pi}{2} }{ \tf{\pi}{2} }$, 
\beq
p(\la) \; = \; \f{1}{2} p\big( -\f{\pi}{2} \big) \, + \,  \f{ \la \, + \, \tf{\pi}{2} }{ 2 \pi } 
\; + \; \f{ 1 }{ 2\pi } \sul{ n \in \mathbb{Z}\setminus\{0\} }{} \f{ \ex{2 \i n \la }  }{ 2 \i n \cosh(n\eta) } \;. 
\enq
Thus
\beq
p\big(-\f{\pi}{2} \big)=0 \quad \e{and} \quad p\big(\f{\pi}{2} \big)=\f{1}{2} \quad viz. \quad
p(z) \; = \; \Int{ - \tf{\pi}{2} }{ z } p^{\prime}(s) \cdot \dd s  \;. 
\enq
As a consequence
\beq
p(\la) \; = \; \f{ \la \, + \, \tf{\pi}{2} }{ 2 \pi } 
\; + \; \f{ 1 }{ 2\pi } \sul{ n \in \mathbb{Z}\setminus\{0\}}{} \f{ \ex{2 \i n \la }  }{ 2 \i n \cosh(n\eta) } \;. 
\enq
In fact, the derivative $p^{\prime}$ can be expressed in terms of elliptic functions as
\beq
p^{\prime}(\la) \; = \;  \f{ 1 }{ 2\pi } \sul{ n \in \mathbb{Z} }{} \f{ \ex{2 \i n \la }  }{  \cosh(n\eta) }
\; =\; \sul{n \in \mathbb{Z} }{}  
			\f{ 1 }{ 2\eta  \cosh\Big[\f{\pi}{\eta}(n \pi - \la) \Big]}  
			\; = \; \f{ 1 }{ 2\pi } \pl{n \geq 1}{} \bigg\{  \f{ 1-q^{2n} }{ 1+q^{2n} }\bigg\}^2	
\cdot \f{ \vartheta_3(\la\mid q)  }{  \vartheta_4(\la\mid q) }		\;. 
\label{ecriture rep eq pour p prime}
\enq
The second representation for $p^{\prime}$ ensures that $p$ is strictly increasing on $\intff{-\tf{\pi}{2} }{ \tf{\pi}{2} }$.
Furthermore, it is also easy to deduce from it  that $p^{\prime}$ is $\pi$-periodic and $\i\eta$
anti-periodic. Above, we used the following convention for the $\vartheta$ functions of nome $q=\ex{-\eta}$
\beq
\label{JacobiThetaFunctions}
\begin{split}
& \vartheta_3(\la\mid q) \; = \; \pl{n \geq 1}{} \Big\{ \big( 1-q^{2n} \big)\cdot 
			\big(1+2q^{2n-1}\cos(2\la) + q^{4n-2} \big) \Big\} \\%	 \qquad \e{and} \qquad 
& \vartheta_4(\la\mid q) \; = \;	\vartheta_3(\la-\tf{\pi}{2}\mid q) \;, \quad
  \vartheta_1(\la\mid q) \; = \;	- {\rm i} {\rm e}^{{\rm i} \lambda - \eta/4}
                                        \vartheta_4(\la+\tf{{\rm i}\eta}{2}\mid q) \;, \quad
  \vartheta_2(\la\mid q) \; = \;	\vartheta_1(\la+\tf{\pi}{2}\mid q) \; .	
\end{split}
\enq
We stress that the second equality in \eqref{ecriture rep eq pour p prime}
follows from the Poisson summation formula whereas the third one follows from the fact that 
$\la \mapsto p^{\prime}(\la)$ is a $\pi$-periodic, $\i \eta$ anti-periodic function whose only simple pole in the fundamental 
simplex is located at $\tf{\eta}{2}$ and has residue $2 \i \pi $.  

 The function $p$ can be explicitly computed in terms of the semi-infinite product
\beq
(z)_{\infty} \, = \, \pl{ n \geq 0}{} \big( 1-z q^{4n} \big) \qquad \e{and} \qquad
\bigg( \ba{c}  \{a_k\}_1^p  \\ \{b_k\}_1^{s} \ea \bigg)_{\infty} \, = \, 
\f{  \pl{k=1}{p} (a_k)_{\infty} }{  \pl{k=1}{s} (b_k)_{\infty} }
\enq
as
\beq
p(\la) \;= \;\f{ \la \, + \, \tf{\pi}{2} }{ 2 \pi }  \; + \; \f{1}{2 \i \pi}
\ln \bigg( \ba{c} q \ex{-2 \i \la} ; q^3 \ex{2 \i \la} \\  q \ex{2 \i \la} ; q^3 \ex{-2 \i \la}   \ea \bigg)_{\infty} 
\; = \; \f{ \la \, + \, \tf{\pi}{2} }{ 2 \pi }  \; + \; \f{1}{2 \i \pi}
\ln \bigg(  \f{ \vartheta_4\big(\la+\i\tf{\eta}{2} \mid q^2 \big)  }{ \vartheta_4\big(\la-\i\tf{\eta}{2} \mid q^2 \big)  } \bigg)\;. 
\enq
%
%
%
%Note that 
The last  expression represents %representation recasts
 the dressed momentum as a $q$-deformation of $p_0$. 

\noindent It follows %readily, 
either from the integral representation for $p$ or from the above representation, that it is an $\i \eta$ anti-periodic 
function with values in $\tf{ \Cx }{ \big( 2\i\pi\mathbb{Z} \big) }$. For instance, one checks that all terms cancel out in the expression for 
\beq
\exp\Big\{ 2 \i \pi\big[ p(\la) + p(\la-\i \eta) \big] \Big\} \; = \; 1 \;. 
\enq
Finally, a direct integration shows that, given $z=x + \i y$, 
\beq
\Re\big[ 2 \i \pi p(z) \big] \; = \; 2 \eta \ln \bigg[ \pl{n \in \mathbb{Z} }{}  
\f{  \cosh\big[ \frac{\pi}{\eta}(x-n\pi) \big]   - \sin\big[ \frac{\pi}{\eta}y \big] }
   { \cosh\big[ \frac{\pi}{\eta}(x-n\pi) \big]  + \sin\big[ \frac{\pi}{\eta}y \big]  } 
\bigg] \;. 
\enq
Hence, 
\beq
0<\Im(z) < \eta \quad \Rightarrow \quad \Re\big[ 2 \i \pi p(z) \big] \; < \;0  
\qquad \e{and} \qquad
-\eta <\Im(z) < 0 \quad \Rightarrow \quad \Re\big[ 2 \i \pi p(z) \big] \; > \;0  \;. 
\label{ecriture signe Re de dressed momentum}
\enq

A direct manipulation of the linear integral equation driving the dressed energy shows that the latter can be recast as
\beq
\veps(\la) \, = \, \f{h}{2} - 4 \pi J \sinh(\eta) \cdot p^{\prime}(\la) \;. 
\enq
It follows from this representation that the model is massive (\textit{viz}. $\veps(\la)<0$ on $\intff{ - \tf{\pi}{2} }{ \tf{\pi}{2} }$)
if the magnetic field satisfies $0\leq h < h_c$, with the critical field $h_c$ being given by
\beq
h_{c} \; = \; 4 J \sinh(\eta)  \pl{n \geq 1}{} \bigg( \f{ 1-q^n }{1+q^n} \bigg)^2 \;. 
\label{definition chp mah critique}
\enq

\subsection{The resolvent}
\label{Subsection Resolvent kernel}

The resolvent kernel is defined as the solution to the linear integral equation
\beq
\label{resolvent_kernel_ieqn}
R(\la-\mu) \, + \, \Int{-\tf{\pi}{2} }{ \tf{\pi}{2} } \! K(\la-\nu) R(\nu-\mu) \cdot \dd \nu  \; = \; K(\la-\mu) \;. 
\enq
The equation can be solved explicitly in terms of Fourier expansion, which yields
\beq
R(\la) \; = \; \sul{ n \in \mathbb{Z} }{} c_n[R]\ex{2 \i n\la} \qquad \e{with} \qquad
c_n[R] \; = \; \f{  \ex{-2|n|\eta} }{ \pi \big( 1\, + \, \ex{-2|n|\eta}  \big)  } \;. 
\enq
One can recast  the Fourier expansion of $R$ in a form that is more suited to the study of the analytic properties of $R$. 
Namely, one has
\bem
R(\la) \; = \; \sul{ n \in \mathbb{Z}}{} \sul{\ell \geq 1 }{} \f{ (-1)^{\ell-1} }{\pi} \ex{-2|n|\ell \eta +2 \i n\la}
\; = \; \f{1}{2\pi} \; + \; 
\sul{\ell\geq 1}{} \f{ (-1)^{\ell-1} }{\pi} \sul{n \geq 1}{}
\Big\{ \ex{ \big(2 \i \la -2\ell \eta \big)n } \; + \;  \ex{ -\big(2 \i \la +2\ell \eta \big)n } \Big\}  \\
\; = \; \f{1}{2\pi} \; + \; \sul{\ell\geq 1}{} \f{ (-1)^{\ell-1} }{\pi} 
\bigg\{  \f{ \ex{ 2 \i \la - 2\ell\eta} }{  1- \ex{2 \i \la - 2\ell\eta}}  \; +\; 
\f{ \ex{-2 \i \la - 2\ell\eta} }{  1- \ex{- 2 \i \la - 2\ell\eta}}  \bigg\} \;. 
\end{multline}
%
%
%
%It is readily seen on the basis of
The above representation clearly shows that $R$ admits a meromorphic extension to 
$\Cx$ which  satisfies the first order finite difference equation
\beq
R(\la+ \i \eta) \,+ \, R(\la) \; = \;  \f{1}{\pi} \Big( \f{1}{1- \ex{-2 \i\la} } \, - \, 
\f{1}{1- \ex{-2\i \la+2\eta} }  \Big) \;. 
\label{ecriture relation shift resolvent R}
\enq

\subsection{The dressed phase and its homogenised version}
\label{Appendix dressed phase}

\subsubsection{\boldmath The case of `close' auxiliary argument $|\Im(z)|<\eta$}

When $|\Im(z)|<\eta$, it is easily %readily 
checked  that the dressed phase has its Fourier coefficients given by  
\beq
c_n\big[ \phi(*, z) \big]  \; = \;  \i \de_{n,0} \f{\pi - 2 z }{ 2 } \; + \; 
\big( 1- \de_{n,0}\big) \cdot \f{ (-1)^{n+1} + \ex{-2 \i n z - 2|n|\eta}  }{ n \cdot \big( 1 + \ex{- 2|n|\eta} \big) } \;. 
\enq
Above, $*$ refers to the variable in respect to which the Fourier coefficients are computed.

The Fourier series for $\phi(\la,z)$ can, in fact, be re-summed and re-cast as
\beq
 \phi( \la , z)  \; = \; \i \f{\pi + 4\la -2z}{2} \; + \; 
 \ln \bigg( \ba{c} - q^4 \ex{2 \i \la} ; -q^2 \ex{-2 \i \la};  q^2\ex{-2 \i (\la-z)} ; q^4 \ex{2 \i (\la-z) }   \\
 				  - q^2 \ex{2 \i \la} ; - q^4\ex{-2 \i \la};  q^4\ex{-2 \i (\la-z)} ; q^2 \ex{2\i (\la-z) }  \ea \bigg)_{\infty} \;. 
\label{ecriture representation dressed phase comme q produit}
\enq
Indeed, by using that 
\beq
\la \; = \; \sul{ n \in \mathbb{Z}\setminus \{0\} }{} \f{ (-1)^{n+1} }{2\i n } \ex{2\i n \la} \qquad \e{on} \quad \intff{- \tf{\pi}{2} }{ \tf{\pi}{2} } \;, 
\enq
one can %recast 
rewrite $\phi(\la,z)$ as
\beq
 \phi( \la , z)  \; = \; \i \f{\pi + 4\la -2z}{2} \; + \; \sul{\eps=\pm 1}{}\eps \sul{n  \geq 1}{} 
 \f{ (-1)^n+\ex{-2\i n z \eps} }{ n(1 + \ex{-2n\eta} )} \ex{-2n\eta + 2 \i n \la \eps} \;.  
\enq
At this stage it solely remains to observe that, for $ | \Im(z) | < \eta$ and $\la$ real, one has
\bem
 \sul{\eps=\pm 1}{}\eps \sul{n  \geq 1}{} 
 \f{ \ex{-2n\eta } }{n (1 + \ex{-2n\eta} ) } \ex{ 2 \i n (\la-z) \eps} \; = \;  
 \sul{\eps=\pm 1}{}\eps \sul{k\geq 1}{} \sul{n  \geq 1}{}   \f{(-1)^{k-1}}{n}\ex{ 2 \i n (\la-z) \eps} \ex{-2n k \eta} \\ 
\; = \; \sul{ \eps=\pm 1 }{} \eps \sul{ p \geq 1}{} \ln \bigg(   \f{ 1- q^{4p} \ex{ 2 \i  (\la-z) \eps} }
								  {1- q^{4p-2}  \ex{ 2 \i  (\la-z) \eps}  } \bigg) \; = \; 
\ln \bigg(  \ba{c}   q^2\ex{-2 \i (\la-z)} ; q^4 \ex{2 \i (\la-z) }   \\
 		    q^4\ex{-2 \i (\la-z)} ; q^2 \ex{2\i (\la-z) }  \ea \bigg)_{\infty} 
\end{multline}
and then add up the expressions for $z$ general and for $z=\tf{\pi}{2}$.

The representation \eqref{ecriture representation dressed phase comme q produit} %readily 
immediately leads to the identity
\beq
\ex{ \phi( \la , z) \, + \, \phi( \la + \i \eta , z) } \; = \; \f{ \cos(\la) \sin(\la-z) }{ \cos(\la + \i \eta) \sin(\la-z+ \i \eta) }
\quad \e{and} \quad 
\ex{ \phi( \la , z) \, + \, \phi( \la - \i \eta , z) } \; = \; \f{ \cos(\la - \i\eta) \sin(\la-z-\i \eta) }{ \cos(\la) \sin(\la-z) } \;. 
\label{ecriture propriete reduction dressed phase close argument}
\enq
The homogenised counterpart of the dressed phase admits the Fourier series expansion
\beq
    \vp(x,z) \;= \; \i \Big( \f{\pi}{2} + x - z \Big) \; + \,  2 \i \sul{n=1}{\infty}  \frac{ \sin\big[2n (x - z) \big] }{ n \big(1 + \ex{-2n\eta} \big)  } \;.
\enq
By using the method described above, the Fourier series can be re-summed as
%and re-cast into the form
%
%
%
\beq
     \vp(x,z) = \i \Big( \f{\pi}{2} + x - z \Big)
     \; +  \; \ln \Bigg( \frac{\Ga_{q^4} \Big( \f{1}{2} - \frac{\i (x - z)}{2 \eta} \Big) \cdot \Ga_{q^4} \Big( 1 + \frac{\i (x - z)}{2 \eta} \Big)}
     {\Ga_{q^4} \Big( \f{1}{2} + \frac{\i (x - z)}{2 \eta} \Big) \cdot \Ga_{q^4} \Big( 1 - \frac{\i (x - z)}{2 \eta} \Big) }
				    \Bigg) \;,
\label{ecriture homogenised dressed phase small z}
\enq
where $\Ga_q$ is the $q$-Gamma function defined in the whole complex plane
by its product representation
\beq
\Ga_q(x) \; = \;  (1 - q)^{1-x} \pl{n=1}{\infty} \f{1 - q^n}{1 - q^{n + x -1}} \;.
\enq
Using the fundamental functional relation of the $q\Ga$ function, $\Ga_q (x + 1)
= [x]_q \Ga_q (x)$, where $[x]_q = (1 - q^x)/(1 - q)$, we obtain the following
functional relation for the periodised dressed phase,
\beq
\ex{\vp (x, z) + \vp (x \pm \i \eta, z)} \; = \; \Big( \frac{\sin(x - z)}{\sin(x - z \pm \i \eta)}  \Big)^{\pm 1}  \;. 
\enq

\subsubsection{\boldmath The case of `wide' auxiliary argument $|\Im(z)|>\eta$}

When $|\Im(z)|>\eta$, the dressed phase and its periodised version coincide. 
It is readily seen that the dressed phase has its Fourier coefficients given by  
\beq
c_n\big[ \phi(*, z) \big]  \; = \;   \eta \e{sgn}\big(\Im(z) \big)\cdot \de_{n,0} \; + \; 
\big( 1- \de_{n,0}\big) \cdot \f{ 2 \sinh(2n\eta) \ex{-2 \i nz} }{ n \cdot \big( 1 + \ex{- 2|n|\eta} \big) }
\Big( \de_{n < 0} \bs{1}_{\Im(z)> \eta} \; - \; \de_{n >0} \bs{1}_{\Im(z) < -\eta}  \Big) \;.
\enq
The Fourier series for $\phi\big( \la,z \big)$ and $\phi\big( \la, \ov{z} \big)$ can, in fact, 
be re-summed %and re-cast, 
for $\Im(z)>\eta$, as
\beq
\phi\big( \la,z \big) \; = \;  \vp\big( \la, z \big)   \; = \;
 \ln \bigg( \ba{c}  \ex{-2 \i (\la-z)} ; q^2\ex{-2 \i (\la-z)}  \\
 		 q^4  \ex{-2 \i (\la-z)} ;  q^{-2}  \ex{-2 \i  (\la-z)}   \ea \bigg)_{\infty} \; + \; \eta \; = \; \ln \bigg( \f{\sin( \la  - z)}{\sin(\la  - z +\i \eta)} \bigg) \:,
\label{ecriture propriete reduction dressed phase wide argument 1}
\enq
while, for $\Im(z)<-\eta$, it can be re-summed as
\beq
 \phi\big( \la, z \big)  \; = \;  \vp\big( \la, z \big)  	\; = \;
 \ln \bigg( \ba{c}  q^{-2} \ex{2 \i (\la-z)};  q^4 \ex{2 \i (\la-z) }   \\
 			q^2\ex{2 \i (\la-z) } ;  \ex{2 \i (\la-z)}   \ea \bigg)_{\infty} \; - \; \eta
\; = \; \ln \bigg( \f{\sin(\la  - z -\i \eta)}{\sin(\la  - z)} \bigg)	\;. 
\label{ecriture propriete reduction dressed phase wide argument 2}
\enq

\end{document}